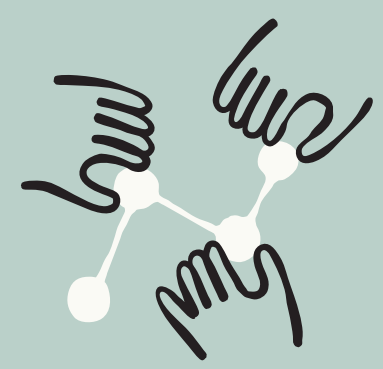

ANTHROP\C

# An evidence review of worker retraining

**Authors**
David Roodman, Independent
Maxim Massenkoff, Anthropic

## Acknowledgements

We thank David Fein, Richard Hendra, and Evan Rose for invaluable comments on an earlier draft and Shad Ahmed, Jeremy Avins, Jon Baron, Michael Bernick, Adam Farina, Martha Gimbel, Rebecca Hiscott, Anders Humlum, Benjamin Hyman, Tamara Johnson, Molly Kinder, Sang Lee, Frieda Molina, Kathleen Nazar, Laura Peck, Kerry Persen, Sarah Pollack, Mette Rasmussen, Santi Ruiz, Szymon Sacher, Chase Sackett, Kelsey Schaberg, Monika Tuchowska, Garrett Warfield, Kim Withee, Heather Whitney, Nathan Wilmers, Peter McCrory, and Jack Clark for materials, conversations, and feedback.

Code and data for this report are at github.com/droodman/job-training-meta-analysis. A web interface onto many more meta-analytical results is at droodman.github.io/job-training-meta-analysis.

**Abstract**: We review the evidence on subsidized job training programs in industrial countries. We provide critical, narrative summaries of randomized studies of major job training in the US; of two "judge randomization" studies in the US and Denmark; and of previous meta-analyses. We then perform a meta-analysis of 56 randomized trials of job training in the US since 1973, the most exhaustive meta-analysis in this niche. Here, average impacts are positive but modest: employment rises 1.7 percentage points in years 3–5, and annual pre-tax earnings by roughly $800 per person offered training. At an average cost of $13,598 per participant, programs roughly break even under (debatable) assumptions about how long benefits persist. They recoup about three-quarters of their cost for government through higher tax revenue and lower benefit use. Evaluations of the largest federal efforts—JTPA, the Workforce Investment Act, and Job Corps—return small or null effects. One cluster of programs stands out: "sector programs," which screen applicants intensively, design curricula with employers, and track local skill demand. They raise earnings roughly ten times as much. They have proven difficult to replicate.

# Executive summary

Among experts and the public, worker retraining is the most popular policy for mitigating labor market disruption from AI.[1] In this report, we ask if these efforts would work. We present a targeted review of the evidence on job training and a brief history of key programs.

We screened the literature for high-quality randomized trials, drawing on 146 impact estimates from 56 randomized trials conducted in the United States from 1973 to the present. The programs trained job seekers—mostly low-income adults and youth—for positions such as nursing aides, IT support technicians, and welders. On average, they run six months and cost roughly $13,000 per person.

Overall, we find:

- **Job training programs have small impacts**. Across randomized trials in the US, programs on average increased employment by 1.7 percentage points, compared to a baseline employment rate of 63% in the control group. They increase earnings by $800 per year on average. These figures are per person *offered* treatment, which matters because compliance in these studies is often low. The programs pass a cost-benefit test, but this relies on uncertain assumptions about how long the benefits last.

- **Scaled programs have also been lackluster.** Large programs in the US—with the scale required to buffer widespread labor market disruption—have been rigorously evaluated. They mostly come up short. An evaluation of the main federal training system in the 1980s yielded impacts comparable to the averages above. An evaluation of its successor and the youth-targeted Job Corps found essentially no impact.

- **Some "sector programs" boost earnings 10 times as much.** A few programs have lifted pay by $5–10,000 per year by intensively screening applicants, involving employers in deciding what to teach, and tracking local trends in the demand for skills. From the government's perspective, these programs clearly pay for themselves through higher tax revenue and lower use of public benefits.

- **Sector programs can contribute to a crash response to AI disruption, but face an uphill battle**. The impact of sector programs is subject to several qualifications:

    - *They filter out >80% of applicants*. Their admissions processes select for workers that employers would probably not find on their own, but who have the grit and smarts to succeed.

---

[1] Karger et al. (2026), Figure 16.

- *They impart skills teachable in months*. Speed holds down costs and gives managers more certainty about what skills will be in demand at graduation. The programs are not designed for skilled professionals who may need years of retraining to recover their previous incomes and who already know how to navigate the job market.

    - *Attempts to set up sector programs often fail*. It can take years to cultivate relationships with local employers and develop competence in tracking regional trends in the demand for skills. A speedy expansion could copy the form and lose the essence.

- **Scaling and randomizing the most promising programs would bring needed lessons.** In applying these lessons to AI, we are, of course, forced to extrapolate into a highly uncertain world. We don't know if or when technological advances will cause a surge in unemployment; it's unclear which jobs would be most affected, and the unemployed population may look different from the subjects in these studies (mostly low-income or persistently jobless individuals).

  But it seems doubtful that we currently have programs capable of meeting the moment. None of the large government-led training programs have produced large and durable benefits, and the few shining sector programs—a likely option given the excitement around them from top experts[2]—arise from a bundle of deep relationships with employers, selectivity, and high upfront costs. A "fire drill" evaluation meant to rapidly scale and test leading job retraining programs could pay lasting dividends.

[2] See, e.g., Katz et al. (2022).

# Overview

Americans' top concern about AI is job loss (Anthropic 2026). If current data on AI usage are any guide, college-educated workers—software developers, paralegals, accountants—are most at risk (Massenkoff and McCrory 2026). But this is highly uncertain. Fast advances in robotics and autonomous vehicles would affect different jobs. And the employment of less-educated white-collar workers, like those in data entry or customer service, might fall sooner (Richmond 2026).

In the event of widespread labor market disruption, displaced workers, regardless of their occupation, would need to quickly find jobs outside of their experience and training. An ambitious government, acting swiftly as in the COVID pandemic, would likely consider expanding retraining programs. Indeed, in a recent survey of economists, AI experts, and superforecasters, "Retraining support" was the most favored policy response to AI-driven job loss, beating out enhanced unemployment insurance, a jobs guarantee, and universal basic income (Karger et al. 2026).

This report takes this possibility seriously by reviewing what we know about how well training has worked in the past. We hope to help calibrate people's expectations for a large scale-up of retraining programs in the event of rapid and disruptive AI diffusion, and to identify the most pressing areas for research.

We find a substantial body of training initiatives to study. Since the Kennedy Administration, the federal government has spent billions of dollars yearly on job training. In 2017, the US government spent $14 billion per year on job search assistance, counseling, and worker retraining, and reached an estimated 10.7 million people (GAO 2019, Figure 4).

The track record for this spending, as we shall see, is mixed. A recent White House report described these programs as "largely ineffective" (CEA 2019). But a few private programs have achieved remarkable results by working closely with employers, and these results have been validated by high-quality, randomized studies.

The body of evidence and experience is of course an imperfect match for an AI-disrupted future. If layoffs mainly affect white-collar workers, the new raft of long-term unemployed would be the most educated and highly paid in history. These workers on average have more liquid savings and know how to navigate the job market (Manning and Aguirre 2026). If they do return to school, it could take them years to reskill. The evidence gathered here is about training programs that last weeks or months and mostly target low-income people on the margins of the job market. Retraining for the highest skilled workers may look more like a compressed mid-career master's degree, and future research should consider how traditional degree programs could be expanded and repurposed for this scenario. Still, we see durable lessons in the programs we study here—and the bulk of retraining efforts may yet concentrate on workers in exposed lower-wage roles.

This project is not purely a systematic review, nor more specifically a meta-analysis. It was not structured from the start to mechanically filter and synthesize findings—though we do ultimately conduct a meta-analysis. Our primary approach is to make intensive use of our judgment to think critically about the findings and claims in individual studies and the portrait they paint. Alongside deep assessments of the findings in the best studies in this area, we provide a history of job training initiatives and evaluation in America. By carefully digging into the details of program implementation, we draw lessons about program design and challenges that go beyond headline effect sizes. In particular, in this review we:

- *Impose a high evidentiary standard*. We favor randomized trials, along with studies that make prima facie plausible claims to frame compelling natural experiments. Designs in the latter category include regression discontinuity and "judge randomization."[3]

- *Aggressively search for studies meeting this standard*.[4] We find 7 randomized or discontinuity-based studies of interventions in Europe involving training; a pair of judge-randomization studies from the US and Denmark; and 56 randomized studies in the US.

- *Synopsize the US experiments—146 impact estimates from 56 studies—with a new meta-analysis*. Since most of the experiments are documented by long reports submitted to government agencies, Claude is used to extract traits and impacts. No previous meta-analysis has included nearly so many US-based randomized studies.

- *Individually review the most important national studies in the US*, along with the European and judge-randomization studies, to surface key challenges in program implementation and assessment.

- *Engage in more depth with the research on "sector programs,"* which intensively screen applicants, work closely with employers on curriculum design, and have some of the most promising results—in some cases boosting earnings by $5–10,000/year.

---

[3] We do not claim to have perfectly applied our "high evidentiary standard." Notably, we do not cover any matching studies even though some are arguably as credible as research we include. Circa 2000, a lively debate played out among economists over the reliability of matching methods (Dehejia and Wahba 1999, 2002; Smith and Todd 2005). Both sides analyzed job training. All agreed that a matching study is *more* credible when it passes a placebo check, estimating null impacts on pre-treatment outcomes; and that this is especially the case when the pre-treatment time series is long. Still, the literature continues to question placebo-passed matching (Arceneaux, Gerber, and Green 2010; Griffen and Todd 2017). A good recent example of such matching, which we do not include, is the Rothstein et al. (2022) evaluation of training programs in California. Such studies arguably deserve as much weight as Hyman (2018), which we cover in §5.1, and whose (partial) credibility rests in part on a placebo check.

[4] "Aggressive" searching included scanning reports that were not previously online, at the Library of Congress and the Department of Labor library; contacting staff at organizations that produced certain reports, in search of copies; tracking down authors now in retirement; scouring the 2005 *Digest of Social Experiments*; and following up on studies listed in earlier reviews as in-progress.

The main findings follow.

### 1.1 In the US, average impacts are positive but not transformative.

In the randomized studies in the US, the impacts on employment and earnings range from negative to positive, but tilt positive. Not all of the studies included in our meta-analysis deliver training exclusively; some bundle multiple features. The programs in which training was a *primary* component lifted employment by an average 2.8 percentage points in the second year after randomized assignment to treatment, and 1.7 points in years 3–5. The impacts on pre-tax earnings are $1,139 and $791 per year (in 2025 dollars). All these averages are highly significant, statistically.[5] See Table 2 in §6.5.2. (Impacts are expressed per person *offered* training, rather than per person trained, the two differing when some treatment group members drop out or control group members get into the program.) These and many more meta-analytical results can be explored graphically at this GitHub page.

The raw data on earnings impacts is depicted in Figure 1, which plots impacts in dollars per year against the number of years since the start of an experiment. From each study write-up, the point estimates are extracted at the finest time resolution provided, usually monthly, quarterly, or yearly. Interventions in which training was a primary component—“training-primary” interventions—are given solid circles and the rest, hollow. All estimates are plotted in grey except for a few programs whose dots are connected and colored for prominence—blues for the large federal programs and magentas for the high-impact sector programs; they will be introduced below. One can see that more earnings impacts are positive than negative, and that most studies stop following subjects by year six.

### 1.2 Monetary benefits exceed costs on average, under reasonable but debatable assumptions.

Among the 67% of training-primary interventions whose write-ups include cost information, costs average $13,598 per treatment group member. Fully accounting for the benefits requires projecting the long-run impacts, since most studies observe subjects for just a few years. We estimate that the net present value (NPV) of impacts on earnings, tracking a stream of benefits that lasts to retirement, is $14,146. It rises to $19,525 when including employer-paid taxes and other fringe benefits. (See §6.5.3 and Appendix G.) In light of the great uncertainties in the projection of impacts, this result should be read as a little better than break-even. Interestingly, training programs ultimately generate flows to government equal to about three-quarters of their costs, through higher tax revenue and lower use of public welfare. So when government funds them, the long-term fiscal cost is only ~24% of the upfront cost.[6]

[5] Each of these results comes from a somewhat different set of underlying studies, since not all report impacts for both outcomes and both timeframes. The averages are random-effects means. The choice of timeframes is adapted from Card, Kluve, and Weber (2018), which splits follow-up into year 1, year 2, and beyond. Year-1 effects are omitted for concision because they often contain artifacts of being in training as distinct from having been trained.

[6] The payback is substantially lower if Social Security contributions are not viewed purely as taxes, but as securing future benefits for which the government is liable.

### 1.3 The impacts of large-scale US programs are as small.

The US government has commissioned randomized evaluations of three national workforce programs:

- In the late 1980s, the Job Training Partnership Act (JTPA) lifted employment among low-income adults by about 2.3 points, from a base of about 70%.[7] It lifted annual pay by $1,100 in 2025 dollars, from about $13,000 for women and $17,500 for men. There were no clear benefits for low-income youth (GAO 1996).
- The Job Corps is an intensive residential program for young people that was enacted as part of the 1960s "war on poverty." It did not affect employment or earnings in follow-ups extending 20 years, except for a transitory 1–2 percentage point employment bump in years 3–5 (Schochet 2021).
- An evaluation of the Workforce Investment Act, the successor to the JTPA, returned essentially zeros for the programs for low-income adults and for dislocated workers (Fortson et al. 2017). Estimates were as often negative as positive and lacked statistical significance. One challenge was a low uptake differential: many people offered training did not take it, while many in the control group found training anyway. As a result, the treatment-control difference in participation was only 15 percentage points.

The earnings impact estimates from these studies are highlighted in blues in Figure 1.

### 1.4 Average effects are higher per trainee than per person offered training.

The evidence we draw from is muddied by non-compliance, which is when treated people drop out of the program or the control group receives training. In our collection of trials, treated individuals are 66 percentage points more likely to receive the *targeted* training and only 28 percentage points more likely to receive *any* training compared to the control group.

Our main estimates give the average impact of an offer, regardless of whether people complied. This tells us about a program's effectiveness in its context. It's arguable though that we should imagine a world with fuller compliance: much higher participation among those with offers, and much lower participation among those without due to a shortage of programs. We can calculate impacts per trainee from the existing studies by scaling up the estimates with the fraction impacted by the offer. We find that this estimate, likely an upper bound, is about 2–4 times higher than our main estimates, although it does not change the benefit-cost picture.

### 1.5 The evidence on programs to help dislocated workers—whose unemployment or low pay is more attributable to large-scale economic changes such as deindustrialization—is sparser and mixed.

- A Denmark study that exploits arbitrariness in assignment to caseworkers—by day of month of the client's birth—finds that an average 52 days of classroom-based training helped people change occupations; the effect was possibly larger for people coming out of occupations exposed to offshoring. The study does not detect similar effects for on-the-job training (Humlum, Munch, and Rasmussen 2025).

---

[7]Figures approximate average results for years 3–5. See Figure 4.

- A similar study of Trade Adjustment Assistance (TAA) in the US, but with a weaker claim to arbitrary assignment, reports that workers who could access the support did go on to work more and earn more. Sometimes they did so by changing occupations or locations. Since the assistance included both training and an extension of unemployment insurance, the effects of training per se cannot be distinguished (Hyman 2018).

- As mentioned, the evaluation of the 1998 Workforce Investment Act's (WIA) dislocated worker program essentially returned zeros. The experiment did not include people eligible for TAA (Fortson et al. 2017).

- Randomized studies of pilot programs for dislocated workers in New Jersey (Corson et al. 1989) and Texas (Bloom 1990) find that a combination of job search assistance and training was no better than job search assistance alone for reducing unemployment within 12 months.

We also find suggestive evidence that some programs that bundle other features alongside training perform well. Trainings backed by a hiring commitment from employers increase the earnings of admitted students an order of magnitude more than the average program, although this is driven by just three programs.

### 1.6 Some "sector programs" stand out, with large and sustained benefits.

In the evidence base on US programs, one cluster stands out with programs that easily pay for themselves. These programs work closely with employers, screen job seekers, teach them occupational and work-readiness skills, and place them in quality jobs in specific high-demand industries. Experts have converged on "sectoral strategy" and "sector program" as labels for these training approaches (Conway and Giloth 2014, Katz et al. 2022).[8] Some of them, like Year Up (Fein and Dastrup 2022) and Per Scholas (Maguire et al. 2010; Schaberg and Greenberg 2020), are marked in shades of magenta in Figure 1.

---

[8] It is difficult to offer a pithy definition. As Conway and Giloth (2014) observe, "the range of tactics and the variety of institutions that may be involved in an initiative make drawing a tight boundary around the field of sectoral employment development challenging." A helpful characterization comes from Katz et al. (2022): "Sectoral employment programs train job seekers for 'high-quality' employment in specific industries and occupational clusters that are believed to have strong current local labor demand and opportunities for longer-term career advancement… The programs attempt to forge strong employer relationships, do some upfront screening of applicants, combine soft skills (or work-readiness) training with occupational skills training, are involved in job development and placement, provide wraparound support services to help participants complete the program, and often include follow-up services to participants after program completion and to employers after job placement." Hendra et al. (2016) are more specific about employer involvement: "To qualify as a sector program, an initiative must bring together multiple employers in a given field to collaborate on developing a qualified workforce." In our meta-analysis, we use employer involvement with the training curricula as a key distinguishing factor.

Our sample of randomized studies covers the sector programs marked in magenta as well as others that have not performed as well, at least within the evaluation periods. As a group, the sector programs cost no more than training-primary programs—$11,602 versus $13,598 per treatment group member, on average. Yet the discounted present value of their impacts on pre-tax earnings comes to $60,319 instead of $14,146. The 7x benefit-cost ratio is not as large as that for the most successful ones. But it is still enough that if the government successfully ran such programs, they would pay for themselves through higher tax revenues and lower use of public benefits.

Sector programs trained people for a wide range of jobs, including IT support and computer repair, software development, medical assisting, nursing, medical billing, construction, building maintenance, manufacturing, accounting, and bookkeeping. While more research is needed to determine which traits are key to their success—and not all sector programs have produced such impressive results—several factors probably matter:

- The programs are selective. An evaluation of Per Scholas and three other programs reports that the programs take about 1 in 5 applicants (Hendra et al. 2016, p. 34). Demanding application processes, with multiple interviews and/or take-home work, select for people who have the need, grit, ability, and family support to throw themselves into this challenging experience.

- Over timescales measured in months, the programs teach both vocational skills, such as would be needed for entry-level work at data centers and hospitals, and “soft skills” in comportment in job interviews and the workplace.

- The programs work hand-in-glove with employers to determine what skill gaps they need to fill now, and even to set up hiring pipelines. The Wisconsin Regional Training Partnership in Milwaukee (Maguire et al. 2010), for example, ran 2- to 8-week trainings in response to specific requests from employers.

- All are privately run. Among other things, this makes it easier for them to screen applicants and adapt curricula to changing market conditions.

Successful sector programs should be seen in part as *arbitrageurs* or matchmakers. They achieve outsized returns by connecting employees and employers who would not otherwise have connected: employees who might not have imagined that they could work at JPMorganChase or Amazon, employers who have not committed the resources, or developed the skills in-house, to effectively recruit from low-income demographics.

For arbitrage to pay off takes both skills and opportunity. The skills are uncommon: the best industry-partnership organizers blend a social worker’s desire to help the unemployed with an HR professional’s responsiveness to the needs of a large employer (Dedrick 2014). As for the opportunity, it arises when there exist employees and employers who *can* work together *and yet* would not find each other on their own.

Even with similar approaches, the conditions for success are not universally present. An attempt to replicate one strong program, the CET in San Jose, failed at all 14 trial sites, including 4 run by the CET itself outside of San Jose (Miller et al. 2005). Only two replications of the sector approach have boosted clients' incomes in a randomized trial: the expansion of Year Up outside its Boston home (Fein and Dastrup 2022); and St. Nicks Alliance in New York, one of three new sites in the WorkAdvance demonstration (Yusim et al. 2025, Table 3).

**Figure 1. Earnings impacts in 56 randomized evaluations of job training programs in the US**

*Notes: All impact estimates are extracted and plotted at the finest temporal resolution in the primary sources, and expressed in 2025 dollars per treatment group member per year. Results from federal programs are colored in blues and certain selected high-impact "sector programs" in magentas. Within each color family, the shades serve only to distinguish programs. All other results are plotted in grey. Interventions in which training was a primary component get solid circles, and the others, hollow. Each colored series is labeled directly at its end. Per Scholas was the subject of two studies. "JTPA" is the Job Training Partnership Act. "WIA-low income" and "WIA-dislocated" are two service tracks under the Workforce Investment Act. The JTPA impacts for young people and the WIA impacts are not statistically significant. For legibility, a high value from the seminal Wildcat experiment in the 1970s is omitted.*

### 1.7 The sector approach can probably help society respond to AI's disruption of work. But the approach is not formulaic and will probably fail if treated as such.

The challenge of replicating sector programs would surely be compounded in any effort to scale them up on a crash basis, whether led by public or private funders. In pondering the failures to replicate the CET, the evaluators noted that "CET-San Jose [has] grown organically over 20 years, with an unusually committed founder and staff, [and] very strong ties to the local community." As a result, "perhaps a homegrown model like CET cannot be easily exported in a top-down way to other areas" (Miller et al. 2005).

Properly executed, an effort to expand sector programs could explore these possibilities, working with organizations with established competence. Scaling any training program is not guaranteed to work. But that risk implies that we should supercharge effective experimentation now, rather than wait for an employment emergency. If effective training requires organizational legitimacy and deep ties to employers, it is better to start scaling successful programs now than wait for a single, top-down federal response. A rigorous approach would pay dividends even by cataloging the failures.

The next section of this report provides background information. Section 3 then delineates some cross-cutting critical themes. Section 4 gives special attention to a historical sequence of four major American randomized evaluations of job training, which provide most of our high-credibility evidence on training at scale. Section 5 turns to certain non-randomized studies that make prima facie claims to high-credibility impact estimation. Section 6 reviews past meta-analyses and then introduces a new one, of randomized studies in the US. Section 7 covers the history and evidence on sector programs. Section 8 concludes with thoughts on how to extrapolate from this evidence base to a future in which the world of work is disrupted by AI. In Appendices C and D, we review the limited high-credibility evidence from Europe.

# 2 Background

## 2.1 Types of training

Governments and private organizations have employed many techniques to help people get jobs, and get better jobs. Among them:

**Career Navigation**

1. Maintenance and sharing of job listings.
2. Assistance with the mechanics of searching for a job.

**Direct workforce training**

3. Teaching "soft skills," such as how to dress for and comport oneself in a job interview and in the workplace.
4. Classroom teaching in academic fundamentals (adult basic education).
5. Teaching vocational skills.

**Work-based learning**

6. On-the-job training.
7. Apprenticeships.
8. Subsidies for private-sector employment.
9. Direct public employment of people who need work.

We will generally limit our review of job training programs to items 3–6. However, this conception contains a few complications:

- In practice, items 6 and 8 are hard to distinguish. Commenting on an earlier draft of this report, Richard Hendra of MDRC told us that most "on-the-job training" in the US rarely involves formal training. Of course, even without a curriculum, subsidized employment *can* educate people and move them into new careers.
- Most workforce programs mix approaches—e.g., perhaps job search assistance coupled with a payment for training. In the meta-analysis in §6.5, we will distinguish roughly between "training-primary" programs, in which training is a primary feature according to Claude's reading of reports, and training-secondary ones.

In general, we are interested in programs to help people move to new occupations, usually by investing in their skills in some way. Several of the left-out items certainly deserve further attention. The US apprenticeship program is large, with 678,014 apprentices active in 2025 (ETA 2025). But in our initial searches, high-quality evidence on their effectiveness was scarce. Also, cutting-edge job matching approaches might be a crucial (and low-cost) tool for dealing with labor market disruption.

# 3 Critical themes

Most of the studies we cover exploit randomization in order to rigorously estimate the impact of job training on employment and earnings. But even the methodological armor of randomization has chinks. In learning from the research, it is important to think critically about the ways that studies may fail to measure what they aspire to measure (what is called internal validity) and how well their results generalize (external validity).

Before exploring the landscape with a magnifying glass, this section therefore inventories some cross-cutting, critical themes from a greater distance. We review more related themes in Appendix A, including whether participants' gains come at the expense of other workers (A.4) and whether training pays off by building new skills or by "signaling" pre-existing talents (A.6).

## 3.1 The importance of randomization

To know if a social program works, you need to somehow measure how participants would have fared in the absence of the program. Often, researchers compare people's incomes before vs. after the training. This is fraught since people might sign up specifically because their earnings have taken a hit; most of them will see their income recover regardless of training. (This confound is called Ashenfelter's dip. We provide a brief review of this intellectual history in Appendix A.2.)

The most powerful fix is the simplest in concept: randomization. If an experimenter flips a coin to decide who gets access to a training program, then the treated and control groups have matching past trajectories on average. Experience has shown that without this feature, even studies with aggressive controls and cutting-edge empirical strategies can lead readers astray. For years, cardiologists took vitamin E because careful observational studies found a persistent correlation between dietary vitamin E and heart health; randomized trials ultimately ruled this out (Lee et al. 2005; Abner et al. 2011). The negligible benefits found in a recent large-scale study of guaranteed income were startling in part because a myriad of non-randomized studies had found that just a little cash would go a long way (Miller et al. 2024). And even one of the shining sectoral programs we review has put out annual reports with non-causal estimates suggesting an 8x larger wage gain than was recorded in its randomized trial.

This is why we devote most of our attention to randomized evaluations, and a few observational studies with a claim to pseudo-randomization. There are some newer programs with suggestive observational results. But without randomized evidence, and given the long trail of initially promising failures, we might default to "ineffective until credibly proven otherwise."

While randomization is a powerful tool for causal inference, even randomized studies need to be read and interpreted with a critical eye. Much of the rest of this section reviews critical themes that apply even to the best job training evaluations.

## 3.2 Imperfect compliance

Random assignment to job training does not guarantee perfect compliance, when everyone in the treatment group participates and no one in the control group does. Instead, an offer might have increased participation from, say, 25% (because one quarter of the control group sought out the program) to 50% (because half of the treatment group ignored the offer), just a 25-percentage-point increase in participation.

This complicates things. Without full compliance, researchers fall back to two notions of causal impact: the intention-to-treat (ITT) effect or, with additional assumptions, the local average treatment effect (LATE; Imbens and Angrist 1994).

The ITT is the simple difference in outcomes between those who did and did not receive the offer, regardless of participation. So all else equal, when the impact on participation—25pp in the above example—is smaller, so is the ITT. In a market saturated with effective training programs, the ITT may be small just because the control group sought other similar programs.

The LATE measures the impact on people whose participation was determined by the offer, calculated by dividing the ITT by the difference in participation. So if our experiment increased participation in training by 25pp, and its ITT was a $500 increase in income, the LATE is ($500/25pp =) $2,000. Assuming the offer could only help you by exposing you to training, this

wedge of people must have been driving the $500 increase, so their benefit must have been $2,000.[9]

Which is more helpful for understanding these impacts under labor market disruption? ITTs might be the best measure of a program's overall impact in its context: if a program fails to recruit more people into a training, it may well be ineffectual (especially if the program and its competitors are about equally effective).[10] But the LATE could indicate a program's potential to help in a context where it is more needed and wanted (provided benefits do not fall much as participation broadens to a larger swath of people). This could make the LATE a better fit for modeling a world of AI-driven job displacement: one could imagine a limited supply of training and much more readiness to train among those offered seats. On the other hand, this readiness might also increase training receipt and general resourcefulness among those not offered spots.

A final complication of the LATE is that sometimes study subjects participate in a separate training program, so that there are different ways to define it: the impact of the targeted training or the impact of any training at all (Kline and Walters 2016).

We default to the ITT in what follows since it's easier to define and seems just as likely to fit the future labor markets we imagine given the high uncertainty. But we also report the average increase in participation to convey how available training was in these different contexts, and so that interested readers can estimate the benefits in a world of fuller compliance.

## 3.3 Attrition

One econometric bugaboo not banished by randomization is the possibility of *selective attrition*—i.e., disappearance of subjects in a way that throws off the comparison between treatment and control groups. For example, some studies have tracked how much people work and earn over the years simply by asking them in follow-up interviews. But if people are less able or willing to talk to surveyors when they are out of a job, or in jail, or relapsing, then the people doing better in the labor market will figure disproportionately in the data. The unplanned narrowing of the sample could also narrow the apparent differences between treatment and control—an instance of negative attrition bias. One can also spin scenarios leading to positive attrition bias; it depends on who responds to the surveys. Thus, attrition bias is a known unknown. We don't know how things went for those who exited the study early, so we don't know how their presence would have affected the results.

Fortunately, in job training research, one substantial remedy is available: government databases. In the US, state-level unemployment agencies, along with the Internal Revenue Service and the Social Security Administration, levy taxes on wage income, which come out of paychecks. And since the mid-1990s, the National Directory of New Hires has aggregated such data in order to track people who owe child support. Researchers frequently (and confidentially) link subjects' social security numbers to such databases, with minimal attrition.

[9] This causal estimate only applies to a subset called the compliers—not people who would have always or never participated regardless of the offer. Their benefits might differ. Also, the LATE requires an assumption that the offer only affected outcomes through its impact on training.
[10] See also Heckman et al. (1999) Section 5.2.

While wage databases cover nearly everyone, they don't paint a complete picture of a person's work life. Some pay passes under the table. Gig workers don't pay unemployment taxes so they don't show up in unemployment databases. But people probably leave out at least as much information when being interviewed. Overall, linking to government data substantially dispels the worry about attrition bias (Ashenfelter 1978), but many studies—including half of the sector program evaluations—are limited to surveys.

## 3.4 Academic vs non-academic publication

It is well recognized that high-powered incentives in academia distort the findings that make it into journals (Christensen and Miguel 2018). In extreme cases, there is fraud. Probably more common are the phenomena of *p*-hacking (Simonsohn, Nelson, and Simmons 2014) and publication bias (Sterling 1959). Each is a filtration, one before submission to a journal and one after.

The evidence reviewed here is relatively undistorted, for two reasons. First, most of the studies are of randomized trials. Their mathematical simplicity—comparing averages in the treatment and control groups—reduces (but does not eliminate) the discretion in running the numbers, thus the scope for mining for statistically significant results. In addition, randomized experiments are more expensive to run than desk-based analysis of downloaded data sets, and the sunk cost reduces the chance that the results will wind up in a file drawer.

A final reason for low filtration is related: most studies reviewed here were funded by government agencies or foundations and carried out by non-academic research firms; for these entities, academic publication is secondary. Indeed, for the large majority of the studies covered here, the terminal publication is an evaluation report that was posted online or submitted to a government agency.

## 3.5 Success is ephemeral

Unlike, say, unemployment insurance, training programs constitute a variegated and nebulous class. They differ in where they operate, how they recruit, who they teach, what they teach, how they teach, how long they teach, and who is teaching.

Organizations that run training programs are diverse too, in ways just as subtle. Reviewing research on attempts to replicate workforce programs that were deemed to "work" in their original instantiations, King (2014, pp. 211–13) lists 16 factors that can make or break a replication. Many cannot be summoned by bureaucratic fiat: "strong, shared leadership," "a record of commitment to developing and implementing innovative practices," "strong advance planning and ongoing communication among partners." Organizations are in turn embedded in contexts that bring to bear certain incentives—political, administrative, financial. This is one reason that scaling a program to serve more people can completely change how it functions.

In the realm of job training, an example of the power of incentives comes from the history of the Job Training Partnership Act of 1982. As will be described in §4.2, the JTPA delegated the design of workforce programs to local actors while holding them accountable to performance standards. The strategy distorted implementation and fostered confusion among those overseeing it from afar. An act of Congress could not overcome Goodhart's Law: when a measure becomes a target, it ceases to be a good measure.

During the experimental evaluation of the JTPA, the research team discovered that sometimes after a person was randomized into the treatment group, the local agency, called an SDA, would hold off recording them as "enrolled" in the JTPA until they had found a classroom or workplace training spot, and sometimes even until they had demonstrated good attendance in the training. "Many SDAs have believed that they have discretion in defining the point at which individuals 'count' in their performance measures and have responded to the system's incentives by delaying enrollment" (Orr et al. 1994). And if staff responded to incentives after enrollment, surely they did before. With limited funding (Orr et al. 1994, p. 55), SDAs were subject to a powerful incentive to engage in "creaming," as Senator Pete Domenici predicted during the Senate debate over the JTPA. "That is they will spend the resources provided by the Federal Government to work with people who are temporarily unemployed but who have good skills or significant work experience and could obtain other jobs without retraining." Yet, when the JTPA was criticized in the press—or when the White House asked questions behind the scenes—Department of Labor officials would cite the performance statistics flowing up from the SDAs to show that the program was working.

Empirically, researchers have found that scale-ups (Allcott 2015; Bold 2018) and government-run programs (Vivalt 2020) tend to do worse. In a more recent study, Bhatt et al. (2021) revisit the stunning reductions in crime attained by the Becoming a Man program for at-risk youth (Heller et al. 2017). They conclude: "In the later years with substantially larger program participation levels, we can no longer reject the null hypothesis that impacts equal zero." In our report, we found three promising programs that were deliberately replicated or scaled and subjected to randomized evaluations: the Center for Employment Training (CET), replicated at 14 sites; Per Scholas, expanded through the WorkAdvance demonstration; and Year Up, scaled from three cities to eight. CET failed at every site and WorkAdvance's new sites did worse than the original; only Year Up showed consistent gains.

Thus, it is a live question for this review: if private charities, working with corporations and foundations, strike upon a promising system, can governments bring it to far more people without denaturing it? If key traits for success are not only ephemeral but fragile, then they can be distorted away by a new administrative and incentive regime. Yet there is a place for funders to monitor performance and base funding on it. The challenge is to determine which broad program features matter for success, and then to track and reward local agencies for implementing those features while allowing flexibility on specifics.

# 4 Major American randomized evaluations

For 50 years, the US government has led the way in generating high-quality evidence on the impact of job training. This section reviews a sequence of four major studies that form the spine of this history. The first, with massive federal funding, sought to demonstrate and test the "supported work" approach to helping people enter the workforce. The other three evaluated ongoing federal programs; they speak to how job training has worked at scale (see Appendix A.1).

## 4.1 The National Supported Work Demonstration

### 4.1.1 The initial study

Workforce programs have a storied history in economics. Probably no other subject in economics has inspired as much randomized research. And no other has contributed as much to the "credibility revolution" in economics (Angrist and Pischke 2010), which has elevated norms about evidence when studying causal relationships. (See Appendix A.2.)

The first major randomized experiment in workforce development was the National Supported Work Demonstration (NSWD). It was conceived as a prudent steppingstone between a small but promising program in New York City and a potential nationwide rollout. The method of "supported work" had been refined by the Vera Institute of Justice in the 1970s through years of experimentation—and the world's first randomized trial of job training. The project offered participants 12–18 months of paid work in small peer crews, under close supervision, followed by help with transitioning into unsubsidized work (Vera Institute of Justice 1976).[11] Typical occupations included construction, building maintenance, and daycare.[12]

With leadership from the Ford Foundation, five departments of the federal government joined the foundation in pouring a remarkable $82.4 million into the demonstration and randomized evaluation of supported work (four times that in today's dollars; MDRC 1980, pp. 15, 19; Auletta 1982, p. 47). To execute the demonstration, the funders created a new non-profit, the Manpower Demonstration Research Corporation. In turn, MDRC recruited 14 organizations to carry out the project in their service areas. Most service areas were major cities, including Atlanta and Philadelphia; a few were rural, such as north-central West Virginia. Four were dropped for not implementing the model fully enough. The demonstration ran from March 1975 to December 1978 (MDRC 1980, pp. 22, 30).

[11]The origin of supported work is itself a remarkable story. See the *New York Times Magazine* story, Porter (1974), and the Wikipedia article on Herbert Sturz.
[12]See also Gary Walker, testimony of March 12, 1980, in U.S. Senate Committee on Labor and Human Resources (1980, 342).

*Vocational training* was not a major feature of the program.[13] However, the New York site (the original model) routed some participants into classes in typing and basic workplace skills. Students in one of these cohorts would become the main characters in Ken Auletta's three-part series for the *New Yorker*, and his 1982 book,[14] *The Underclass*.

A separate entity, Mathematica Policy Research, ran the experiment and analyzed the data. The computerized coin tosses sending subjects to the treatment or control arm took place only after they had expressed interest in the program and cleared initial screening. Across 10 sites, 6,616 people entered the experiment, evenly split between treatment and control (MDRC 1980, p. 4). The people assigned to the treatment arm constituted a third of the 10,043 participants in the demonstration. Subjects were purposively drawn from four tough-to-serve populations: women who had been on Aid to Families with Dependent Children (AFDC) for at least 30 of the past 36 months and whose youngest child was at least 6; addicts recently in treatment; ex-offenders recently released from prison or jail; and teenagers who had recently dropped out of school.[15] The last three were 80–94% male. All groups were about 80% black and 10% Hispanic; their members had spent an average of 10 years in school (MDRC 1980, p. 32).

Mathematica tracked employment, wages, and other outcomes by repeatedly surveying the subjects—at baseline (the moment of entry and randomization) and in follow-ups out to 36 months. Several practicalities complicated the evolution of the samples. Not all programs launched at the start date, March 1975, with enough immediate scale to justify flying in an interview team. So the experiment began later in some places. To achieve statistical power, and to avoid an exclusive focus on the programs' first year, when staff are learning rapidly through trial and error, the experiment continued taking in new people through July 1977. Since follow-up interviews largely ceased on March 30, 1979, the samples shrink as one moves to longer follow-ups. The programs that got off to a fast start, notably in Philadelphia and Jersey City, account for more of the 36-month follow-up. In addition, there was attrition (see §3.3): some people didn't show for interviews, didn't agree to be interviewed, or couldn't be found through phone calls or door knocks. This happened more among people in the control group, which is unsurprising since they were less connected to the program. 84% of women in the AFDC group were interviewed at 18 months, and about 70% of participants in the other groups were (MPR and SSS 2004, pp. 21–22).

It appears from partial information that most subjects adhered to their treatment assignment. 95% of women in the AFDC treatment group accepted the offer of supported work. Only 11% in the AFDC control group managed to obtain it; likewise only 3% of the ex-addicts and 4% of the youth (Heckman et al. 2000, Table I). As a result, the LATEs of program participation are close to the ITTs (see §3.2). No information is available on how many control subjects found training elsewhere, making it impossible to calculate LATEs for training per se.

---

[13]"Supported work guidelines did… permit 25 percent of paid time to be used for support services… defined as "work-related." These included orientation, skill training, and job readiness and placement activities. But generally, local programs made only limited use of this feature." (MDRC 1980, p. 24)
[14]Auletta (1981a, 1981b, 1981c).
[15]Precise group definitions are in MDRC (1980), Table 2-2.

Figure 2 and Figure 3 tell the evaluation's main statistical stories: how employment and earnings evolved in the treatment and control groups. In all the plots, the control group curves rise with time since randomization. Attrition could help explain this seeming improvement.[16] The country's improving economy may also have played a role: between quarter 1 (Q1) of 1975 and Q1 of 1979 the unemployment rate for blacks fell from 14.5% to 12.6%.[17] And earnings and employment rise with age in early adulthood. But probably the largest factor is regression to the mean: in a group of people who are down enough on their luck to qualify for the program at a particular moment, it would not be surprising if more of them saw their circumstances improve than not, even absent supported work.[18]

[16]Attrition cannot explain the whole rise. If *none* of the 23% of control subjects who attrited by Q9 (MPR and SSS 2004, Table 5) were employed at that time, this would make the true Q9 employment rate 26.8%, still well above the starting level of 19.4%.

[17]U.S. Bureau of Labor Statistics (n.d.), quarterly averages of the seasonally adjusted monthly series.

[18]One variant of this story is that those in the control group improved their lot on average through *other* employment programs, such as the government's Work Incentive (WIN) and the Comprehensive Employment Training Act (CETA) programs. However, the control group members reported no rise in their participation in these programs (MDRC 1980, p. 56).

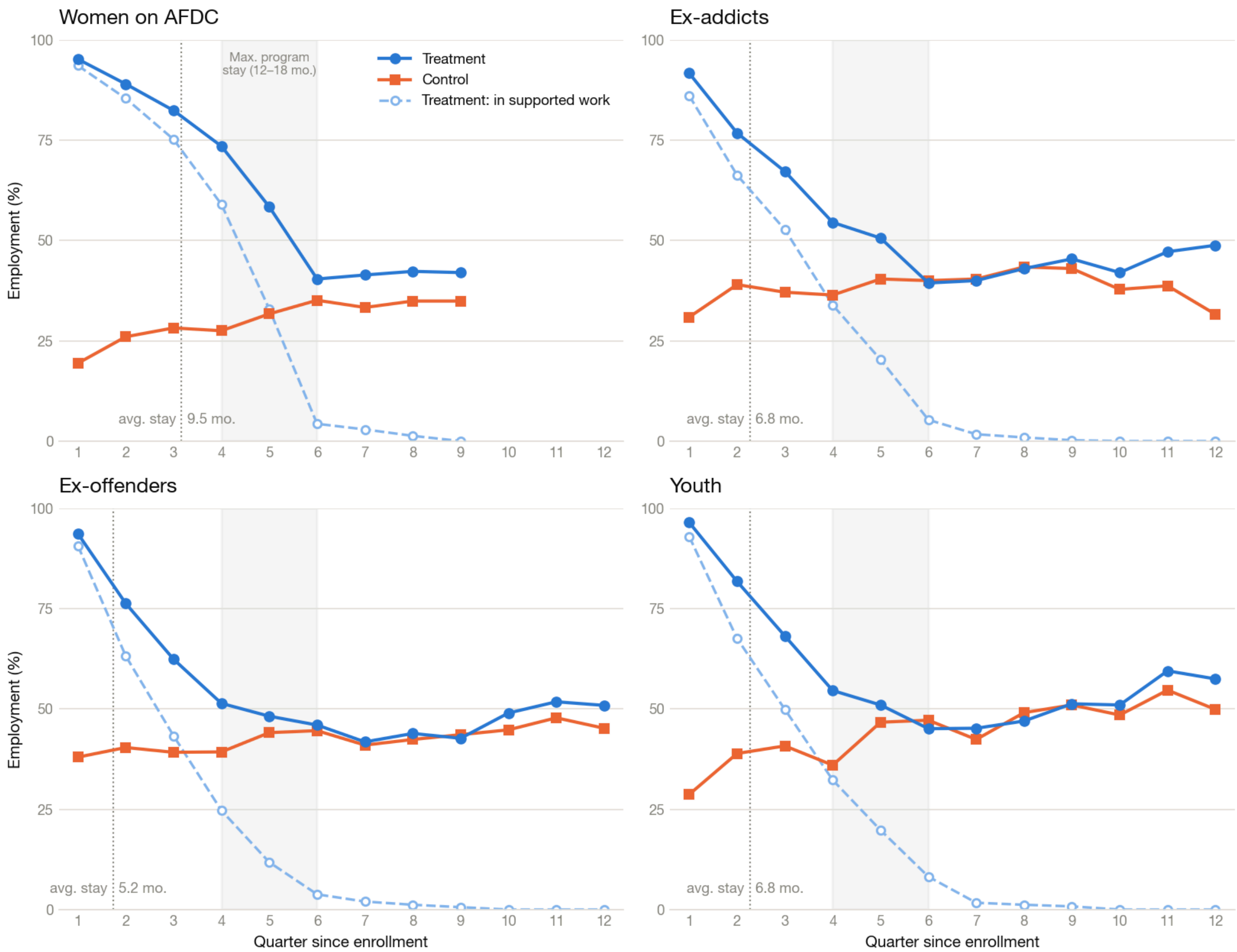

**Figure 2. Employment in National Supported Work Demonstration (1976–79)**

*Notes: This figure shows the share employed in the treatment and control groups in the National Supported Work Demonstration by quarter after enrollment, separately for the four target groups. The dashed line shows the share of the treatment group employed in a supported-work job. Estimates are regression-adjusted for age, sex, race, education, prior work experience, household composition, site, and length of site operation. The shaded band marks the maximum permitted stay in supported work—12 or 18 months, varied experimentally across sites. The dotted vertical line marks each group's average actual stay. AFDC follow-up interviews ended at 27 months (quarter 9); the other groups were followed to 36 months. Source: employment rates from MDRC (1980), Tables 4-1, 5-1, 6-1, and 7-1; maximum stay from p. 35; average length of participation from Table 2-7.*

The solid blue curves in the figures show how the treatment groups fared. Nearly all treatment group members were employed at the start because nearly all were engaged in supported work (dashed light blue lines). But many people left the program within months: they resigned in dissatisfaction, were fired or incarcerated, had to deal with family emergencies, or got a regular job.[19] Yet while the fraction in *supported* work plunges within 18 months, the share with *any* work plateaus higher. For women on AFDC (upper left of Figure 2), the employment curve levels off just above 40%. The treatment-control difference—the ITT—settles at 7.1 percentage points

[19]MDRC (1980), Table 2-7 presents statistics on departure by type for the entire program population, not just the experimental subsample.

(statistically significant at $p < 0.1$). The rest of Figure 2 shows employment results for the other samples. While a delay in the experiment's start for the AFDC group meant that follow-up only extended for nine quarters, the other groups are followed for up to 3 years.[20] All three of these other groups exhibit a peculiar pattern: as people leave the program, the treatment group's curve descends all the way to the control group's—implying no immediate impact—before rising above it somewhat in the final quarters. The MDRC report judges the gap for ex-addicts—17.2 percentage points in the last quarter—to be statistically significant. It considers the impact for ex-offenders "marginal," and perceives no positive result for youth. However, the commonality of the pattern across the three groups suggests that it is more than noise.

The earnings patterns, in Figure 3, are similar. For women on AFDC, supported work boosts pay by $75/month in the final quarters of tracking, or $900/year. Earnings impacts are comparable for ex-addicts and even ex-offenders too, though not with as much statistical significance.

[20]After the main report was released, MDRC did follow up on the AFDC group at 36 months. The data are in public use files (MDRC 2009, DS7) but we have not yet found a write-up, and cannot access the data since we are not affiliated with an ICPSR member.

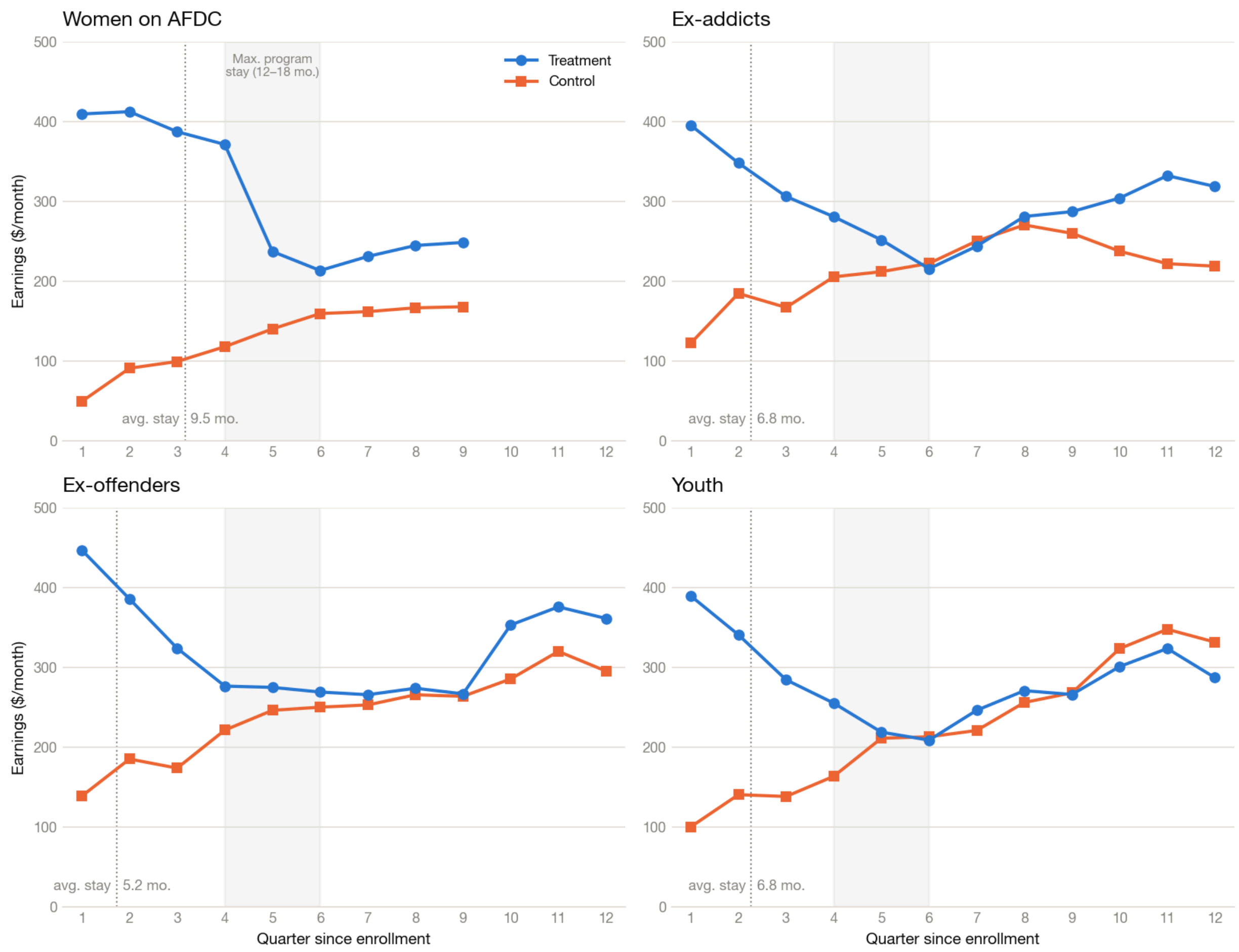

**Figure 3. Earnings in National Supported Work Demonstration (1976–79)**
*Notes: This figure shows earnings in the treatment and control groups in the National Supported Work Demonstration by quarter after enrollment, separately for the four target groups. The estimates are regression-adjusted; see caption to Figure 2. Source: earnings from MDRC (1980), Tables 4-3, 5-3, 6-3, and 7-3; maximum stay from p. 35; average length of participation from Table 2-7.*

If the narrowing, then widening, of the impact in the ex-addict and ex-offender samples is signal, not just noise, what does it signify? It might mean exactly what it seems to mean: supported work helped the predominantly male groups too, if with a one-year delay after exit from the program. But MDRC (1980, pp. 85–88) is unable to confidently explain what would delay the benefit. A second possibility is a compositional effect: perhaps the local programs that increasingly dominated the sample in the final quarters of follow-up, such as Jersey City and Philadelphia, were also more effective. However, MDRC (1980, Figure 5-3) shows that among ex-addicts, the group in which the signal is strongest, the results do not change much if one eliminates compositional effects by restricting to people who completed all interview rounds. A third possibility is attrition bias. The clearest warning sign here would be a sharp drop in survey response between Q9 and Q12. But there was no such plunge in that period (MPR and SSS 2004, p. 21).[21]

With some perplexity, we lean toward the first explanation: the delayed benefits found for ex-addicts and ex-offenders are real.

Overall, the evaluation suggests that as participants exited supported work, a modest but detectable fraction of them obtained better work than they would have otherwise. The impact is clearest for mothers.

### 4.1.2 A later follow-up

After the main NSWD report was published in 1980—and after the Reagan administration all but shut down the demonstration in 1981 (Auletta 1999, p. 316)—one researcher studied its longer-term impacts by linking to Social Security data (Couch 1992). The data set had problems. Some people's earnings information appeared to have been missing in 1980–81, missingness misleadingly coded with zeros (Couch 1992, p. 383). And the data tape Couch received covered only the AFDC and youth samples. Still, the linkage to government data appears to have been relatively free of attrition.

Couch (1992) finds that wage impacts for AFDC women persisted through 1986, with some decay. The non-impacts on youth persisted too. For the women, the earnings bump averaged $441/year (in 1978 dollars) in 1982–86, with high statistical significance (Couch 1992, Table 1). That is about half the $900/year MDRC found through surveys circa 1979.

### 4.1.3 The impacts in perspective

By one standard, the NSWD was a smashing success: it produced new, credible information about who benefits from supported work. The results would not necessarily have been predicted with confidence, so they constituted genuine learnings (Ginzberg, Nathan, and Solow 1984, p. 312).

In a different respect, the demonstration failed, because it helped almost no one. Notice how, in Figure 2, the blue curves for the treatment group drop rapidly from the start: that is people leaving the program without completing it. Even the sustained 8-point treatment-control difference for the AFDC women is overstated in a certain (tendentious) sense, for it pertains to the subset of women who chose to *apply* to the program and thus enter the randomization lottery. A passing statement in the final report (MDRC 1980, p. 43, note 1) suggests that roughly a third of women eligible for the program applied. Thus, the job impact among *eligibles* was about 8/3 ≈ 2.5 points.[22]

---

[21]MDRC (1980, pp. 48–49) says that "Sophisticated statistical tests were applied to the results to check for any bias because of differential rates of response to the interviews. (The tests indicated that the findings were substantially free from response bias.)" This appears to refer to Heckman selection modeling reported in Appendix A of the 18-month report (Maynard, Brown, and Schore 1979). While the analysis is intelligently done, we do not put much stock in it. It only extends to the 18-month follow-up. More important, for reliability, the Heckman model requires that variables be identified that affect survey participation and not the outcomes of interest. But all the baseline variables found to affect survey response, such as being over 35 and not having been incarcerated in the last six months (Maynard, Brown, and Schore 1979, Table A-4), also plausibly predict outcomes. On the importance of exclusion restrictions even when the model is formally identified without, see Roodman (2011, §2.2).

Ken Auletta's *The Underclass* imbues these small numbers with living texture, by sharing what he saw in supported-work classes in New York:

> In the first several weeks of the training regimen, the enrollment… dwindled from twenty-two to sixteen. Ronald Brooks was arrested; Phillip Rivers, Earl Billings, Larry Pearl, and Stanley Lawrence were dismissed from the program for excessive absences and lateness; Liza Lance quit, because, Smith said, "she doesn't feel she can make it—she feels that everyone in the class is smarter than she is." [p. 26]

Later, Auletta updates us on the academic survival rate. Five more left because of "poor attitudes." "In each case, lack of patience or motivation or an enfeebled sense of self played pivotal roles" (p. 276).

The balanced take here is that a) boosting employment 8 points in a group chosen precisely because it was hard to help is an achievement and b) the intervention did not come close to working for everyone.

### 4.1.4 Benchmarking non-randomized evaluations of the NSWD

In the early 1980s, Senator Dan Quayle spearheaded, and Ronald Reagan signed, the Job Training Partnership Act, whose evaluation will be discussed just below. Like predecessor laws, the JTPA ordered the Secretary of Labor to evaluate the programs authorized by the act (Public Law 97-300, §454). For guidance, the Department of Labor (DOL) assembled an advisory panel, which commissioned papers from academics and professional evaluators (JTLSRAP 1985; Stromsdorfer 1987). The inquiry focused on whether the DOL ought to randomize its evaluation of the JTPA.

The data from the supported work demonstration became a touchstone in the debate. In turn, that debate ultimately influenced our decision to heavily favor randomized trials in this review.

Several researchers performed studies that analyzed the data *as if* there had been no randomization. In particular, Fraker and Maynard (1984) and LaLonde (1986) construct control groups from data outside the NSWD. Then they benchmark the non-randomized estimates against the more trustworthy randomized ones. For example, according to Fraker and Maynard (1984, Table IV.2), where the randomized study puts the impact on youth earnings in 1979 at $7, various non-randomized estimates range between –$617 and –$1,982. In a joint summary of

[22]The footnote in the MDRC report states that if they had randomized within a group of known eligibles rather than among applicants, they would have needed a sample of 20,000 rather than 6,600 to preserve statistical power.

the parallel projects, LaLonde and Maynard (1987) concludes that "the current skepticism surrounding the results of nonexperimental evaluations is justified."

While Fraker and Maynard (1984) was written on contract and delivered to an office of the DOL, LaLonde (1986) appeared in a top economics journal and exercised vastly more influence.[23] In an essay that now stands as a shorthand for the credibility revolution, Angrist and Pischke (2010) record:

> Accounting for the origins of the credibility revolution in empirical economics is like trying to chart the birth of rock and roll. Early influences are many, and every fan has a story. But from the trenches of empirical labor economics, we see an important impetus for better designs and more randomized trials coming from studies questioning the reliability of econometric evaluations of subsidized government training programs. A landmark here is Lalonde (1986)….

In the 1980s, as part of the DOL's advisory panel, James Heckman swam against the tide of skepticism of non-randomized evaluation. He and junior coauthors penned a retort to Fraker, Maynard, LaLonde, and others: "The recent denunciation of nonexperimental methods for evaluating manpower training effects is premature…. This article… presents a more optimistic—and realistic—statement about the value of nonexperimental methods in analyzing the effects of training programs on earnings" (Heckman, Hotz, and Dabos 1987). Their argument, developed further in Heckman and Hotz (1989), was not that the DOL should blithely rely on non-experimental evaluations. It was that non-experimental ones are often all that are available, in part because of their advantages in speed and cost. And when reviewing non-randomized research, there are ways to separate the wheat from the chaff. The main winnowing tool is now called a "placebo check." A reliable evaluation method, when used to estimate impacts on a *pre*-intervention outcome, such as earnings before entering supported work, should usually return a null result.

The debate continued for decades, as researchers devised complicated new methods for synthesizing control groups in the absence of randomization. Along the way, the scholarly discourse produced many estimates of the impacts of the NSWD. A recent essay, Imbens and Xu (2025), reviews the 40 years since LaLonde (1986). The retrospective concludes that non-randomized methods have improved. And it concurs with Heckman and coauthors on the importance of placebo checks.[24] To illustrate, Imbens and Xu (2025) checks a host of complex, modern methods using two versions of the NSWD data set. All of the non-randomized methods fail the placebo checks (Imbens and Xu 2025, Figure 4). All estimate that supported work

---

[23]We posted Fraker and Maynard (1984) on archive.org.

[24]Imbens and Xu (2025) seems not to appreciate that Heckman, Hotz, and Dabos (1987) and Heckman and Hotz (1989) make this point. "Heckman and Hotz (1989) responded to LaLonde's critique of nonexperimental evaluation methods by advocating the use of specification tests to rule out particularly poor estimators. However, this approach did not offer a clear way to distinguish among the many estimators that fit the data reasonably well but rely on different identification assumptions. As a result, it gained limited traction in subsequent work." It seems to me that one of the "specification tests" was a placebo check.

*reduced* the earnings of men in 1975 *before* they entered the program. In contrast, the simple, randomized treatment-control comparisons are near zero, as they should be.

In effect, Imbens and Xu's retrospective validates the DOL's decision in the late 1980s: after hearing out the economists' debate, it launched a randomized evaluation of the JTPA.

## 4.2 The Job Training Partnership Act study

The JTPA established, reorganized, or overhauled a string of workforce programs for veterans, Native Americans, people with disabilities, people dislocated by mass layoffs, and low-income workers of the types targeted in the NSWD. In the late 1980s, the DOL commissioned a randomized evaluation for the last category (Title II-A of the act), in particular, the programs that ran year-round, as distinct from the summer programs for young people.

The legal framework of the JTPA delegated responsibility for these training programs to the states. The states in turn subdivided their territories into Service Delivery Areas (SDAs), and for each created a Private Industry Council (PIC) with a majority of members from industry, a minority from labor and education organizations, and none from government. The PICs contracted with local entities to carry out trainings. In return for the freedom to devise local solutions, PICs were subject to performance standards enforced by the threat of funding cuts. This structure embodied the philosophy of "new federalism": it was better for the federal government to minimize direct involvement in administration, grant freedom to states, and hold providers accountable for results (Courty and Marschke 2011). In practice "results" were not impacts—which are hard to measure—but outcomes, such as the percentage of enrollees exiting a training program into full-time employment.

To put a going, national program to the randomized test was to our knowledge unprecedented. And the decentralized administration of this national program fundamentally affected the meaning of the results. Beyond the control of the researchers, different SDAs could recruit different clients and offer them different services. "It was not possible," wrote the researchers, "both to achieve our mandate to examine the impact of JTPA programs as they were being operated at the time… and to isolate the effect of receiving a particular program service" (Orr et al. 1994). At the same time, the evaluation could give a sharper answer to a different question: what happens when the federal government does training in a nationwide, decentralized way?

The study team had hoped to run their experiment at a random subset of the country's 649 SDAs. This would have made the results more nationally representative. That proved impossible, for most SDAs were not interested in randomly rejecting a slice of their recruits, for the sorts of reasons outlined in Appendix A.5. "In the end, it was necessary to approach over 200 training centers in order to find 16 willing to take part in the experiment" (Doolittle and Traeger 1990). And to control costs, the team generally had to exclude both the smallest SDAs and the largest—the smallest because they would not recruit enough subjects and the largest because there, many separate organizations ran intake and recruitment (in Los Angeles, more than 50; Orr et al. 1994, p. 41), complicating the logistics of the evaluation. Still, the final sample of 16 SDAs included the urban areas of Oakland and Jersey City, and nearly matched national averages for unemployment, wage earnings, and percent employed in manufacturing, mining, or agriculture (Orr et al. 1994, Exhibits 3.1, 3.3).

Random assignment took place between November 1987 and September 1989. The study followed a substantial 15,981 subjects through the 30-month anniversary of their treatment assignments. Subjects were split 2:1 between treatment and control. This tilt reduced statistical power but nodded to the SDAs' reluctance to turn away promising applicants. Results were tracked for four subgroups: male and female, youth (aged 16–21) and adults (22+).

The experiment generated a modest but detectable differential between the treatment and control groups in the amount of treatment. As in the supported work experiment, randomization took place after someone had been recruited, screened, selected, and assigned to a service such as classroom or on-the-job training. Even then, however, only 65% of people randomly assigned to treatment officially enrolled in the offered services.[25] Meanwhile, some people in the control group found similar services elsewhere, also often subsidized (Heckman et al. 2000, Table II). And after 18 months they were allowed to cross over, i.e., to participate in the JTPA. The partial uptake and the cross-over meant that the treatment-control differences in JTPA participation were about 65 percentage points and those for participation in *any* training program were about 25 percentage points (Orr et al. 1994, Exhibits 3.15, 4.2, 4.13).

Understanding the JTPA's impact on employment outcomes is surprisingly hard, because at least four substantial analyses were performed:

- The official study finds that JTPA raised earnings by $539 for adult women and $550 for adult men over 18 months. For reference, this is a 4.5% increase for adult men; mean 18-month earnings in the control group were $12,306 (Bloom et al. 1993, Exhibit 5.3). The numbers for female and male youths are –$182 and –$854. The report declares the first and last of these four numbers significant at $p < 0.1$ (Bloom et al. 1993, Exhibit S.1).

- The 30-month report (Orr et al. 1994) reflects considerable effort to understand the negative 18-month impact for young men. It splits their sample according to whether they had ever been arrested. For non-arrestees, the 30-month cumulative impact is –$589, now without significance at conventional levels. For arrestees, two values are reported. The first uses data from Unemployment Insurance (UI) programs while the second relies on self-reports in interviews. The contrast between the two is extraordinary: –$4 versus –$4,209. Despite much investigation, Orr et al. (1994, note 33) finds no convincing explanation for the latter. Perhaps it is best seen as a data point on the unreliability of self-reported income.

  The impact estimates for the other demographics do not shift so radically between the

[25] These figures are probably low, in a technical sense. Orr et al. (1994, p. 59) estimates that half of reported non-enrollees did enroll but ultimately participated little or not at all and were not recorded as enrolled because of pay-for-performance incentives. See §3.8.

18- and 30-month surveys: the total earnings gains are now $1,837 and $1,599 for adult men and women (both significant at $p < 0.1$) and $104 for female youths (insignificant). (Orr et al. 1994, Exhibits 4.6, 4.15)

- A sensitivity analysis by Heckman and Smith (2000) focuses on JTPA's impacts on youth earnings in the first 18 months. The thrust is that the results are sensitive to various reasonable changes to the analysis, and therefore that "The claim that experiments are superior to nonexperimental methods because they produce 'one number' is false." But while the paper's title contains the word "sensitivity," the text does not precisely define the term. And many of the methodological modifications shift the impact estimates by amounts that, while large in absolute terms, are only fractions of the standard errors reported for the original estimates, making them unsurprising.

- A long-term follow-up by the General Accounting Office (GAO 1996) discards the self-reported earnings data in the original studies and links instead to social security records. We prefer this report because it introduces an objective data source on earnings, follows up longer, and estimates impacts on employment as well.

The GAO's earnings and employment ITTs are gathered in Figure 4 and Appendix Figure B1. Here, the eye is a good guide to statistical significance. The treatment-control differences are reasonably significant for adult women between years 1 and 3 after treatment begins, and for years 3 and 4 for men. For both, the difference fades in year 5. The significant impacts amount to 3 percentage points. The JTPA experiment boosted the wage earnings of adult women and men by $500–600/year, with some tapering in year 5. There are no clear impacts for young people.

There are contrasts and continuities between the NSWD and JTPA findings. In both, the strongest results are for adult women, and the weakest for youth. While the JTPA raised women's employment less, 3 points versus 8, it did so from a higher base, leaving less room to rise: the control group's employment rate in the years of highest impact was about 75%, as against 35% for AFDC women in supported work. The lower impact may also have reflected lower investment in training. JTPA cost about $1,000 per enrollee (Orr et al. 1994, Exhibit 6.2) versus $8,000 for supported work (MDRC 1980, Table 8-1, adjusting for 59% inflation between 1980 and 1990).

A final cause of the JTPA's low impact bears highlighting: the low differential between treatment and control groups in the take-up of training. This owes substantially to control group members finding training through other routes—often the same training, if with less subsidy (Heckman et al. 2000, p. 664). In contrast with high-cost supported work, JTPA services were probably fairly representative of training services in general. So it is reasonable to expect that the alternatives had similar impacts. Substitution toward them raised the benchmark outcomes in the control groups and lowered the apparent impact of treatment. Dividing by the typical 25% impact on take-up of any training, mentioned above, produces LATEs for any training four times the ITTs—about 12 percentage points of employment for adults and $2,000–2,400/year in earnings.

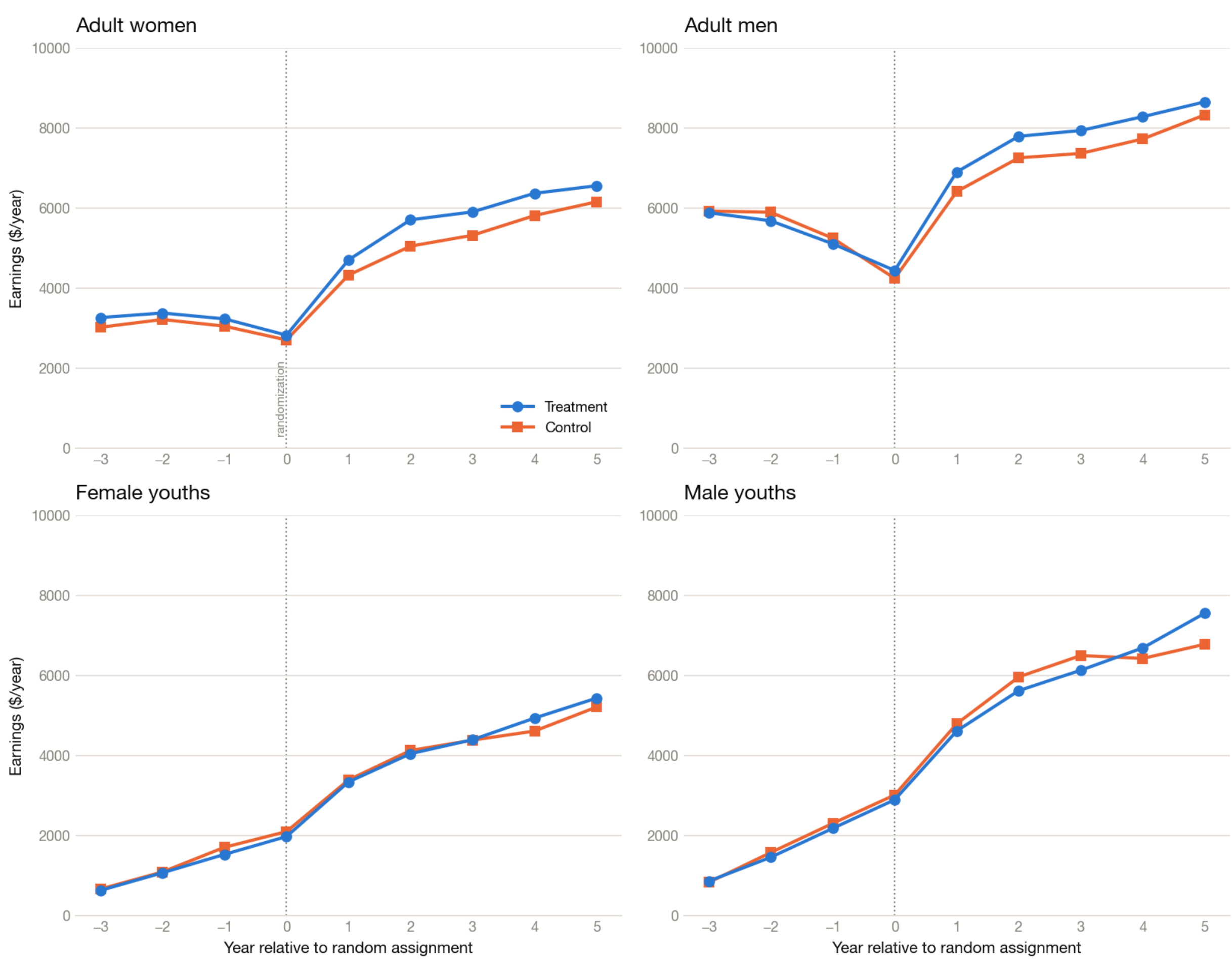

**Figure 4. Employment in National JTPA Study by year before/since randomization**
*Notes: Source is GAO (1996), Tables II.5–II.8.*

## 4.3 National Job Corps Study

The first salvo in Lyndon Johnson's War on Poverty was the Economic Opportunity Act of 1964. It created several social programs. Among them was the Job Corps, which to this day serves teenagers and young adults who have dropped out of school or are having trouble entering the workforce. In the early 1990s, when the program was 30 years old, the Department of Labor decided to find out if it works, by putting it to the randomized test. DOL officials were no doubt mindful of previous disappointing results for youth-focused programs. As we have seen, the National Supported Work Demonstration and the JTPA did little for youth. And a DOL-funded evaluation of 13 local training programs for school dropouts, called JOBSTART, had recently found just one that made a significant difference (see §7.1 below). Among the 12 disappointments were the Job Corps operations in Atlanta, Phoenix, and Los Angeles (Cave et al. 1993).

Evidently, the theory of change of Job Corps is that immersing young people in a supportive and structured environment will divert them onto a different track. As of the mid-1990s, when the National Job Corps Study was carried out, 87% of students *lived* at their training centers. (Many of the buildings were once schools or military installations; Schochet (2021), p. 4.) “While at centers,” explained the Mathematica researchers,

> participants receive intensive vocational training, academic education, and a wide range of other services, including counseling, social skills training, and health education…. Job Corps offered vocational training in more than 75 trades, and a typical center offered 10 or 11 trades. The vocational curricula were developed with input from business and labor organizations, and emphasize the achievement of specific competencies necessary to work in a trade (Schochet, Burghardt, and McConnell 2008).

Participation was voluntary but required an application. Students also had control over when they exited. In the study, the average stay was 8 months. About a quarter participated less than 3 months and another quarter more than 12 (Schochet 2021, p. 4).

The National Job Corps Study was formidable. Qualitative as well as quantitative methods were deployed. The research team produced a dozen reports extending into hundreds of pages.[26] And it was the first randomized study of a job training program—and perhaps any social program—whose results were by design nationally representative. Nearly the entire applicant pool between late 1994 and late 1996 went through the randomization sieve. 5,977 applicants were assigned to the control group, 9,409 to the treatment group, and the remainder of the approximately 81,000 applicants were placed outside the study—like the treatment subjects, free to participate, but not followed up on. Since most applicants did not participate in the study, this design minimized the disruption of ongoing programming, and thus any observer effects (Appendix A.3).

One complication the researchers faced lay in the measurement of employment and earnings. §3.3, in explaining attrition bias, discussed the tradeoff between relying on self-reported data and government databases. A summary paper, Schochet, Burghardt, and McConnell (2008), henceforth SBM, draws on both. On the government side, it uses two databases, for unemployment insurance and Social Security. A later article, Schochet (2021), adds an IRS database and with it follows subjects for 20 years, the longest in this review.

Results from all data sources, each available over a distinct year range, appear in Figure 5.[27] The right panel, which is easier to read, shows that young people self-reported earning twice what their employers reported to Social Security. The three official sources are more consistent with each other. (The IRS data used here, which is what could be extracted from Schochet (2021), includes only wage income disclosed on W-2 forms, not self-employment income.) Aside from a quick, self-reported surge in earnings in the treatment group relative to the control group, the curves for treatment and control are nearly indistinguishable throughout.

[26]The reports are indexed in the ERIC database (Institute of Education Sciences) under the phrase “National Job Corps Study.”
[27]Results are reported by calendar year rather than years since the date of randomization, which is different for different subjects. For reasons of confidentiality, the agencies holding the data limited researchers’ access to it. This may explain why the researchers could only extract results on a calendar-year basis.

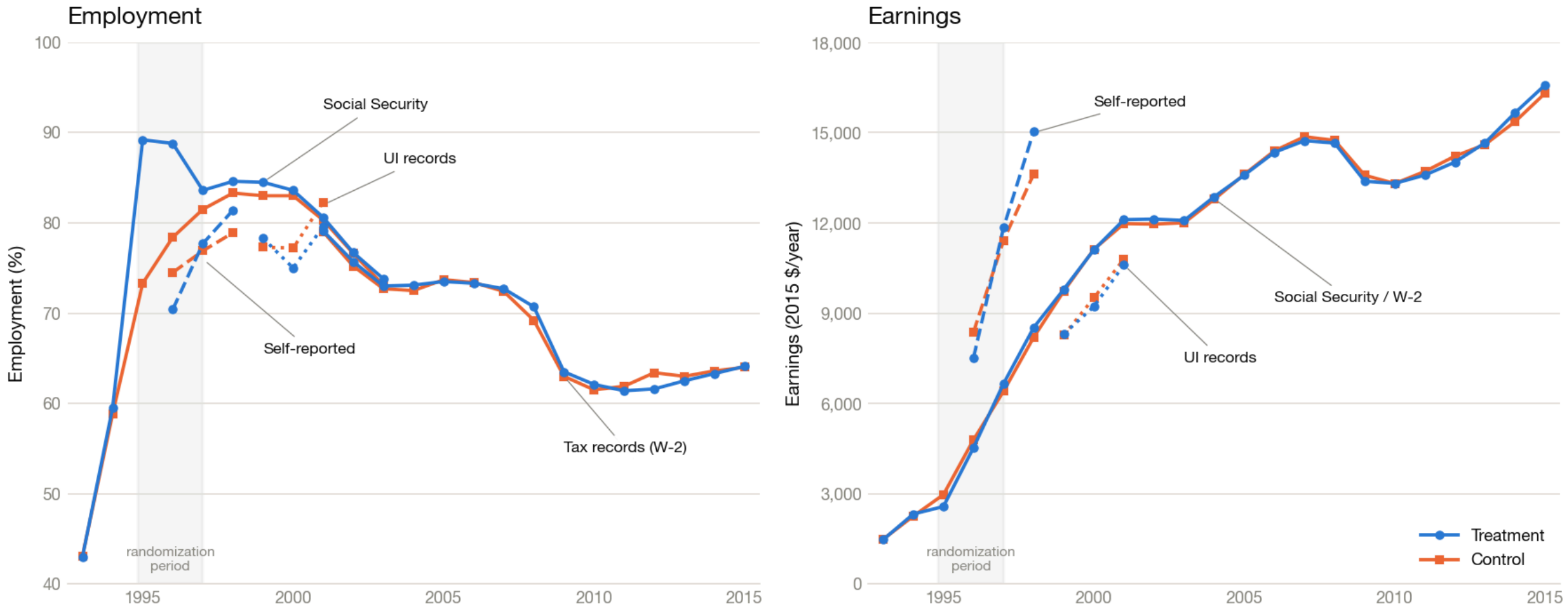

**Figure 5. Employment and wages in the National Job Corps Study, all ages**
*Notes: Based on Schochet, Burghardt, and McConnell (2008), Table 4, and Schochet (2021), Tables A.1–A.2.*

The story is about the same for whether someone has a job (left panel of Figure 5). The one departure comes during and right after randomization, when treated subjects were employed more, presumably in internships or other work arrangements that were part of Job Corps.

Because, in the right panel, self-reported income in 1996–98 rises especially steeply in the treatment group, the 1998 treatment-control difference is much larger going by self-report than by Social Security data, at $972 versus $218. The difference between these two differences is itself highly significant, statistically (SBM, Table 4). SBM (§7) tackles the important question of how to reconcile these findings. Two mechanisms could be at work: differences in *whose* information is counted (attrition bias) and differences in the information that is counted. As a first step, SBM recomputes the Social Security–based averages just for the self-reporting sample. That nearly doubles the Social Security treatment-control difference, to $393, and strongly suggests that attrition bias is present in the self-reports. Further investigation produces evidence of additional factors: survey respondents may have overestimated their *hours*, thus their total pay given their wage rate. And, as noted, some income is not reported to Social Security. Both effects could work as multipliers, scaling the absolute numbers and thus the gaps between them.

When the 20-year follow-up with IRS data, Schochet (2021), was published, it triggered controversy. The ultimate cause may be the paper's tendency to highlight statistically significant, positive findings. One can see the filtration playing out in sequence. First, where the nine-year follow-up (SBM, Table 5) computes impacts across several kinds of subgroups—defined by age, gender, race, or whether a student lived at the training site—Schochet (2021) revisits just the analyses across age groups. Schochet (2021) explains that for privacy reasons, Treasury officials had to carry out the analysis on the IRS data. And "because of… concerns about data disclosure and associated staff time in assessing associated risks, impact estimates could not be obtained for other subgroups examined in previous analyses." This implies that Schochet (2021) exercises some discretion in prioritizing analysis by age. Second, among the three age groups—16–17, 18–19, and 20–24—only impacts for the last are graphed. Third, after checking for an impact in that age bracket for each year in 1993–2015, and separately for employment and earnings, the narrative zeroes in on the last year, in which the *employment* impact is significant at $p < 0.1$. Fourth, the abstract mentions this positive finding, but not the null findings for the full sample (as seen in the IRS curves in Figure 5).

This last omission aroused the ire of Straight Talk on Evidence, a project of Arnold Ventures. The project warned that busy policymakers will read only the abstract, and be misled.[28]

Now, the full paper is clear about the null results for the full sample (Schochet 2021, pp. 12–16). Moreover, in effective reply to the concern that the paper is selectively highlighting the most promising findings, which thus might be no more than filtered noise, Schochet (2021) implicitly offers several rebuttals. All are conceptually grounded in Peter Schochet's own scholarship on multiple hypothesis testing in social policy evaluations (Schochet 2009).[29] The age group analysis, along with the others in SBM, was pre-planned 20 years ago. The classificatory jumps from ages 16–17 to 18–19 to 20–24 are intrinsically more consequential than one might think. For, at entry into Job Corps, the oldest group had worked more than the youngest (90.4% ever worked vs. 68.4% at baseline), earned more ($5.47 vs. $4.71/hour), and was more likely to have children (34.3% vs. 8.6%) (Schochet 2021, Table 1). Circumstantial evidence corroborates the finding that older students benefited more: they stayed in the program longer, they devoted more time to educational and training activities, and they were more highly motivated according to a separate survey of program staff (SBM, pp. 1876–77). We add that, against a historical backdrop of training programs failing to help young people, it would be least surprising for Job Corps to help its least-young students.

In presenting and assessing the evidence, to avoid inadvertently filtering for promising or significant results, we treat the two outcomes and three age groups symmetrically, showing results for all. Within the confines of published data, the way to do that is with TOTs, since those are printed in Schochet (2021).

[28]Straight Talk on Evidence (2020).
[29]The Straight Talk piece also includes a reply from Peter Schochet.

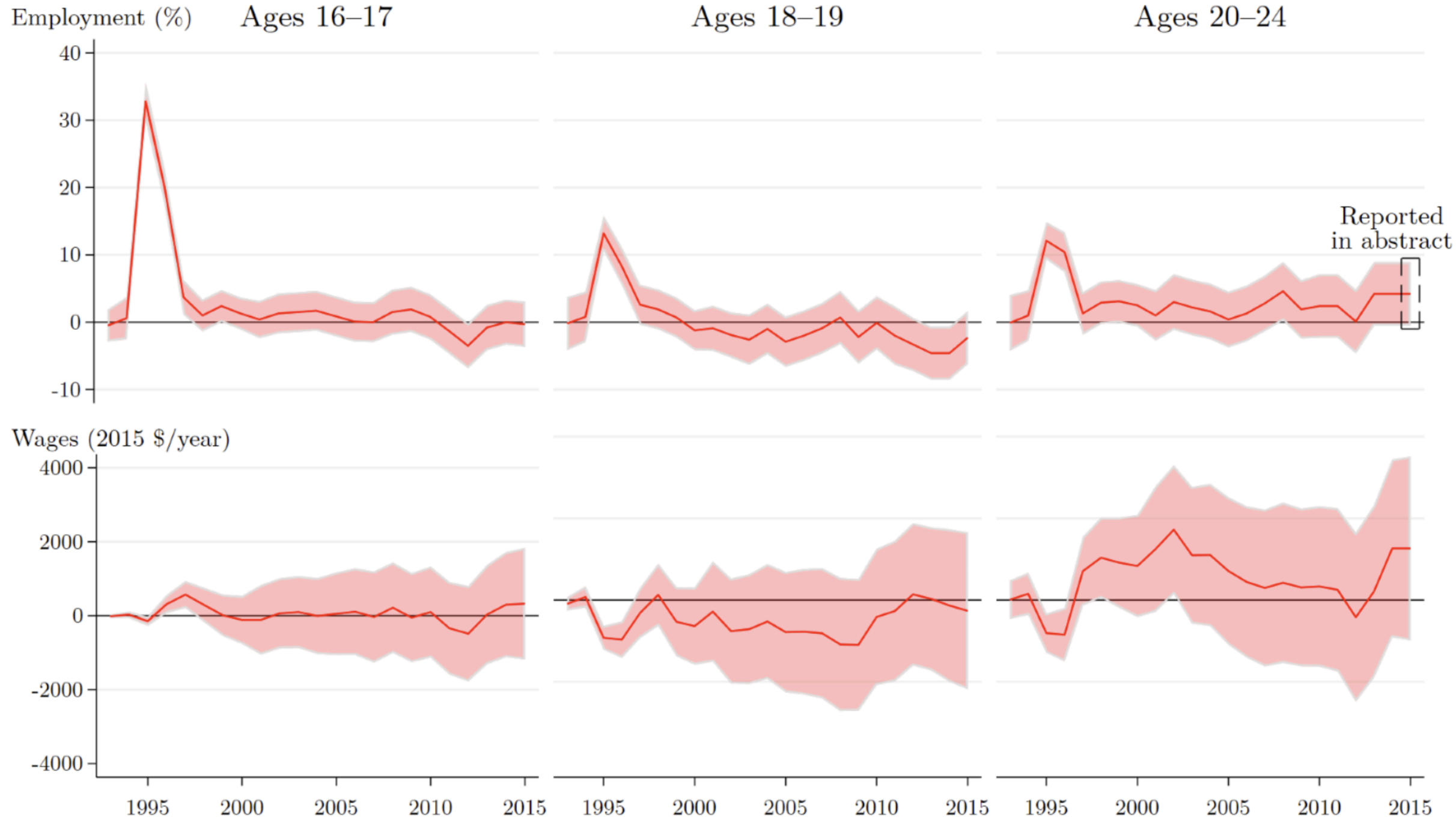

**Figure 6. Estimated impacts of treatment on employment and wages in the National Job Corps Study, by age group, from Schochet (2021)**

*Notes: Based on Schochet (2021), Tables A.3–A.4. Depicted are estimated impacts of treatment on the treated (TOT) along with 95% confidence intervals. TOTs are computed by dividing the treatment-control difference in an average outcome by the treatment-control difference in average uptake of treatment. Data are from the Social Security Administration through 2000 and the IRS after.*

Figure 6 plots TOTs of Job Corps by year and age group. All the underlying data come from government agencies—from the Social Security Administration through 2000 and the IRS thereafter. 95% confidence intervals are plotted too. Bear in mind that these intervals in a sense understate the precision of the impact estimates, because they pertain to each year in isolation. Just as combining two polls increases precision, so would computing the average impact over, say, 2011–15, instead of year by year.[30]

In the upper-right panel of the figure, at the rightmost edge, one can see the statistic highlighted in the Schochet (2021) abstract: the employment TOT for the 20–24-year-old Job Corps entrants was 4.2 percentage points in 2015, with statistical significance. However, across the whole follow-up period, this employment TOT is generally not as large and statistically significant. Zooming out more, the impact on earnings in this same age group—the labor indicator emphasized in the 2008 write-up (SBM)—is harder to distinguish from zero. And the most statistically significant result in the figure is the 4.6-percentage-point employment reduction for the 18–19-year-olds in 2013 and 2014. That result is in itself hard to explain, and is not highlighted.

[30]The effect is smaller than when combining independently sampled polls because impacts in successive years within the same cohort are not statistically independent.

We therefore do put significant weight on the possibility that the deviations from zero in the results by age group are mostly noise. The curve for employment in the youngest group hovers around zero, that for the middle group runs slightly below, and that for the oldest somewhat above. If we are to chalk up the negative result on 18–19-year-olds to noise, then the same cause vies to explain the positive impact in the oldest group.

Unfortunately, the strong signal from Figure 5—essentially no impact overall—looks like the main verdict from the Job Corps evaluation.

## 4.4 Workforce Investment Act Gold Standard evaluation

In 1998, Congress once more overhauled the federal government's array of workforce programs, with the Workforce Investment Act (WIA). The law lessened the historical focus on low-income people; partially deemphasized job training in favor of helping people market the skills they already have; consolidated the customer interface into one-stop employment offices; gave clients more control over the choice of services through a system of vouchers called Individual Training Accounts; and consolidated various programs into three service streams, for Youth, Adults, and Dislocated Workers (Blank, Heald, and Fagnoni 2011, p. 49; Bradley 2013, p. 4). The Youth track included the Job Corps. The Adult track continued the historical focus on low-income people. The Dislocated Worker track covered people in various difficult circumstances: loss of work because a plant downsized or closed, falling demand for one's skills because of broad economic changes, and more (Fortson et al. 2017, p. xvi).

Like the JTPA before it, the WIA authorized the Department of Labor to conduct evaluations. In time, that led to another national, randomized evaluation of ongoing workforce activities—in particular of the Adult and Dislocated Worker programs. The study was dubbed the WIA Gold Standard Evaluation, and fairly lives up to the name. The main report devotes some 50 pages of appendices to methodological issues such as observation weighting and imputation of missing values (Rotz et al. 2017). It arrives unflinchingly at disappointing conclusions.

As in the Job Corps evaluation, the 28 study sites were chosen randomly. To further enhance national representativeness, site selection was *stratified* across 6 regions, with a pre-set number chosen in each region (Fortson et al. 2017, p. 13).[31] Unlike most studies reviewed here, this one had three treatment arms, which corresponded to the three levels of service under the WIA (Fortson et al. 2017, Figure 1):

1. One group was offered only "core services," which "consisted mainly of information and online tools to help customers plan their careers and find employment."

[31]There were small deviations from perfection. Offices with fewer than 100 clients/year, which together served 2% of clients nationally, were excluded for reasons of cost. Two initially chosen sites declined to participate and were replaced.

2. The second group got access to those along with “intensive services,” which “included assessments, workshops, job search assistance, development of career and service plans, one-on-one career counseling and case management, placement in work experience positions, and short-term prevocational training.”

3. The “full-WIA” group got access to all of the above as well as training services “designed to prepare them for jobs in high-demand fields,” whether provided by vocational training organizations, community colleges, or employers (on-the-job training).

Randomization ran from late 2010 to early 2012. In the wake of the financial crisis, unemployment was falling but still high, at 8–9%. 2,974 people entered the experiment’s Adult track and 1,983 the Dislocated Worker track (Fortson et al. 2017, Table III.2).

As in the JTPA evaluation, forces conspired to limit the treatment differential. Budget cuts meant that training sometimes wasn’t available (Fortson et al. 2017, p. xxi). Subjects could also opt out of training as they decided how to spend their vouchers. And for many people, the WIA was not the only option for job training: private sources, charitable or commercial, were also available. For all these reasons, for the first three quarters of the study period, the differentials in participation in job training across the three arms were modest: 34% of the core group got some training, while 41% of the intensive group and 50% of the full-WIA group did. In the full-WIA group, training spending averaged $3,223, with $1,521 coming from the government and the rest from participants (Fortson et al. 2017, Table VIII.1). Adding $888 for other services brings the government’s cost to $2,409 per treatment subject, which is far below the average reported for smaller programs (see Table 1 and §6.5.1 later in this document).

Figure 7 pulls together the main results. It takes employment and earnings data from the official National Directory of New Hires (see §3.3). It shows that as the economy recovered, employment and earnings rose in all three arms of the adult and dislocated worker experiments. The full-WIA group—with access to WIA-supported job training—had lower employment and earnings in the first two quarters, presumably because some subjects were in training. But that investment of time did not clearly pay off over the three years of follow-up. In the last year, for dislocated workers, there is a mild suggestion that the full-WIA subjects earned more than the group a step down in services, core-and-intensive (bottom right panel). The positive impact might still deserve some weight if the four graphs told a consistent story of positive impact across outcomes or study populations. But they do not.

The WIA evaluation unfortunately turns out to have little power to detect the effects of job training. The treatment-control difference in training take-up was modest, 16 percentage points at best. And relatively little was spent on training. The delivered training *may* have boosted employment and earnings. But when the ITT is statistically indistinguishable from zero, so is the LATE.

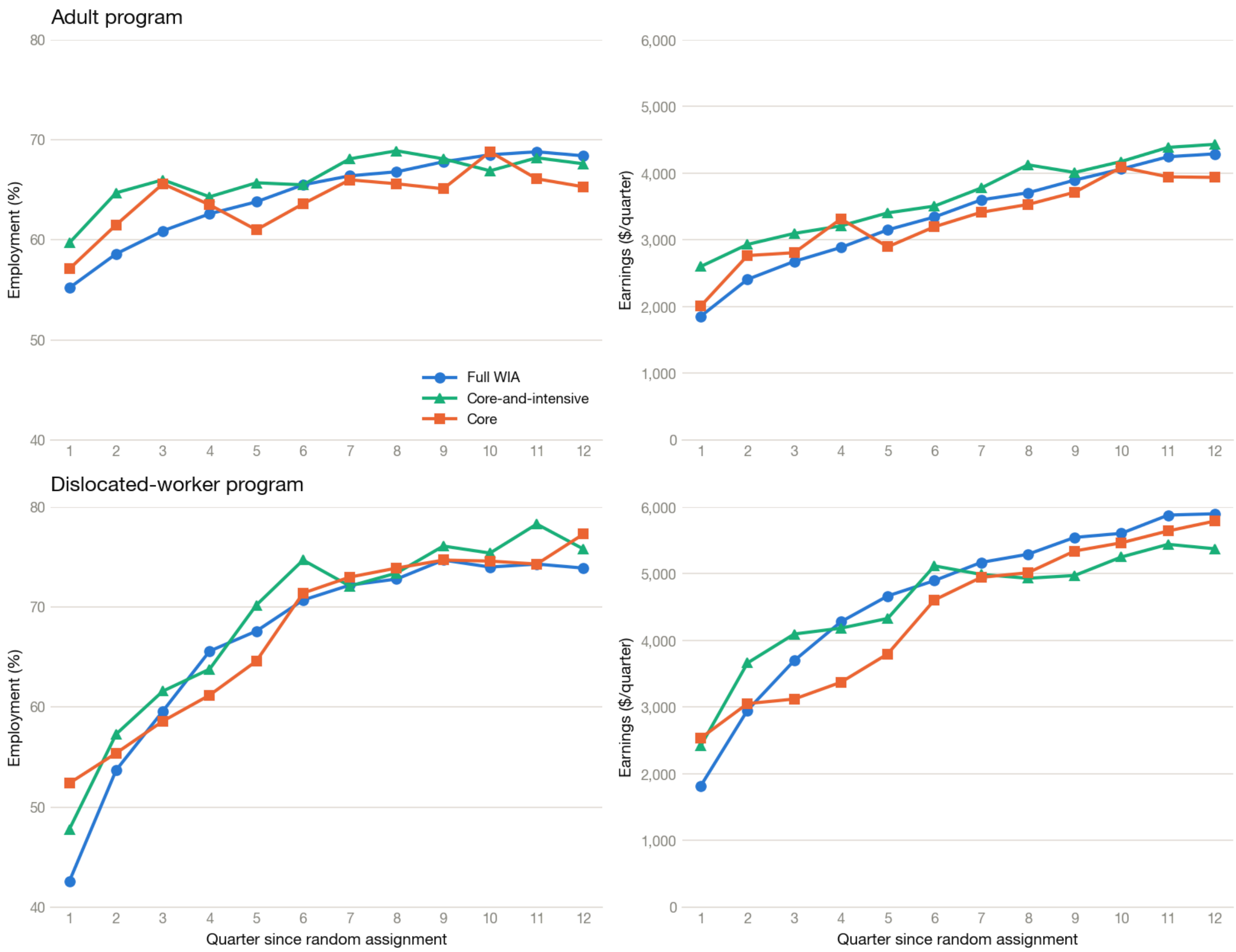

**Figure 7. Employment and earnings by quarter in the three arms of the Workforce Investment Act Gold Standard evaluation, in the adult and dislocated-worker programs**
*Notes: Based on Rotz et al. (2017), Tables G.VI.1, G.VI.2, H.VI.1, and H.VI.2. All data from official sources. Only the full-WIA group was offered job training under the Workforce Investment Act. Some in the core-and-intensive group were offered short-term prevocational training. The core group was not offered training.*

## 4.5 Major American randomized evaluations: summary

The story of American leadership in seeking empirical truth about workforce programs is one of ambition and disappointment. We are aware of no other instance of a national government committing this much energy to rigorous assessment of whether a family of social programs works. Unfortunately, the answer, at least from the national trials of the JTPA, Job Corps, and WIA, is that the main-line programs have not moved the needle much on labor market outcomes. Alone, the JTPA produced positive impacts on some of its study populations, the adults. Especially in the WIA study, low uptake made it harder to detect any actual benefits for those who did get training. But low uptake is not a failure mode we want to abstract from.

As we will see in §7, some smaller, privately run programs have done much better. This will raise questions about how much those successes can be expanded, with what sort of funding, and with what sort of administration.

# 5 "Judge randomization" studies

Some researchers on workforce interventions have borrowed a trick from criminologists. It is called "judge randomization" (Martin, Annan, and Forst 1993; Kling 2006) even when no judges are involved. In our context, the idea (as an example) is that people participating in an unemployment program are arbitrarily assigned to a caseworker within a given employment office; and if some caseworkers are more apt to assign clients to on-the-job training, that constitutes a natural experiment in on-the-job training. Here, the propensity of a caseworker to assign a particular treatment is the *instrument*. It replaces the coin toss in a randomized experiment.

This section reviews the two judge randomization studies of training that we are aware of. One is set in the US, the other in Denmark. As will be discussed, it is not clear *a priori* that the US one approaches the ideal of randomized treatment. In many studies involving actual judges, assignment really is random, if with some scope for human beings to override the assignment (e.g., Kling 2006; Green and Winik 2010; Loeffler 2013; Aizer and Doyle 2015; Bhuller et al. 2020). Deliberate scrambling is rare in job training. Evidently governments take more care that justice be blind than that unemployment caseworker assignment be blind.

## 5.1 Hyman (2018), "Can displaced labor be retrained? Evidence from quasi-random assignment to trade adjustment assistance"

For decades, it was US policy to increase economic integration with other countries while aiding people at home who lost jobs to foreign competition. Starting in 1962, Trade Adjustment Assistance (TAA) covered the cost of job training for such dislocated workers. TAA also paid for unemployment insurance during training, extending the usual 26 weeks of coverage to up to three years, and provided assistance with job search and relocation (Collins 2014). (The program has been in legislative limbo since 2022, when Congress failed to reauthorize it. At least for now, the political compromise that undergirded trade promotion has broken down.)

The study's largest limitation from our point of view is that the extension of unemployment insurance was contingent upon being in training, making it impossible to statistically distinguish the effects of training and income support.

While local government entities administered much of the program, an office in the Department of Labor determined which workers qualified, i.e., which plausibly lost their jobs to foreign competition or offshoring. In fact, the DOL did not make the decision worker by worker, but plant by plant. If a single worker laid off from a plant won TAA assistance, then *all* subsequent applicants from the same plant within the next three years did too (Hyman 2018).

Hyman (2018) studies whether this assistance helps people recover from layoffs. The premises of the analytical strategy are that 1) among the several hundred DOL investigators who decided which layoffs and closures owe to foreign competition and therefore merit TAA, some were more lenient; and 2) the assignment of cases to investigators was effectively arbitrary, or could be cast as such with proper econometric framing.[32] The first assumption is fairly demonstrable, as will be discussed. The second can be backed only by circumstantial evidence, for the assignment process was somewhat opaque, and there is no suggestion that it was a lottery. Hyman quotes this description from the DOL:

> TAA cases are assigned… primarily based on investigator caseload, as well as previous experience with a company or industry. Staff leave or other scheduling issues can be a factor as well.

Assignment based on who has bandwidth and who is not on vacation could be effectively random: two similar cases could easily go to different investigators. Assignment based on previous experience is not random since it could induce sustained associations between what industry a person worked in and which caseworkers got their cases.

To carry out the analysis, Hyman (2018) links two data sets. He obtains data on all petitions for TAA filed between 1974 and 2016. For each petition, the data set names the (former) employer claimed to have been pressured by foreign competition and the DOL investigator who adjudicated the case. Hyman links the petitions to the Longitudinal Employer-Household Dynamics (LEHD) database maintained by the Census Bureau, which aggregates records from state unemployment insurance programs. This allows Hyman to track whether a worker held—and then lost or left—a job at a plant that was the subject of a TAA petition; whether the plant's former workers won TAA; and how much recipients and non-recipients worked and earned in the years that followed. For this project, 24 states and the District of Columbia granted access to their contributions to the LEHD data set. The resulting sample includes some rust belt states (Illinois, Indiana, Pennsylvania) as well as South Carolina, where the textile industry was once a major employer. The LEHD snapshot used in the study starts between 1985 and 2002, depending on the state, and stops in 2011.

Hyman (2018) exemplifies a nearly universal pattern in empirical economics: because the treatment variation is not random, analytical complications are introduced in the hope of effectively pushing it in that direction. Concretely, once a company learns that, in response to a petition from a former worker, a DOL investigator has certified its entire workforce as TAA-eligible, the company might feel freer to lay off more workers. And the workers laid off later might differ systematically from those laid off earlier: perhaps they are more skilled and more valuable. But their counterparts in the control group—people who also escaped initial layoffs, but at plants that did not get TAA assistance—might not lose their jobs, at least not in the same numbers. To avoid introducing such a mismatch, Hyman (2018) limits its sample to people who had *already* been laid off in the 12 months preceding the official decision on TAA support.

[32]As is common in judge randomization studies (e.g., Aizer and Doyle 2015), the instrument is the leave-one-out approval rate, recomputed for each application. This "jackknifing" should reduce the endogeneity of the instrument to application-specific circumstances.

To further narrow the comparison groups and arguably make them more meaningful, Hyman adds many controls: for the year and quarter when the TAA case was filed; for the industry (important in light of the statement quoted above that investigators may tend to specialize in certain industries); for the type of organization that filed the petition (a company, union, worker group, or state career office); and for whether the application was for special handling in connection with the North American Free Trade Agreement (NAFTA). As well, Hyman includes controls for demographic traits such as race and sex, and aspects of education and employment history such as high school completion and salary before being laid off.

The hope is that as more controls are added and the domains of comparison are narrowed, some variation in the probability of access to TAA will remain, and it will be tantamount to random. The hope, in other words, is that as one shifts to comparing the careers of two white women with high school degrees who had stopped working at different car plants in Indiana just before TAA applications were filed for the plants in the summer of 1998, any other differences in the women's lives *not* controlled for did not affect their work careers *even as*, because of differences in the leniency of investigators, one was more likely to get the TAA package. Then, any difference in their subsequent careers is attributable to TAA.

It is not obvious that this works, i.e., that adding lots of controls will expunge most non-treatment differences but preserve arbitrary differences in treatment, forming a natural experiment. Hyman (2018) conducts several statistical tests to validate or rebut the strategy. One test is for *relevance* of the instrument—whether, even with all the controls, an investigator's *average* approval rate still helps predict their *individual* decisions. It does: each 10 percentage point increase in an investigator's average approval rate lifts the probability of approval in individual cases by 6 points, and with great statistical significance (Hyman 2018, Table 4, first row).

Then there are two tests that speak indirectly to *excludability*, that is, whether after incorporating all the fixed effects and other controls, an investigator's leniency has no statistical connection with post-treatment employment and wage earnings *except* via the TAA treatment. When this condition holds, leniency indeed plays the role of a coin toss in a classical experiment. The first test starts with the observation that if the leniency of the assigned investigator were truly random, it would be uncorrelated with all the other variables in the study. Technically, this property need not hold for the impact estimates to be unbiased; still, circumstantial evidence that it does is reassuring. In the event, Hyman (2018, Table 3, cols. 5–6) finds that leniency is all but unrelated to demographic traits, but is higher for cases where employees have longer tenure or higher pay, where the county unemployment rate is higher, where the investigator is less experienced, or the plant is in an industry more exposed to foreign competition. Hyman (2018, p. 20) argues that even the associations that are statistically significant are not practically significant. For example, an extra decade of experience only bumps an investigator's approval rate by 0.004 percentage points (in a multivariate regression, Table 3, col. 6).

When testing for correlations with many variables, a few will be reported as statistically significant by chance. For this reason, Hyman (2018, Table 3, col. 6) runs an F test for the hypothesis that *all* of the correlations are actually 0. It returns a *p* value of 0.18. This is above the conventional maximum of 0.05 for statistical significance. But for checks such as this one, low thresholds are the opposite of conservative (Roodman 2009, p. 142). There is only an 18% chance that this much correlation would be measured if the treatment were unrelated to all of the controls, which is not strongly reassuring.

Still, in a context where perfection is neither expected nor formally needed, the tests for effective randomness produce somewhat reassuring results. Some correlations are probably real, but small.

The second indirect test of excludability is a placebo check of the sort Heckman championed (see §4.1.4). If the investigator leniency variable generates a valid natural experiment, it should only appear to affect post-treatment outcomes, not pre-treatment ones.

That placebo check, along with the study's main results, is depicted in plots gathered in Figure 8. The blue lines show the estimated impact of TAA being made available to former employees of a given plant, on both the number of quarters per year in which they work, and on how much they earned per year. In the ten years before a plant's access to TAA is adjudicated, the blue lines do not stray far from zero; indeed, most of the 90% confidence intervals, shown with dashed lines, easily embrace 0. The treatment does not appear to affect outcomes before it could. Then, in the first two years after the TAA decision, employment and earnings drop in the treatment group; that is consistent with entering a classroom rather than immediately looking for, and sometimes finding, a new job. After, the impacts turn positive, suggesting that TAA helps people find more and/or better work. However, the benefits fade by year 10. (Top row of Figure 8.) Hyman (2018, p. 24) calculates that workers net an extra 20 months of work and $50,000 in earnings over the 10 years.

The plausibility of the overall pattern—the roughly flat stretch, the short, sharp fall, the marked rise, then the long-term decay—adds credibility to the results. If the apparent benefit of TAA is statistical noise, this particular pattern would be surprising.

Having established the basic result, Hyman (2018) slices the data in a few ways in pursuit of insight into which groups of people are most helped and how. One major question is about the impact of the *training* part of TAA. The data do not reveal how long individuals spend in TAA-funded training. However, they do provide averages by state and quarter. Splitting the sample by whether a person worked in a state that was above or below the 50$^{th}$ percentile just before being laid off, it emerges that earnings do *not* decay for those from states and quarters where people spent more time in training (Hyman 2018, Figure 10). Whereas, in the complementary group, earnings initially shoot higher—consistent with spending less time in training—and then fall. This characterization of the ups and downs is not formally tested.[33] And it

[33]Another impediment to assessing the strength of this finding is that confidence intervals are omitted for legibility from Hyman (2018), Figure 12. For standard errors, the figure's notes refer the reader to an

is based on an endogenous splitting of the sample: perhaps it mostly shows that where new jobs were harder to find, people opted more for training. On balance, it hints mildly that the training phase of TAA support deserves some credit for TAA's apparent benefits. However, since the extension of unemployment insurance is contingent on participating in job training, the check still cannot distinguish the effects of training from the effects of the concomitant income support.

Also important—and related—are the results on mobility across space and industry. Dorn (2009) divides the United States into "commuting zones" that represent local labor markets. The North American Industry Classification System, at the 2-digit level, categorizes major industries such as construction and retail. Hyman (2018), Figure 12, here copied into the second row of Figure 8, finds that people who could tap TAA assistance were 20–40 points more likely to move to a new commuting zone. Impacts were similar for moving to a new industry. These findings are big if true: transitional assistance indeed helps people transition to better work settings after economic disruption.

How much should we believe these conclusions? There are reasons to doubt and reasons to trust. Reasons to doubt include that this is a large, complicated, opaque, non-experimental study. Complexity brings more opportunity for bugs and increases the degrees of freedom in defining the sample and running the numbers, which widens the space for *p*-hacking, conscious or unconscious. Treatment was not randomized. The study has not been published, and the follow-up data set is confidential, which makes it difficult to peer inside the black box.

On the other hand, the checks for excludability and relevance return fairly reassuring results. And the time profile, including the lack of impact in the pre-treatment period and the sharp but temporary drop, is hard to explain as mere artifacts of reverse causation or third-variable confounding.

We view the topline results as plausible but are unsure how much to attribute them to training rather than the income support during training.

---

online appendix, which does not appear to be available. Significance stars are in the corresponding Table 8, but that table leaves out the last year—year 10—whose low point estimate for the below-median group especially influences the degree of downward trend in this group.

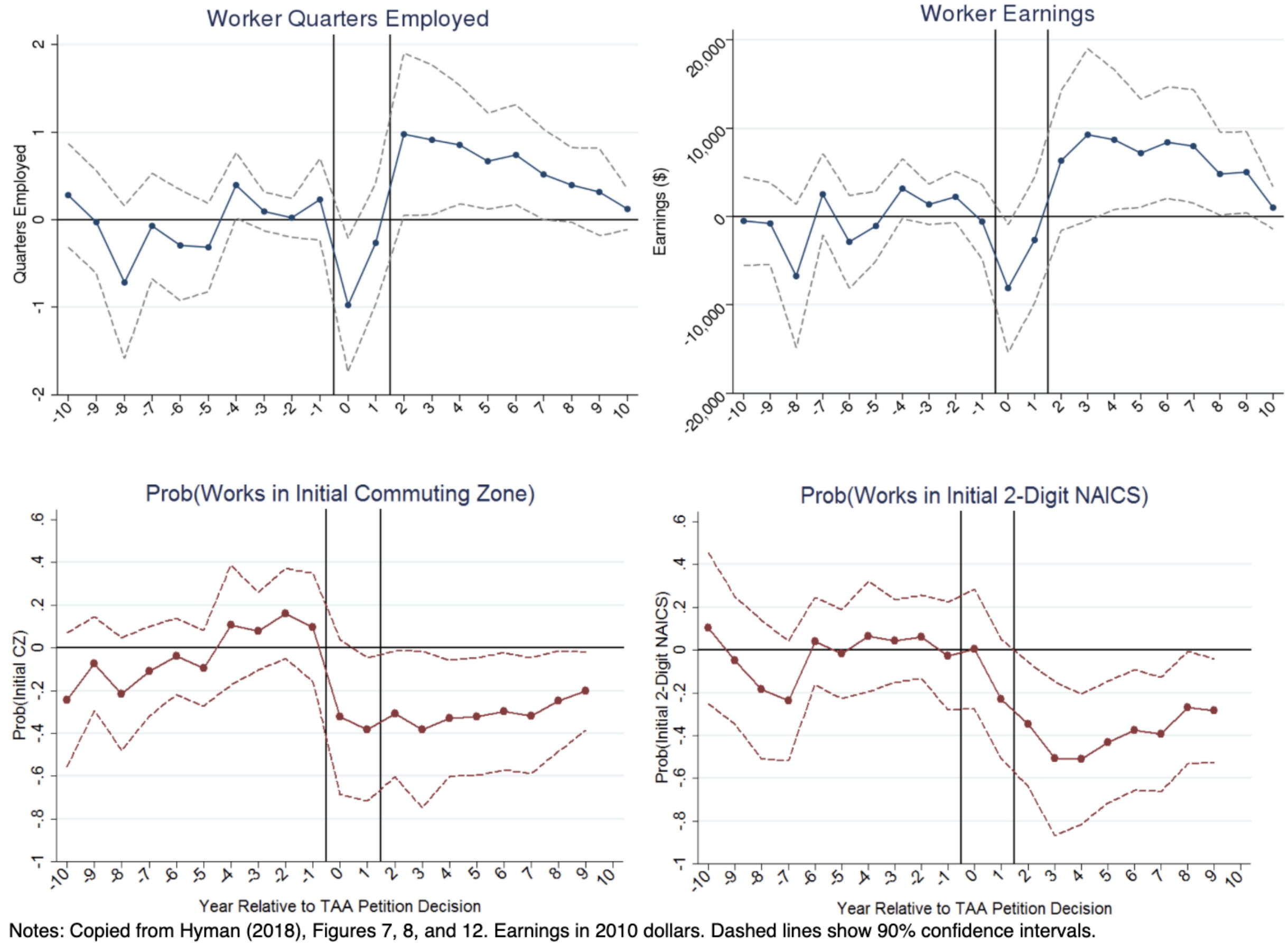

**Figure 8. Estimated impacts of availability of Trade Adjustment Assistance, from Hyman (2018)**

*Notes: Copied from Hyman (2018), Figures 7, 8, and 12. Earnings in 2010 dollars. Dashed lines show 90% confidence intervals.*

## 5.2 Humlum, Munch, and Rasmussen (2025), "What works for the unemployed? Evidence from quasi-random caseworker assignments"

The other judge randomization study in this review differs from Hyman (2018) in a few ways:

- It is set in Denmark.
- The decisions about training options are made on an individual basis, not for whole companies.
- The decisions are made by caseworkers in local job centers rather than at a central agency.
- The caseworkers do not merely authorize funding; they design a program for each client, which in about half of cases includes classroom-based education, on-the-job training, or both (Humlum, Munch, and Rasmussen 2025, Table 2). This variation in two treatments allows Humlum, Munch, and Rasmussen—henceforth “HMR”—to estimate the impact of each.
- The paper makes a more airtight argument that the assignment of cases to caseworkers contains an isolatable component that is as good as random.

Despite the prima facie credibility of this natural experiment, the study’s results taken together are surprising, which causes us to put somewhat less weight on them.

Citing Danish law and norms (HMR, §2.2), HMR splits the training treatment into two main types. Classroom training encompasses ordinary education, basic skills in job search, and “wrap-around” services to help people choose new kinds of jobs and find employers. Classroom training is also defined to include vocational teaching. Meanwhile, on-the-job training consists of internships and subsidized fixed-term contracts.

The paper leverages a quirk in the administration of the Danish unemployment program: within an employment office, clients were often assigned to caseworkers based on the day of month of their birth. For example, in one local office, clients born in the first week of the month might go to one caseworker, those born in the second to another, and so on. HMR provides an illustration using simulated data, which is copied below as Figure 9. In the simulation, 58% of cases follow the week-based assignment “rule.” This generates distinct bulges for each caseworker in the distribution of cases by birth day of month. In the real data, which privacy restrictions prevent us from seeing, the picture would sometimes be neater than depicted here, sometimes less so.[34]

[34]HMR borrows this strategy from Cederlöf, Söderström, and Vikström (2025), which is similar in many ways, and set in Sweden. The latter, however, focuses on estimating the effectiveness of caseworkers in reducing clients’ time on unemployment and helping them get higher-paying jobs, as distinct from the impacts of training. It does report that caseworkers who assigned training more were *less* effective on average (Cederlöf, Söderström, and Vikström 2025, Table 8, “Supportive” row). However, this finding is in itself non-experimental.

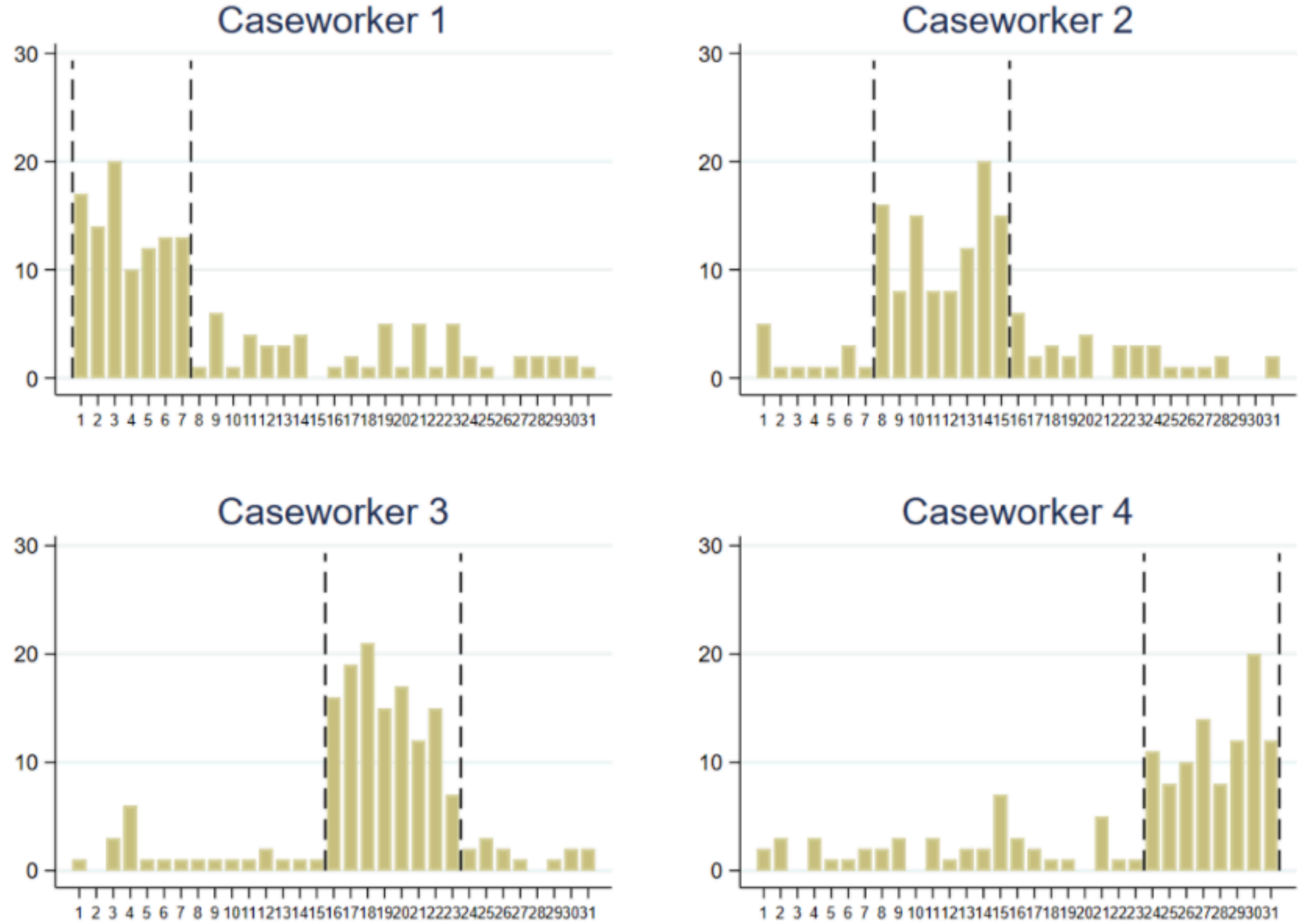

**Figure 9. Simulation: Number of clients assigned to four caseworkers in a job center, by day of the month of birth, from Humlum, Munch, and Rasmussen (2025)**
*Notes: Copied from Humlum, Munch, and Rasmussen (2025), Figure 1.*

Since it is hard to see how being born on the 7th of the month would affect one's career differently than being born on the 8th *other* than via assignment to caseworkers with different propensities for recommending training, this set-up bids fair to produce a strong natural experiment.

A wrinkle is that assignment by birth date is far from universal, and the specifics of the practice are not documented—how much various job centers practiced it, and how they partitioned the days of the month into blocks. HMR therefore combines automatic and manual techniques to estimate the rules actually used. For each job center—or unit within the job center where it appears that different age groups are handled separately—HMR provisionally assumes that whichever caseworker gets the most people born on, say, the 12th of the month, is in fact the caseworker for that day. Once these guesses are made, plots like those in Figure 9 are drawn. If, as in the figure, a blocking pattern is evident, the office is kept in the sample and the date-based assignment is inferred from the graph. After this construction process, the HMR sample consists of 167,222 unemployment episodes experienced by 127,713 people, who saw any of 536 caseworkers between 2012 and 2017 (HMR Table A.1).

This construction process leaves some scope—probably modest—for unobserved exercise of discretion. And to the degree the researchers' judgments are wrong—again, probably modest—they erode the relevance of the birth date instruments, i.e., the tendencies of the *predicted* caseworker to assign classroom and/or on-the-job training. Finally, even with a perfect understanding of the caseworker assignment process, impacts in this design could be driven by something else about the caseworkers who happen to prefer certain programs, violating the exclusion assumption. But HMR presents a battery of fairly convincing tests suggesting that exclusion is met.

Just as in Hyman (2018), the authors also carefully check the relevance and independence of the instruments. As for relevance of the instruments: After controlling for client demographics and time and region fixed effects, clients whose cases are *predicted*, by virtue of when they were born, to be handled by a worker with 10 points higher propensity to require classroom training indeed are assigned more classroom time—3.8 points more. This association between classroom treatment and its instrument is highly significant, statistically. For on-the-job training, the association is half as big—2.1 points—but still significant by conventional standards (HMR, Table C.2).

As for independence of the instruments: Like Hyman (2018), HMR checks whether the instruments are related to demographic traits and other variables that are controlled for—which, recall, is not technically required to avoid bias, but can strengthen the view that the instrument is as good as random. Since 65 controls are listed in the table of results from these checks, under the null hypothesis that the instruments are related to none of them, we should see ~6.5 correlations significant at 0.1 for each instrument. There are 5 for the classroom training instrument and 12 for the on-the-job training instrument.

HMR's major finding is that of the two types of treatment—classroom and on-the-job—the first appears to have helped people find work. See Figure 10 below, which is adapted from HMR Table D.1. Being assigned classroom training increases employment by about 26 hours/month in the second year after entering the unemployment system (HMR, Table D.2). Dividing that by a typical 160 hours/month yields a substantial 16% increase in employment, which is among the very largest in this review. The story is the same if one takes employment (having a job) or earnings as the outcome, as distinct from number of hours worked. They find that being assigned to the classroom training increases employment seven quarters after the start of the unemployment spell by about 20 percentage points (HMR, Figure D.2).

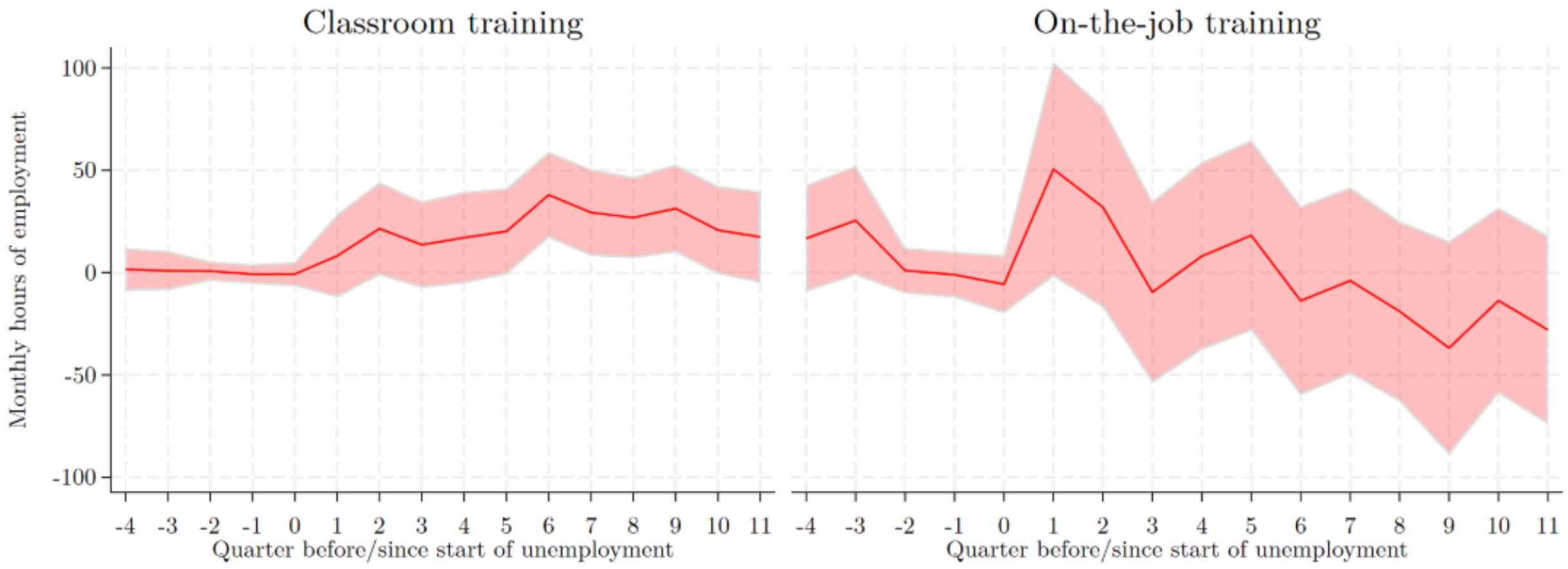

**Figure 10. Impact of assignment to classroom training or on-the-job training on work hours per month, by quarter before/since start of unemployment episode, from Humlum, Munch, and Rasmussen (2025)**

*Notes: Based on Humlum, Munch, and Rasmussen (2025), Table D.1. Sample consists of those who entered unemployment between 2012 and 2017. Because follow-up stops after 2019 (at the onset of COVID-19), unemployment spells that began in 2017 start to drop out of the sample after quarter 7.*

But for on-the-job training, the confidence intervals are about twice as wide and the central impact estimates are in general statistically indistinguishable from 0. Despite this imprecision, there is enough statistical power to reject the hypothesis that classroom and on-the-job training have about the same impact. The *p* values for this hypothesis range between 0.03 and 0.22 for quarters 6–11 (HMR Table D.1), which is low. So it is unlikely the two treatments are the same, and that the impact of one is merely hard to detect because of low statistical power.

For insight into how classroom training helps, HMR, like Hyman (2018), subdivides the main analysis along several dimensions. A distinction is drawn between returning to one's old occupation and finding a new one. Meanwhile, classroom courses are split into "job search courses" (which averaged 19 days) and "skills and wrap-around courses." The latter subdivision puts wrap-around services such as job counseling in the same grouping as vocational training. HMR explains that "While 'job search' courses are easy to distinguish from other classroom courses, the dividing lines between 'skills' and 'wrap-around' courses are less clear." This is surprising because *a priori*, skills training looks like the odd one out relative to job counseling and job search assistance. HMR goes on to explain that the combination of skills and wrap-around "also helps support the statistical power of our analysis." This trades off precision in the conception of what is being measured for precision in its measurement.

The product of these two two-way splits—new or old occupation, job search courses or other classroom subjects—is four results. HMR finds that about a year after unemployment starts, both job search courses and skills-and-wrap-around-courses increase reemployment in new occupations. Neither reliably increases it in old occupations (HMR, Figure 9).[35] This suggests that the classroom training helped Danish workers recover from job loss by putting them on new professional paths.

To further explore this story of resilience, HMR analyzes whether training especially helped people whose jobs were most exposed to *offshoring*. HMR classes as least exposed those whose old jobs required on-site work or face-to-face interaction with customers—think of salespeople and car mechanics, not computer programmers and call center workers. The impact estimates for the more-exposed are over twice as large, though the difference is not statistically significant ($p = 0.14$ for hours, weaker for other outcomes).

HMR also decomposes the analysis by stage in the training process. As the effects of classroom training emerge in the months after entry into the unemployment system, clients pass through four stages: not yet being assigned to any kind of training; being assigned; participating; and having finished, after either dropping out early or completing the course. HMR (Figure 6(b)) shows that the largest aggregate employment boost accrues to those who have finished training, and it appears after 14 months. This broadly corroborates the hypothesis that classroom training works, once it is completed. But other results from this decomposition fit the theory less well. Dropping out of a class early, and never starting one's assigned class, have roughly the same per-person impact on unemployment as completing the class.[36]

The HMR results taken together are a bit hard to explain. The estimates are large, matched only by the LATE estimates from the sectoral training programs discussed below. And why would classroom training be distinctly better than on-the-job training at guiding people to new occupations? HMR (§7.1) suggests that the reason is that a return to the classroom is often needed for fundamental reskilling. However, it is not clear that the Danish classes for the unemployed are so foundational. They last only 52 days on average (HMR, Table 1). And, as just mentioned, the benefits are about as big for those who drop out early, or don't even take their assigned courses. Among the subjects taught in the classroom, *job search* has, if anything, bigger impacts even though it is not about reskilling (HMR, Table H.7, col. 2). Meanwhile, because the analysis amalgamates skills training with job counseling, it is hard to tell whether vocational training indeed has a major impact, as the reskilling theory predicts.

Overall, the study is tricky to evaluate. The methodology looks sound. And the claim to quasi-experimental identification of impacts is stronger than in Hyman (2018), because of the birthday rule. But like Hyman (2018), the study is also intrinsically opaque and complex.

---

[35]Across 8 quarters—0 through 7 after randomization—and two treatments, the impact on reemployment in one's original occupation is significant at $p < 0.05$ in one case: job search services in quarter 0. It is hard to know whether this significant result is signal or noise.
[36]HMR (§7) estimates the impacts at various stages by estimating a set of $\gamma_{1^*t^*}{}^{*s^*}$ parameters for the probability, for someone assigned classroom training, of being in stage *s* in month *t*, as well as estimating quantities we will label $\beta_{1^*t^*}{}^{*s^*}$ for the employment impacts of being in each state. HMR Figures 5 and 6 plot the $\gamma$ estimates and the $\gamma \times \beta$ estimates, respectively. The latter are the aggregate impacts in the sample for compliers in each stage. The impacts per client—the $\beta$'s—can be roughly estimated by eyeballing the figures and taking ratios.

# 6 Meta-analyses

In the US, randomized studies of small workforce programs are too numerous to justify writing—or reading—critiques of each. This section therefore turns to meta-analysis, the discipline of synthesizing estimates from many studies. We first review four previous meta-analyses. Then we report on our own meta-analysis of randomized studies in the US.

The five meta-analyses reviewed here cover distinctive but overlapping swaths of the literature: Smedslund et al. (2006) looks at randomized evaluations of welfare-to-work programs, almost all from the US; Haelermans and Borghans (2012) at non-randomized studies of on-the-job training worldwide; Card, Kluve, and Weber (2018) at a large, international collection of randomized and non-randomized studies of labor market interventions; Peck et al. (2021) at career pathways studies in the US, randomized and not; and our comprehensive meta-analysis of randomized studies in the US that involve training.

The studies concur on the first-order question: average impacts on earnings or employment are positive, though not transformational. There is little consistency on the second-order question of which traits of an intervention are associated with greater impact. One reason is that different studies check different traits. Another may be that these meta-regression findings are fragile given the small samples and intercorrelations among traits. We provide some conceptual preliminaries in Appendix E.

## 6.1 Smedslund et al. (2006), “Work programmes for welfare recipients”

Smedslund et al. (2006) is a product of the Campbell Collaboration, which promotes systematic reviews of evidence on social interventions. Like its health-focused sibling, Cochrane, Campbell leans heavily on randomized trials. The domain of interest for Smedslund et al. is welfare-to-work programs, worldwide. That overlaps with the scope of our review in the manner of a Venn diagram: some welfare-to-work programs do not train; some training programs are not expressly intended for people on welfare.

Because of the American tradition of social experiments, Smedslund et al. is drawn gravitationally to the US. 44 of the 46 studies are set in the country, and the other two in Canada.

Smedslund et al. finds that, on average, welfare-to-work programs produce real benefits that gradually fade. The programs increased employment by 9.7% (not 9.7 percentage points) in the first year, 9.2% in the second year, and 3.7% in the fifth year. All those figures are random-effects averages, and all have great statistical significance. The parallel figures for earnings impacts are 4.3%, 4.4%, and 1.1%, the last being difficult to distinguish from 0 (Smedslund et al. 2006, p. 2).

Smedslund et al. finds “some asymmetry” in funnel plots for employment and earnings impacts, but does not quantify any resulting publication bias. As explained in §3.4, when most of the underlying literature is government-funded randomized trials, we expect less publication bias.

The study does estimate average impacts just of programs with a training component. Here, it does not break out the impact by time since randomization. Overall, programs that included skills training raised employment by 5.6% and earnings by 4.5% (Smedslund et al. 2006, p. 2). Those estimates too are statistically distinguishable from zero. They are statistically *in*distinguishable from the averages for programs that did *not* incorporate skills training. But that statement makes a cross-group comparison (as defined in Appendix E), and so should be interpreted gingerly.

## 6.2 Haelermans and Borghans (2012), “Wage effects of on-the-job training: A meta-analysis”

This meta-analysis draws together 71 impact estimates from 38 studies of on-the-job training. 37 of the estimates come from Europe and 13 from the US. Almost none meet the evidentiary standard used in our review. In particular, none are randomized. One with a potentially compelling quasi-experiment is included: the Leuven and Oosterbeek (2004) study of the Dutch tax credit for training over-40s, which is reviewed in Appendix D.

Across the 71, and across the various follow-up periods, the random-effects mean impact on wage earnings is 3.9%, with great statistical significance.

However, Haelermans and Borghans find strong evidence of publication bias. The study’s funnel plot appears in their Figure 2. Each dot represents an estimate of the impact of on-the-job training, with the standard error (akin to the margin of error in a poll) on the horizontal axis and the impact on the vertical. The suggestion of publication bias is clear: the precise estimates cluster together on the left while the imprecise ones are scattered to the right and swing asymmetrically upward. This means that noisy studies finding large impacts were far more likely to get published than those with negative or small impacts.

Haelermans and Borghans (2012, Table 5, row 1, penultimate column) estimate that adjusting for this publication bias cuts the average wage impact of on-the-job training from 3.9% to 2.6%. Turning that around, in this collection of non-randomized academic studies, publication bias inflates the typical reported impact by 50%.

## 6.3 Card, Kluve, and Weber (2018), “What works? A meta analysis of recent active labor market program evaluations”

In the small fleet of meta-analyses of labor market programs, Card, Kluve, and Weber (2018) (CKW) is the flagship. Its scope extends beyond those of the meta-analyses above, to all individually targeted programs meant to help people into the labor market. That includes job training, job search assistance, subsidies for private employment, and public employment. This second edition of the work gathers 857 impact estimates from 207 papers. Fully 502 estimates come from the Germanic and Nordic countries, and 87 from Anglo-Saxon nations (US, UK, Australia, New Zealand, and Canada; CKW, Table 1). Training programs are best represented, with 418 estimates. Of these, 54 result from randomized experiments (CKW, Table 3).

The big data set gives CKW more statistical power to address some important questions. Which kinds of labor market interventions work best? Do their impacts fade or persist? How does the answer depend on the target demographic? Do the programs work better during recessions?

CKW discovers that while the outcome most often tracked in papers from Anglo-Saxon countries is how much a person earns, in the rest of the world—and thus overall in the sample—it is whether a person has a job. The difference in emphasis may reflect a difference in the structure of labor markets (Krugman 1994), since minimum wages, worker protections, and unemployment have historically been higher in the EU than in the US. The predominance of employment as an outcome leads CKW to focus on it exclusively.

To analyze the various impact estimates, CKW first makes them comparable. In contrast with nearly all the research we have encountered in this review, which estimates simple treatment-control differences at various follow-up dates (ITTs), there are studies, especially non-randomized ones in Europe, that express results in other ways. A typical alternative unit is the *probability per unit time of getting a job if one has not already done so*, which is a "hazard rate." For example, a job training program in Germany might boost an unemployed person's chance of finding a job from 10% to 20% per month. To make all results comparable, CKW coarsens them into a trichotomous indicator, of whether an impact estimate is negative and significant, insignificant, or positive and significant (at $p < 0.05$). Since this quantization makes a serious tradeoff—more studies brought together for meta-analysis but with less precise information from each—CKW also runs the meta-analysis without it, restricting to the studies that report simple impacts on employment.[37,38] The two approaches generate broadly parallel results. We will focus on those from the smaller sample, with actual employment impacts, since they are easier to interpret.

A separate and subtle technicality has to do with how results are averaged across studies. It is neither common-effect nor random-effects. CKW, which operates in the tradition of economics rather than medicine, goes for a mathematically simpler alternative: it weights all studies equally, regardless of precision.[39] This can be thought of as taking random-effects meta-analysis to an extreme in which the impacts vary so much across study contexts that small studies are as informative as large ones. And since labor market programs *do* clearly vary widely in their impacts, CKW's equal-weighting may not deviate far from the balances struck by conventional random-effects methods. Still, so heavily weighting small studies may make the results noisier.

---

[37]The trichotomous variable is modeled as ordered probit.
[38]This quantization also works around another problem that is common in non-academic impact studies: statistical significance is not reported, except through indications that it exceeds thresholds such as $p <$ 0.05. Smedslund et al. (2006) handles this problem by imputing $p$ = 0.005, 0.03, 0.075, and 0.1 for three, two, one, or no stars. We will do nearly the same in §6.7.
[39]Nodding to Solon, Haider, and Wooldridge (2015), CKW accompanies this unweighted OLS with standard errors that are heteroskedasticity-robust, indeed clustered by study.

CKW finds that:

- Job training lifts employment by 1.4–2 percentage points in the first year from the start of training, and 4.0–4.6 points more than that in year 2 and beyond (CKW, Table 5, cols 3 & 4).[40] This makes sense since training can temporarily pull people out of the labor market.

- The average impact of job training on the long-term unemployed is 12 points higher than for "regular" recipients of unemployment insurance (CKW, Table 9, col. 2).

- The average impact is 6 points higher for women (CKW, Table 9, col. 2).

- The average impact is 3 points higher for those younger than 25 than those older (CKW, Table 9, col. 2).

- Across Denmark, France, Germany, and the US (countries with enough data to check) individually targeted programs as a whole (not just training programs) helped people more during recessions. They increased employment by about 3 points extra for each 1-point rise in the national unemployment rate or 1-point drop in economic growth.

- This meta-analysis finds little publication bias. Nor do results that make it into *academic* publications differ on average from those that do not.[41]

- Experimental estimates are about 1 point lower on average than non-experimental ones, but that result has essentially no statistical significance.

Interpreted as statements about causation, most of the results are broadly plausible. Of course, the results come to us through several layers of uncertainty. Most of the studies are not randomized. (In fact, dozens of randomized studies are not included, as we will see in the next section.) The meta-analytical method puts especially heavy weight on small studies. Many hypotheses are tested, without any adjustment for the fact that a certain number of findings will be statistically significant by chance. And, as CKW emphasizes, when comparing impacts across groups, correlation is not causation. We would be surprised if serving women rather than men truly causes programs to lift employment by 6 more points.

We find the results for the time-profile of impacts—low in the first year, then rising—especially plausible. We are less confident in the subgroup findings, causally interpreted.

---

[40]The short-term impact is the constant term in the regression estimate, which is not reported in CKW and is instead extracted from the public data and code.
[41]The CKW (Figure 3) funnel plots are visually and technically complicated, so they are not reproduced here.

## 6.4 Peck et al. (2021), “A Meta-analysis of 46 career pathways impact evaluations”

In recent decades, two major paradigms have emerged in the American community of practice in subsidized job training: sectoral strategies and career pathways. Sectoral strategies stress working with employers in designing training programs; section 7 will give them great attention. Career pathways programs are about supporting participants through steps of graduated difficulty, such as schooling, on-the-job training, and work experience (Fein 2012; Sarna and Strawn 2018, Exhibit 2.1). In the 2010s, parts of the US government invested in expanding and assessing the evidence base on career pathways programs. One result is the Peck et al. (2021) meta-analysis.

The meta-analysis is framed by two questions: What are the average effects of the programs? And which aspects of program design and implementation are associated with success? The study gathers results from 46 career pathways studies, 27 randomized and 19 not.

The answer to the first question is “positive.” Peck et al. (2021, Exhibit 2-1) estimates that career pathways programs increased employment *in the professions they targeted*—by an impressive 72%, from 26 percentage points to 45 points. To this extent, the programs worked as intended. More broadly, the programs increased total employment by about 6 points, from a base of about 60%. However, these statistics are averages across whatever timeframes the underlying studies use in computing impacts (Peck et al. 2021, p. 123), so we cannot tell whether the impacts started high and fell as in the Job Corps evaluation, or the opposite, as CKW found. Impacts on earnings *are* broken out by timeframe; they amount to $1,040/year in the first three years after treatment, over a base of $16,320. The longer-term impacts appear to be similar in magnitude, but the report never expresses them in dollars because they just miss the report’s significance threshold (Peck et al. 2021, Exhibit D-5).[42]

To answer the second research question, about correlates of impact, the study compiles a list of more than 75 program traits, from whether the training lasts more than a year to whether it leads to a college degree. Coding these traits—assigning the 0’s and 1’s for some 75 traits and 46 studies—constitutes a distinctive contribution. It will inform our analysis of sector programs in §7.

Since there are more traits than studies, it is impossible to identify the impacts of each. In fact, a common rule of thumb is that 10 studies are needed for each potential correlate, which here would allow only about 4.6 traits. To narrow the field, Peck et al. groups traits into blocks, with themes such as demographics and eligibility criteria. Then outcomes including employment and earnings impacts are regressed upon each block alone. Only the “survivors” are retained: those statistically significant at $p < 0.1$ and having a standardized coefficient exceeding 0.1 in magnitude.[43] Finally, the outcomes are regressed only on all the survivors. The process is carried out separately for each outcome and timeframe.

[42]The main analysis is Bayesian, so uncertainty is expressed with credible intervals. Peck et al. (2021, Exhibit D-6) estimates the probability of any long-term (beyond three-year) effect at 89%—and then imposes a threshold of 90%. The meta-analysis is conducted on standardized effect sizes. Because this result is discarded as insignificant, it is never translated back into dollar terms.

Peck et al. nicely presents the results on correlates of success in a diagram, which is here excerpted as Figure 11. A handful of traits appear in the rightmost column, having the largest association with positive impact on employment or earnings. Programs *not* led by a community college performed better, as did ones that set minimum-income requirements, or upskilled workers to help them stay employed at the same company, or involved employers in designing curricula. As with the CKW meta-analysis, it is hard to know how much weight to put on these non-experimental correlations, even as they look broadly plausible.

[43]Standardized coefficients are those that arise after standardizing all variables, i.e., dividing them by their sample standard deviations.

| CHARACTERISTICS | NOT ASSOCIATED | SMALLER IMPACTS | LARGER IMPACTS |
|---|---|---|---|
| ADMINISTRATIVE ARRANGEMENTS | ■ School or school district (partner)<br>■ Labor union (partner)<br>■ Any other agency type | ■ Community or technical college (lead or partner)<br>■ Government agency (partner)<br>■ University (partner)<br>■ Other (partner) | ■ Workforce agency (lead)<br>■ Community organization (lead)<br>■ Trade association (partner) |
| ELIGIBILITY | ■ Any other eligibility criteria | — | ■ Program has income requirements |
| INSTRUCTIONAL OFFERINGS | ■ Offers basic skills training<br>■ Offers basic/secondary education, English language acquisition, developmental or remedial education<br>■ Offers alternative times/places | ■ Offers flexible sequencing, hybrid instruction, online courses | — |
| PATHWAYS AND TRAINING | — | — | ■ Offers training above entry-level |
| CREDENTIAL TYPES | ■ State or local licensure<br>■ Certification developed by an employer or industry association | ■ Occupational certificate or technical diploma<br>■ College degree (associate's, bachelor's) | — |
| TRAINING DURATION | ■ All training durations | — | — |
| TRAINING INDUSTRY | ■ Healthcare<br>■ Manufacturing or construction<br>■ Information technology<br>■ Education | — | — |
| EMPLOYER ENGAGEMENT ACTIVITIES | ■ Formal employer partnership | ■ Program convenes an employer advisory council | ■ Offers incumbent worker training |
| EMPLOYER ROLE | ■ Offers work-based learning (paid or unpaid)<br>■ Employer provides resources<br>■ Employer delivers instruction<br>■ Employer provides labor market information | ■ Employers deliver career awareness services | ■ Provide input on curriculum or program development |
| ONE-ON-ONE SUPPORT | ■ Case management<br>■ Career or college navigation<br>■ Staff assistance mandatory/required | ■ Academic advising | — |
| SUPPORT SERVICES | ■ Emergency assistance<br>■ Food assistance<br>■ Internet<br>■ Tutoring<br>■ Child/dependent care assistance<br>■ Transportation assistance<br>■ Connection with benefits and social services<br>■ Job search and placement | ■ Tuition, training cost, other financial assistance | — |
| PARTICIPANT COMPOSITION | ■ Gender<br>■ Education levels<br>■ Race/ethnicity | — | — |
| LOCAL CONTEXT | ■ Unemployment rate | — | — |

**Figure 11. Correlates of positive impacts on employment and earnings, reproduced using Peck et al. (2021)**
*Notes: Reproduced from Peck et al. (2021), Exhibit 3-2.*

## 6.5 A new meta-analysis of randomized job training experiments in the US

To develop a systematic view on the rich evidence from randomized job training trials in the US, we perform an aggressive search for studies and meta-analyze them. We deploy Claude Opus—with engaged oversight—to extract the needed information on each experiment from the thousands of pages of reports.

To build the sample, we search for US-based randomized trials of interventions that included training, in the following reviews: Greenberg and Shroder (2004); Smedslund et al. (2006); Dutta-Gupta et al. (2016); Card, Kluve, and Weber (2018); Peck et al. (2021); and Katz et al. (2022). We added recent follow-ups on the PACE and WorkAdvance evaluations.

The humble warning in Card, Kluve, and Weber (2010)—"There are likely to be measurement errors and errors of interpretation in the extraction of information from the studies."—applies as well to this AI-accelerated effort. The collection and collation of studies involves many judgment calls—some here made by humans and some by AI. Does an intervention involve training enough that it belongs in the sample? Is training primary in the intervention? If an impact is reported for months 7–18 since randomization, is that a short-term (year-1) or medium-term (year-2) impact?

To improve the extraction, we instruct Claude to: a) produce a "reasoning table" documenting sources and logic for extracted values and b) perform an adversarial review of all extractions, by reading the reasoning table and returning to sources. As in analyses done without the aid of AI, as we learn our way around the data and analyze it, we spot-check and revise the extraction.

Another limitation is typical in this literature: many non-academic impact studies do not report continuous metrics of statistical significance, such as standard errors. Instead they only indicate whether significance exceeds certain standard thresholds. When the indicator is binary (treatment take-up, employment) and no controls are included, this is not a problem because a standard formula links point estimates and standard errors. When it is not, then, for binary outcomes, we use that formula anyway, while for the continuous outcome of earnings, we follow Greenberg, Michalopoulos, and Robins (2006, note 9) in imputing $p = 0.005$, 0.03, 0.075, and 0.55 for $p < 0.01$, $p < 0.05$, $p < 0.1$, and $p > 0.1$. These uncertain imputations influence the meta-analytic, random-effects results.

### 6.5.1 The sample

Our sample consists of 56 studies. Of these, 24 report results for several demographics, implementing organizations, or localities. This brings the number of impact estimates to 146.

The sample covers its niche more thoroughly than previous meta-analyses. Thirty-seven of the 56 studies are in Smedslund et al. (2006), zero are in Haelermans and Borghans (2012), 2 are in Card, Kluve, and Weber (2018), and 26 are in Peck et al. (2021). The sample includes the four major US-based experiments reviewed in §4 as well as many studies of smaller programs.

Because workforce programs are diverse and multifaceted, job training features more prominently in some than others, and drawing the line between training and non-training interventions is difficult. We erred toward inclusion, while having Claude assess whether job training was a primary feature in each program. Roughly half the experiments cleared this hurdle: 33, with 78 impact estimates. We meta-analyze the full set of studies as well as this "training-primary" subset.

| | Experiments incorporating training | | | Those with training primary | | |
|---|---|---|---|---|---|---|
| | Mean | SD | N | Mean | SD | N |
| **Setting** | | | | | | |
| Randomization year | 1998 | 12 | 146 | 1998 | 13 | 78 |
| South (%) | 18 | 38 | 146 | 18 | 39 | 78 |
| Northeast (%) | 19 | 40 | 146 | 19 | 40 | 78 |
| Midwest (%) | 22 | 42 | 146 | 17 | 38 | 78 |
| West (%) | 23 | 42 | 146 | 17 | 38 | 78 |
| Urban (%) | 90 | 30 | 146 | 97 | 16 | 78 |
| Rural (%) | 8 | 26 | 146 | 3 | 16 | 78 |
| Unemp. at randomization (%) | 6.4 | 1.3 | 146 | 6.5 | 1.3 | 78 |
| **Target population** | | | | | | |
| Welfare or low-income adults (%) | 62 | 49 | 146 | 54 | 50 | 78 |
| Noncustodial parents (%) | 3 | 18 | 146 | 0 | 0 | 78 |
| Substance abuse (%) | 1 | 12 | 146 | 3 | 16 | 78 |
| Justice-involved (%) | 6 | 24 | 146 | 1 | 11 | 78 |
| Dislocated workers (%) | 3 | 16 | 146 | 4 | 19 | 78 |
| Youth (%) | 23 | 42 | 146 | 36 | 48 | 78 |
| Disability (%) | 1 | 12 | 146 | 3 | 16 | 78 |
| **Demographics** | | | | | | |
| Male (%) | 37 | 34 | 135 | 40 | 31 | 78 |
| Mean age | 29 | 7 | 113 | 28 | 8 | 59 |
| Black (%) | 47 | 27 | 132 | 47 | 29 | 75 |
| Hispanic (%) | 26 | 25 | 117 | 28 | 26 | 70 |
| No high school diploma (%) | 44 | 30 | 117 | 46 | 38 | 65 |
| **Treatment** | | | | | | |
| Classroom training (%) | 77 | 42 | 146 | 90 | 31 | 78 |
| On-the-job training (%) | 39 | 49 | 146 | 29 | 46 | 78 |
| Job search assistance (%) | 83 | 38 | 146 | 85 | 36 | 78 |
| Mandatory (%) | 28 | 45 | 146 | 12 | 32 | 78 |
| Treatment duration (months) | 5.4 | 2.7 | 86 | 5.8 | 2.9 | 61 |
| Cost per treated (2025 $) | 11,402 | 8,629 | 98 | 13,598 | 9,359 | 52 |
| Scaled (%) | 7 | 25 | 146 | 10 | 31 | 78 |
| **Employer engagement** | | | | | | |
| Sector program (%) | 12 | 32 | 146 | 22 | 42 | 78 |
| Curricula adapted for employers (%) | 18 | 38 | 145 | 33 | 47 | 78 |
| Job development/placement (%) | 31 | 46 | 145 | 40 | 49 | 78 |
| Employer input on curriculum (%) | 19 | 39 | 145 | 35 | 48 | 78 |
| Employer commits to hiring (%) | 3 | 18 | 145 | 5 | 22 | 78 |
| **Administration** | | | | | | |
| Funding: public (%) | 77 | 42 | 146 | 64 | 48 | 78 |
| Funding: public & private (%) | 19 | 40 | 146 | 31 | 46 | 78 |
| Funding: private (%) | 3 | 18 | 146 | 5 | 22 | 78 |
| Admin: public (%) | 45 | 50 | 146 | 40 | 49 | 78 |
| Admin: public & private (%) | 18 | 38 | 146 | 15 | 36 | 78 |
| Admin: private (%) | 37 | 48 | 146 | 45 | 50 | 78 |
| **Study characteristics** | | | | | | |
| Sample size | 2,948 | 6,738 | 146 | 2,083 | 4,040 | 78 |
| Outcome data from official records (%) | 70 | 46 | 146 | 59 | 50 | 78 |
| Response rate (%) | 97 | 11 | 113 | 96 | 12 | 53 |
| Academic publication (%) | 7 | 25 | 146 | 3 | 16 | 78 |

**Table 1. Study traits in US-based job training experiments**

*Notes: Full sample is 146 estimates from 56 studies. Averages are unweighted. Sample sizes vary because of missing data. For multi-region experiments, "regional" unemployment is national. Cost figures are per treatment group member.*

Most of the study traits we collect are listed in Table 1 along with their unweighted averages and standard deviations. The studies are overwhelmingly set in cities and spread rather evenly across Census regions. Most aim to serve low-income people, with only 3–4% meant for dislocated workers. Women and blacks disproportionately enter the study samples. Where training is primary, classroom training is a component 90% of the time and on-the-job training just 29%. People spend about 6 months in these programs, at a cost of $13,598 per person offered treatment (in 2025 dollars; not per person trained). Self-described sector programs are a minority, as are, more generally, efforts to engage employers. Nearly all studied interventions were publicly funded, but slightly less than half the training-primary interventions delegated local implementation to private organizations. Few of the experiments are written up in academic journals.

Figure 1 provides a graphical overview of the impact estimates in the data set. It shows all the annual—or annualized—earnings impacts as a function of follow-up time. In fact, the data points differ from those in the meta-analysis because they are kept at the cadence used in the primary sources—monthly, quarterly, yearly, etc. The three federal programs—JTPA, Job Corps, and WIA—are plotted in blues. Some standout sector programs are marked in shades of magenta.

### 6.5.2 Average impacts

| | **Experiments incorporating training** | | | **Those with training primary** | | |
|---|---|---|---|---|---|---|
| | **Control mean** | **Impact** | **N** | **Control mean** | **Impact** | **N** |
| Program take-up (%) | 10.7<br>(2.8) | 58.3<br>(5.8) | 140 | 8.6<br>(3.5) | 66.1<br>(5.7) | 78 |
| Any training take-up (%) | 40.3<br>(2.9) | 19.2<br>(4.4) | 95 | 49.1<br>(2.8) | 28.1<br>(4.9) | 58 |
| Employment, short-term (%) | 56.0<br>(3.0) | 4.5<br>(1.4) | 93 | 59.5<br>(4.3) | 4.8<br>(1.7) | 41 |
| Employment, medium-term (%) | 56.9<br>(2.8) | 2.6<br>(0.4) | 87 | 62.6<br>(4.0) | 2.8<br>(0.6) | 40 |
| Employment, long-term (%) | 59.8<br>(4.4) | 1.8<br>(0.5) | 64 | 63.4<br>(5.1) | 1.7<br>(0.5) | 36 |
| Earnings, short-term (2025 $) | 10,575<br>(1,194) | 640<br>(244) | 119 | 13,834<br>(2,038) | 835<br>(553) | 53 |
| Earnings, medium-term (2025 $) | 12,577<br>(1,020) | 984<br>(255) | 121 | 14,645<br>(1,473) | 1,139<br>(446) | 67 |
| Earnings, long-term (2025 $) | 14,361<br>(1,802) | 657<br>(185) | 81 | 16,148<br>(2,443) | 791<br>(302) | 53 |

**Table 2. Control groups means and impact estimates in US-based job training experiments**
*Notes: Impacts are estimated with restricted maximum likelihood (REML) meta-analysis. Control group averages are computed using the corresponding REML weights. The numbers in parentheses are standard deviations of control-group values and standard errors of impact estimates. Sample sizes vary because of missing data. Short-term outcomes have a ~1-year follow-up, medium-term ~2 years, and long-term 3–5 years.*

Random-effects impact averages appear in Table 2. The outcomes are four: the take-up rate for the tested program; the take-up rate for *any* training; employment; and earnings. Impacts on the latter two are split by timeframe. Following CKW, the timeframes are short-term (year 1 after randomization), medium-term (year 2), or long-term. We implement the latter as years 3–5, subject to data availability, because most studies stop follow-up before year 6, and many well before that.[44] Since some studies produce estimates for multiple populations, which may not be fully independent, we cluster standard errors by study. We use the dominant random-effects estimator, restricted maximum likelihood (REML). The results are therefore geared to represent the impact of the typical study, not the average impact on participants across all studies.

If “work” means to increase employment and earnings on average, and to a statistically significant degree, then job training has worked in the US. In Table 2, 11 of the 12 mean impact estimates are 2–4 times their standard errors, making them highly significant. (All are ITTs, expressed per person offered treatment.) The exception is for the short-term impact of training-primary programs on earnings, which may reflect temporary earnings losses while in training.

However, the average training program does not make a big difference. From a base employment rate of about 60%, it boosts the chance of having a job by 2.6-2.8 points in the medium term (year 2) and 1.7–1.8 points in the longer term (the pairs of numbers being for all programs and training-primary ones). The average program lifts pre-tax earnings by 7.8% in the medium term, meaning by $1,000–1,100/year from $12,500 in the full sample and $14,700 in the training-primary sample. The long-term earnings impacts are 4.6%, i.e., $700–800 from bases of $14,400 and $16,100 (all figures in 2025 dollars).

One cause of the low impacts on labor market outcomes—but also a source of hope in some scenarios—is the low impact on participation in training. In training-primary interventions, the experimental offer of entry is accepted 75% of the time, while only 9% of the control group manages to access the treatment. Dividing that difference, 66%, into the ITTs quoted just above produces what we call “LATE-program” estimates 50% larger. Similarly, the impact on participation in *any* training program is only 28%, because many control group members find alternative trainings. If these substitutes are equally effective, then the impacts of training per se—the values of “LATE-training”—are 1/0.28 ≈ 4 times the ITTs reported in the lower rows of Table 2. If the average training program were hypothetically run in a context where alternatives were scarcer, its training take-up and the ITTs could be up to 4 times higher.

[44]Longer-term estimates are available for a few programs, notably Year Up, for 7 years (Fein and Dastrup 2022), Project QUEST for 11 (Roder and Elliott 2021), and the Job Corps for 20 (Schochet 2021). However, there may be a tendency to follow successful programs longer.

### 6.5.3 Benefit and cost

Many studies in our sample compare the benefits and costs of training. All count the upfront cost of the programs and the impacts on (pre-tax) earnings. They vary in what else they include: impacts on crime, tax transfers, the value of goods and services produced during training, etc.

In the face of this diversity, to perform a benefit-cost analysis of meta-analytical results, we mostly hew to the model of the recent, four-site WorkAdvance demonstration (Schaberg and Greenberg 2020).[45] See Appendix G for details. The analysis distinguishes between the perspectives of the participant, government, and society, the last being the algebraic sum of the first two. It estimates transfers between government and participants—more precisely, treatment group members—associated with taxes and reliance on public benefits such as food stamps. Sales taxes, which in the US are levied by state and local governments, are included, for they are another channel that transfers some of what is earned and spent from trainee to the public fisc. The analysis adds (or subtracts) work-related spending, such as on uniforms and gas to get to work, and the opportunity cost of unpaid time.[46]

We compute five summary statistics on the information arrayed in this framework:

- A narrow benefit-cost (B-C) ratio. The numerator is the net present value of lifetime impacts on total compensation (not just pre-tax earnings). The denominator is upfront training costs.

- A broader B-C ratio, which nets out from the denominator the reduction in spending on other training.

- The marginal value of public funds (MVPF), which is the ratio of the net benefit to the participant to the net cost for governments. Here, government is assumed to fund the training.

- The ratio of net cost to gross cost for government, which indicates how much training pays for itself through higher tax revenue and lower spending on public benefits (again, assuming government pays for it).

- The internal rate of return (IRR) from the societal point of view, which is the discount rate at which costs and benefits balance.

[45]The late David H. Greenberg also coauthored a textbook on benefit-cost analysis.
[46]Two studies in our sample estimate and monetize crime impacts: MDRC (1980) on the National Supported Work Demonstration (ex-addict and ex-convict groups) and McConnell and Glazerman (2001) on Job Corps. We opt not to because the linkage is complex and incompletely understood. For example, successful applicants to sector programs could have much lower criminal propensity than school dropouts, ex-convicts, and ex-addicts. Likewise for AI-displaced workers.

Although the narrow benefit-cost ratio as defined here leaves out tax flows and other factors, it has the virtue of being easy to understand, thus perhaps more salient to policymakers. It is meaningful for private funders too, who cannot net out tax savings.

All dollar figures are expressed *per treatment group member*, not per trainee. In other words, they are on an ITT rather than LATE basis. That distinction, however, does not affect the five summary statistics.

For training-primary programs, we roughly estimate that:

- The average program does a bit better than breaking even, with a narrow B-C ratio of 1.44, a wide one of 1.80, and an MVPF of 2.52.

- The average program, assuming it is government-funded, ultimately covers some 76% of its costs. That figure includes fiscal savings for federal, state, and local governments.

- The societal IRR is 5.97%.

### 6.5.4 Correlates of impact

For extrapolating to an AI-disrupted future, it matters how the impacts differ across subgroups. For example, if training has worked better for mature adults, that will speak (tentatively) to its efficacy for dislocated, midcareer workers. That said, the caveat about cross-group comparisons bears repeating: no matter how rigorous the individual studies, meta-analytical comparisons among them are not as dispositive as to which factors *cause* training to be more effective as distinct from merely being correlated with them.

To prune the list of potential correlates in our meta-regressions, we copy the block-based method in the Peck et al. (2021) meta-analysis (§6.4). In addition, to compensate for missing values in some variables, we run the meta-regressions with and without multiple imputation, which is a method for filling in the missing values while adjusting the standard errors for the uncertainty of those imputations.[47]

[47]We generate 10 imputed data sets, separately for each outcome and timeframe.

| | Employment | | | | Earnings | | | |
|---|---|---|---|---|---|---|---|---|
| | Medium-term | | Long-term | | Medium-term | | Long-term | |
| **Setting (omitted = South, urban)** | | | | | | | | |
| Randomization year | −0.08** | −0.06 | −0.07* | −0.05 | | | | |
| | (0.04) | (0.04) | (0.04) | (0.05) | | | | |
| Northeast | 4.36*** | 4.34*** | | | | | | |
| | (1.28) | (1.30) | | | | | | |
| Midwest | 3.61*** | 3.37*** | −0.04 | −0.29 | | | | |
| | (0.91) | (0.92) | (0.85) | (0.90) | | | | |
| West | 2.20* | 2.13* | | | | | | |
| | (1.15) | (1.26) | | | | | | |
| Unemployment over follow-up (%) | 1.19*** | 1.18*** | | | 267* | 225* | | |
| | (0.36) | (0.38) | | | (155) | (135) | | |
| **Target population (omitted = welfare or low-income adult)** | | | | | | | | |
| Substance abuse | | | 7.12*** | 7.63*** | | | 4468*** | 4468*** |
| | | | (1.43) | (1.08) | | | (114) | (114) |
| Justice-involved | | | | | | | 2701*** | 2701*** |
| | | | | | | | (114) | (114) |
| Dislocated workers | −4.05*** | −3.15*** | −2.57** | −2.65** | | | | |
| | (0.95) | (1.07) | (1.11) | (1.24) | | | | |
| Youth | −2.22** | −3.00*** | | | | | | |
| | (0.96) | (0.97) | | | | | | |
| Disability | 6.05** | 6.28** | 8.18*** | 8.50*** | | | | |
| | (2.18) | (3.11) | (1.32) | (0.77) | | | | |
| **Treatment (omitted = mandatory)** | | | | | | | | |
| Classroom training | | 0.08 | 0.33 | | | | | |
| | | (0.85) | (1.20) | | | | | |
| On-the-job training | | | | | | 690** | | |
| | | | | | | (333) | | |
| Voluntary | | | | −0.83 | | | | |
| | | | | (0.85) | | | | |
| Cost per treated ($1,000s) | | 0.00 | | | | | | |
| | | (0.00) | | | | | | |
| **Employer engagement** | | | | | | | | |
| Employer commits to hiring | | | | | 8830*** | 8569*** | 9161*** | 9161*** |
| | | | | | (1813) | (1735) | (135) | (135) |
| **Study characteristics (omitted = outcome data source: official records)** | | | | | | | | |
| Outcome data source: survey | | | | | | | −315* | −315* |
| | | | | | | | (176) | (176) |
| Multiple imputation | | ✓ | | ✓ | | ✓ | | ✓ |
| $R^2$ | 0.36 | 0.44 | 0.21 | 0.24 | 0.69 | 0.72 | 0.87 | 0.87 |
| Number of experiments | 87 | 87 | 64 | 64 | 120 | 121 | 81 | 81 |

**Table 3. Meta-regressions on “survivors” of initial block-by-block meta-regressions**

*Notes: REML estimates. Employment impacts are measured in percent, and earnings impacts are in 2025 dollars. As explained in text, in each regression, the included regressors are survivors of initial block-by-block regressions. Medium-term impacts are ~2 years and long-term >~3 years. Standard errors in parentheses, clustered by study. Results in MI columns (✓) are pooled across 10 imputations; other columns use complete cases. Each pair of employment regressions produces different results despite the same sample size because missingness in non-surviving variables restricts the samples in some initial regressions that determine survivors. * p < 0.10, ** p < 0.05, *** p < 0.01*

The final meta-regressions—the ones on the "survivors" of block-based selection—are reported in Table 3. For statistical power, the sample includes all studies, not just training-primary ones. Results for short-term (year-1) impacts are omitted since they matter less and may be dominated by temporary effects of being in the programs. In reviewing the results, it is best to focus on consistent patterns across the columns.

We see that the medium- and long-term impacts on employment have fallen by about 0.07 points per year since the 1970s. Programs tested in the South (the reference region in the first panel) did not work as well for employment; employment impacts in the other regions (Northeast, Midwest, West) were 2–4 percentage points higher. Classroom training showed no clear advantage, and it appears harder to help dislocated workers.

Other statistically strong results are associated with small numbers of programs. The sole program with long-term results that expressly targeted people with drug problems or past involvement with the law—the National Supported Work Demonstration—entirely drives the impressive impacts on the "justice-involved." The two programs that served people with intellectual disabilities boosted employment about 8 points in the long term.[48,49] And the apparently powerful impact of an employer committing to hire graduates—$8,000–9,000 in year 2 and beyond—owes to three programs: the Wildcat program that inspired the NSWD, Year Up, and the Wisconsin Regional Training Partnership. Section 7.1 will discuss the second and third.

Intriguingly, the local unemployment rate is also associated with the effectiveness of training programs. To investigate the link to job market conditions, we include a covariate giving the average regional unemployment rate during the follow-up for that study. (The "region" is one of the four Census regions or, for multi-site evaluations, the nation.) Where the unemployment rate is 1 point higher, a training program boosts employment by 1.2 points more in the medium term (year 2 after randomization) and earnings by $250/year. However, the association is much weaker in the longer term (years 3–5). In the initial regression for the block of various variables characterizing the study setting (not shown here), the coefficient on long-term employment is only 0.37 points and is quite indistinguishable from 0, statistically. It does not survive into the long-term regressions reported in the table, so the corresponding cells are blank.

Several theories could explain the medium-term association between the local unemployment rate and the impact of training. When unemployment is high, the marginal participant may be someone who just lost a job, who has a track record of productive work, and who is more able to take advantage of training and the opportunities it leads to. It could also be that mismatch unemployment is higher when unemployment is higher, which increases the returns from switching industries (Şahin et al. 2014). A final explanation is that when unemployment spikes, competition for training increases. That reduces the alternatives available to the control group, against whom results in the treatment group are benchmarked. In effect, each training program becomes more needed and less redundant. As discussed in section 3.2, that could lift the effect of *offering* training (ITT) even without any change in the *impact* of training on the trained.

---

[48]The Structured Training and Employment Transitional Services trial (Kerachsky et al. 1985) and the Transitional Employment Training Demonstration (Decker and Thornton 1995).
[49]The relatively low impact on programs for dislocated workers solely reflects the null results of the WIA (§4.4). Two other programs in the sample—the New Jersey Unemployment Insurance Reemployment Demonstration (Anderson, Corson, and Decker 1990) and the Texas Worker Adjustment Demonstration (Bloom 1990)—also targeted this group. But the studies only reported short-term impacts.

Evidence not shown in the table supports the latter theory. In the initial block-by-block regressions, a 1-point increase in unemployment is associated with 3.4 more points of program participation; the standard error is 2.8, meaning that the finding, while suggestive, does not meet the statistical significance “survival” criterion applied here. But when we switch from *participation in the program of interest* to *participation in any training program*, the medium-term association with the unemployment rate becomes even bigger and quite statistically strong, at 4.7 points (standard error 1.8). Indeed, when jobs are tighter, offering another training program increases the number of people getting any training.

A distinct puzzle is how to interpret the qualitatively different associations between impact and unemployment over the medium and long term. The weak association in the long term might just owe to the shrinkage of sample and the loss of statistical power. Or it may be that training’s direct impact on employment comes mainly in the months after exiting a program—roughly, the medium term. That effect may persist over years, as those with jobs retain them, but may depend little on the contemporaneous unemployment rate. In Appendix F, we report results from meta-regressions like those in Table 3, with one addition: the impact of training in the previous timeframe is controlled for. We expect that its coefficient will be smaller in the medium-term impact meta-regressions because there the persistence term is for the short term (year 1 after randomization), and contains transient artifacts of being in training. Sometimes training takes people out of the job market. Sometimes it guarantees them a job, temporarily.

This “persistence” term, shown in the top row of the table in Appendix F, is statistically strong throughout. It is smaller in the medium term for employment (as expected), but not for earnings (surprisingly). In the long-term regressions without multiple imputation, the medium-term impacts gain coefficients of 0.48 and 0.52.[50] This suggests that the medium-term associations with unemployment mentioned above—1.2 points of employment impact and $250/year of earnings impact—persist into years 3–5 at half those levels.

Overall, the results relating to the unemployment rate suggest that training helps people bounce back faster after losing their jobs. However, the lack of longer-term association suggests that local job market conditions do not matter for whether training puts people on a permanently better work track.

Forest plots showing the impacts of individual studies along with these bottom lines are viewable through a web interface at [droodman.github.io/job-training-meta-analysis](https://droodman.github.io/job-training-meta-analysis). Also posted there are tables reporting the block-by-block meta-regressions from which the survivors are extracted.

[50] The multiple imputation regressions are somewhat less reliable here because they assume that the missingness in long-term impact estimates is unrelated to the new regressor, medium-term impact. In fact there is some tendency to follow successful experiments longer.

### 6.5.5 Randomized job training experiments in the US: summary

The meta-analysis of randomized trials in the US supports these conclusions:

- Job training programs on average boost employment and earnings.

- The ITT impacts are small, at about 2 points of employment, and $1,000–1,100 in pre-tax earnings in year 2 and $700–800/year beyond. This is enough to offset the costs, we roughly estimate.

- Programs backed by a hiring commitment from employers increase the earnings of admitted students an order of magnitude more. The result is driven by a small number of studies but is intriguing in light of the excitement about "sector programs," to which we will soon turn.

# 7 Zeroing in on sector programs

## 7.1 History

As the US government lumbered through iterations and evaluations of its billion-dollar workforce programs, social entrepreneurs, foundations, and state and local governments continued experimenting at smaller scales. As mentioned, one paradigm that emerged was the "sector strategy," programs that worked closely with employers, screened job seekers, taught them occupational and work-readiness skills, and placed them in quality jobs in specific high-demand industries. In a sort of convergent evolution, elements of the approach appeared in the 1980s, in spots across the country (Mangat 2007; Conway 2014; King and Prince 2015). Eventually observers noted the similarities and gave the family a name.[51]

While experts may disagree on nuances, all stress that sector programs actively intermediate between potential employers and potential employees. This stance contrasts with the implicit "train and pray" approach of most schools and community colleges. Intermediaries who work closely with major local employers can keep tabs on which skills are in demand, incorporate that into curricula, arrange internships, and earn employers' trust that their graduates make good workers. King and Prince (2015) explains that the sector approach breaks from a common pattern:

> …in which multiple training providers, to degrees varying between "hardly at all" and "effectively," identified the skills in demand, created curricula… and then competed… to have their trainees hired. Duplication of effort, inconsistency in

[51] See footnote 6 for more on the definition.

> training standards, and the occasional fly-by-night training providers all contributed to employers' suspicion…. [E]ducation and training institutions have little incentive to engage employers because their funding is based on enrollment in, and sometimes completion of, classes rather than on job placement.

Most of the trial, error, and learning that forged the sectoral strategy took place without recourse to randomized trials. But the example set by the National Supported Work Demonstration *did* play a role, as private foundations sought to rigorously test claims, and positive results stoked optimism and funding. Those encouraging results are precisely why this section is devoted to sector programs.

When thinking about how to build on the success of some sector programs, it is important to recognize that how they work is intertwined with how they came to be. Embedding the model in another institutional context, such as a federal program, might lose the essence, depending on how done. We therefore begin by recapping the history while emphasizing the role of randomized research.

### 7.1.1 The Center for Employment Training in San Jose

In the 1980s, after the federal government terminated the NSWD, the Rockefeller Foundation funded a set of four randomized studies called the Minority Female Single Parent demonstration. One site stood out: the Center for Employment Training (CET) in San Jose. In the last year of the 2.5-year follow-up, CET boosted employment from 57% to 66%, and monthly pay from $405 to $506 (in 1986 dollars; Burghardt et al. 1992, Table IV.1).

Meanwhile, with funding through the JTPA, MDRC launched a 13-site demonstration called JOBSTART. It tested ways to help young school dropouts, the group that the NSWD had helped least. Once more, one program stands out in the results tables: CET. In the last two years of follow-up, CET's treatment-control difference averaged $250/month. (Cave et al. 1993, Table 5.12)

What made CET different? It was *not* that the students spent six months in the classroom learning occupational skills. More distinctive was that CET shared features with what would later be called sector programs, "including a focus on good jobs and links to employers" (Giloth 2010, p. 20).

Impressed by the findings, the Department of Labor launched a project in 1992 to replicate CET's model at 14 locations across the country and evaluate it through randomization. It did not go well. Only four sites were deemed to have copied the model with high fidelity—and those were CET chapters outside San Jose. No site produced sustained, positive, significant results (Miller et al. 2005, Table 3.2). The evaluators tentatively attributed the lack of impact to several factors: the youth who were served by the replications were slightly better off; the economy was doing better; and control group members had other options for training (Miller et al. 2005, p. xi). They also suggested that the secret sauce consisted of bonds to local actors, built over years:

> CET-San Jose is unique in so many ways, having grown organically over 20 years, with an unusually committed founder and staff, very strong ties to the local community, and a tradition of political advocacy on behalf of the local Hispanic community. Perhaps a homegrown model like CET cannot be easily exported in a top-down way to other areas (Miller et al. 2005).

### 7.1.2 Project QUEST: Hint from non-randomized research

Noticing patterns, private funders settled on the term “sectoral strategy” (Conway 2014). To learn more about programs under this umbrella, the Mott, Ford, and Casey foundations funded two collections of non-randomized evaluations, which merely checked whether people earned more after going through the program than before (Zandniapour and Conway 2002; Roder 2008; Conway 2014, p. 50).

One should expect regression to the mean to bias such studies, known in labor economics as “Ashenfelter’s dip” (Ashenfelter 1978; see Appendix A.2). People whose fortunes oscillate are more likely to enter a training program just after their latest downswing and then, in many cases, at least partially recover even without training. As with CET a decade earlier, one program appeared in both collections and was the standout in each. Project QUEST in San Antonio, Texas, produced the largest earnings rise among six programs analyzed in Zandniapour and Conway (2002, Figure 2), from $5,367/year before entry to $24,907/year after. It ranked first among the nine programs in Roder (2008, Figure 1) as well. The program did not run trainings. Instead, after consulting local healthcare employers about their staffing needs, it worked with community colleges to structure and run certificate programs for its recruits. Its graduates earned qualifications as radiology technicians or registered nurses, among other professions.

### 7.1.3 The Sectoral Employment Impact Study and a randomized evaluation of Project QUEST

The organization that ran that last study, Public/Private Ventures, embarked on randomized evaluations of a trio of mature sector programs in the mid-2000s. P/PV also initiated a trial of Project QUEST, which another group, Economic Mobility Corporation (“Mobility”), would complete. The four are about as diverse as four sector programs can be (Maguire et al. 2010, p. ii; Roder and Elliott 2018, pp. 26–27):

- Project QUEST, in San Antonio.
- The Wisconsin Regional Training Partnership (WRTP) is an association of unions and employers in Milwaukee that ran 2–8-week trainings in response to specific requests from employers. The employers committed to hiring the graduates.
- Jewish Vocational Service in Boston (JVS-Boston) conducted intensive, five-and-a-half-month courses that prepared people to work in medical billing and accounting, among other activities.

- Per Scholas in New York was a “social venture” that refurbished old computers, distributed them to low-income people, and used these activities as a training conduit.

The programs boosted earnings by $2,000–5,000/year (Maguire et al. 2010, Tables 11, 15, 17; Roder and Elliott 2018, Figure 6). This is a remarkable set of results—four for four—and was immediately recognized as such.

### 7.1.4 Year Up: Early randomized evidence

In 2008, a non-profit called Year Up, which works with young adults and forms tight partnerships with major corporations, retained Mobility to run a small evaluation of its operations in Boston, New York, and Providence. Initial earnings impacts were similarly impressive. That appears to have led Mobility to publish the pilot and obtain funding to follow subjects longer (Roder and Elliott 2014, p. 1).

### 7.1.5 Pathways for Advancing Careers and Education (PACE)

As part of the federal interest in career pathways programs, the US Department of Health and Human Services launched a family of trials called PACE (Pathways for Advancing Careers and Education). Recall that career pathways programs support participants through schooling, training, and work steps of increasing difficulty. While such sequencing need not come at the expense of collaborating with employers—one plank of the sectoral strategy—in practice it tends to (King and Prince 2015, p. 202).

PACE evaluated several programs that arguably merited both labels, including the Valley Initiative for Development and Advancement (VIDA) in the Lower Rio Grande Valley of Texas, which was modeled on Project QUEST; and Year Up, this time with a larger sample across eight cities. Once again, the repeat performer performed by far the best: Year Up boosted incomes by an extraordinary $8,000/year, with the impact showing no diminishment after seven years (Fein and Dastrup 2022, Exhibit 2-1), although this also had standout costs of around $30,000 per participant. The other eight PACE evaluations produced no clear impacts on employment or earnings.

### 7.1.6 WorkAdvance

Starting in 2011, MDRC fielded a quartet of evaluations called WorkAdvance. The “advance” in the name nods to an emphasis on supporting and coaching people after they have found better jobs, so that their trajectories continue to climb after they make the leap to steady work. Yet again the repeat performer—Per Scholas (see above)—was the standout. It unambiguously lifted wages. The other three did not at first (Towards Employment, St. Nicks Alliance, and Madison Strategies Group; Schaberg and Greenberg 2020, Tables 2.1–2.4). However, 7–10 years after randomization, the impact of St. Nicks rose to $8,000/year, with statistical significance (Yusim et al. 2025, Table 3).

### 7.1.7 Katz et al. (2022): Academic recognition

In 2020, academic economists Lawrence Katz and Jonathan Roth collaborated with MDRC's Richard Hendra and Kelsey Schaberg to analyze why some sector programs are so effective. The paper marshals the evidence for the trio evaluated by P/PV, as well as Project QUEST, Year Up, and the WorkAdvance programs. Katz et al. (2022) became the leading summation of the evidence of the effectiveness of sectoral programs.

## 7.2 Some takes on the history of research on sectoral programs

We distill this history with a few observations:

- The diversity, within a large country, of implementers and funders; the flexibility of private organizations to experiment; the mechanisms of community to share hopes and discoveries; and the collaboration between doers, funders, and researchers have led to slow collective learning.

- The placement of randomized studies is indeed non-random. Several programs got evaluated more than once. Evidently, their managers were committed enough to the scientific method to collaborate with researchers wanting to double-check startling research findings. In one case, Year Up, results from a pilot study were apparently impressive enough to raise the probability that they reached our eyes, and to motivate the launch of a larger study. In another, Project QUEST, strong results attracted funding to continue tracking impacts (Roder and Elliott 2021).

- Notwithstanding the selection and self-selection, it is clear that a) certain programs are genuinely making a difference for their enrollees and b) they share traits, notably a commitment to collecting and incorporating knowledge about the demand for skills among local employers. The earnings impacts of a few of them are highlighted in Figure 1.

- From the point of view of evaluators, stars are generally found, not made. One reason appears to be that it takes programs time to mature. Ten of the 14 cases in the CET replication were classed as low- or medium-fidelity. The newer programs in WorkAdvance did not perform as well as Per Scholas, at least for the first 5 years. The generalization here is not neat, however. Even CET's high-fidelity replications of its San Jose program did not raise earnings over the long term. The most successful replication evidenced in the research is Year Up's expansion outside of Boston (Fein and Dastrup 2022, Exhibit 2-9).

- It is appropriate to theorize post hoc about why certain programs work. But it is also tricky. The sample of studies is small. It is tempting to explain away failures as not being true sector programs. Comparisons across trials are subject to the caveat that correlation is not causation.

## 7.3 Revisiting the sample of studies in Katz et al. (2022)

Katz et al. (2022) is not a meta-analysis or systematic review. Its purpose is to use the results from certain encouraging evaluations to test economic theories about why they work. Nevertheless, because Katz et al. has become a proof point for the idea that sector programs work, it is worth thinking critically about how well its 9 studies of 8 programs represent the body of evidence on sector programs.

Revisiting the selection of studies leads to a question of definition. What is a sector program? Does the CET count even if it predated the term? What about its low-fidelity replications? “Sector program” turns out to be a bundle of loosely defined potential attributes. The most systematic effort to concretize “sector program” is in the Peck et al. (2021) meta-analysis reviewed in §6.4. Among the more than 75 program traits that review defines, 15 or so pertain to working with employers. Examples: whether programs convene advisory councils to solicit high-level input on program design; whether companies promise to hire graduates, or provide mentors. These traits are listed on the left of Table 4, below; the table also shows how Peck et al. scores the 8 programs in Katz et al.

| | | PACE | Sectoral Employment Impact Study | | | WorkAdvance | | | |
|---|---|---|---|---|---|---|---|---|---|
| Trait | Project QUEST | Year Up | JVS-Boston | Per Scholas | Wisconsin Regional Training | Madison Strategies Group | Towards Employment | Per Scholas | St. Nicks Alliance |
| *Program efforts to engage employer* | | | | | | | | | |
| Convene employer advisory council | | | ✓ | | ✓ | ✓ | ✓ | ✓ | |
| Curricula adapted for employers' needs | ✓ | ✓ | ✓ | ✓ | ✓ | ✓ | ✓ | ✓ | ✓ |
| Formally partner | | ✓ | | | | | | | |
| Invite to events | | ✓ | | ✓ | | | | | |
| Offer incumbent worker training | | ✓ | ✓ | | ✓ | | | | |
| Other | | ✓ | | | | | | | |
| *Employer contributions* | | | | | | | | | |
| Input on curriculum or program development | ✓ | ✓ | ✓ | ✓ | ✓ | ✓ | ✓ | ✓ | ✓ |
| Input on type of applicants | | | | ✓ | | ✓ | ✓ | ✓ | |
| Instruction or instructors | | ✓ | | | | | | | |
| Career awareness services | | ✓ | | ✓ | | | | | |
| Paid work-based learning | | ✓ | | | | | | | |
| Unpaid work-based learning | | | | | | | | | |
| Mentors | | ✓ | | | | | | | |
| Financial aid to students | | | | | | | | | |
| Information on labor market demand | ✓ | ✓ | | | ✓ | ✓ | ✓ | ✓ | |
| Commit to hire graduates | | ✓ | | | ✓ | | | | |

**Table 4. Coding of employer engagement traits by Peck et al. (2021) of programs in Katz et al. (2022)**

*Notes: Peck et al. (2021) codes the programs in 46 studies on more than 75 traits. Shown here are the codings related to employers' role. Per Scholas was evaluated two times, and was given somewhat different codings in each case. Program efforts to assure that "curricula adapted for employers' needs" and employer contributions of "input on curriculum or program development" are in practice hard to distinguish. The eight studies shown are those covered in Katz et al. (2022). PACE is the Pathways for Advancing Careers and Education evaluation program.*[52]

[52] Source: public-use dataset accompanying Peck et al. (2021) (U.S. Department of Labor, Chief Evaluation Office, n.d.).

When it comes to collaborating with employers, Year Up checks by far the most boxes. It is one of two in which corporate partners "commit to hire" graduates—or, more precisely, to offer six-month internships after the six months of training (Fein and Dastrup 2022, p. 4). Only Year Up arranges for the partners to supply mentors.

When the table is scanned by rows rather than columns, two nearly synonymous traits stand out as universal: "curricula adapted for employers' needs" and employers having "input on curriculum or program development." We therefore provisionally equate "sector program" with *employer involvement in curricula*. That resonates with the core idea of forging a strong link between pedagogy and workplace needs. As mentioned in §6.4, Peck et al. finds that this trait is one of the strongest correlates of positive wage impact.[53] (In our meta-analysis in §6.5, this explanatory factor is outcompeted by whether employers commit to hiring graduates.)

With this working definition of a sector program in hand, we ask which sector programs with randomized studies are *not* covered by Katz et al. Table 5 addresses this question. It is built starting from the list of programs with randomized evaluations that Peck et al. rates as having either of those two nearly indistinguishable traits. Three programs are then dropped for lack of strong textual evidence that the programs were designed to substantially involve employers in curriculum development, or for lack of medium- or long-term impact estimates.[54] On the other hand, CET and its replications, which are excluded from Peck et al. for not being "career pathways" programs, are added.

[53]Technically, this statement refers to the second trait, employers having input. But the two are correlated 0.58 (Peck et al. 2007, Exhibit E-9), so their associations with outcomes are hard to distinguish.
[54]Green Jobs and Health Care in Minneapolis, IBEST, and YouthBuild are excluded on the first criterion and Career Pathways in Louisiana on the second.

| Program | Evidence employer involvement was achieved? | Last follow-up | Outcome source (response rate) | In Katz et al. |
|---|---|---|---|---|
| Center for Employment and Training (San Jose) | Strong: "CET programs involve employers in the design of their programs and as reviewers of training curricula" | Q7–10 (adults), Q9–16 (youth) | Self-report (80%, 84%) | No |
| CET replications | Mixed: 6 programs rated High, 5 Medium, and 1 Low on "employer involvement in design and training" | Year 5 | Self-report (77%) | No |
| *Sectoral Employment Impact Study* | | | | |
| WRTP | Strong: Employer-union committees develop training; employers "ordered up" training | Q5–8 | Self-report (87%) | Yes |
| JVS–Boston | Strong: Employer advisory boards "helped craft occupational skills training curriculum" | Q5–8 | Self-report (73%) | Yes |
| Per Scholas | Strong: Employers "advise about the curriculum" | Q5–8 | Self-report (78%) | Yes |
| *Green Jobs and Health Care* | | | | |
| Bakersfield, CA | Strong: Employers gave "guidance on course content"; curricula adjusted for employers | Q5–6 | Official | No |
| Gainesville, TX | Weak: "For some programs, instructors engaged a few employers for input on aligning curricula." | Q5–6 | Official | No |
| Grand Rapids, MI | Moderate: Employers initially developed curricula, but "role of employers lessened" | Q5–6 | Official | No |
| Project QUEST | Strong: Staff embedded at college to redesign diesel mechanics curriculum "in consultation with local employers"; LVN curriculum altered to add geriatric care based on employer demand. | Year 11 | Official | Yes |
| Accelerated Training for Illinois Manufacturing | Mixed: "Some regions succeeded in getting employers to provide input" | Q5–8 | Official | No |
| Accelerating Connections to Employment | Strong: Employers advised on curriculum and project design; tailored content to employer needs | Q5–8 | Official | No |
| *WorkAdvance* | | | | |
| Per Scholas | Strong: Employers "advise about the curriculum" | Year 6 | Official | Yes |
| St. Nicks Alliance | Weak: Input limited to career readiness; staff "never truly committed to serving employers" | Year 6 | Official | Yes |
| Madison Strategies | Strong: Advisory groups; employer quote on framing curriculum; courses changed on feedback | Year 6 | Official | Yes |
| Towards Employment | Moderate: Curriculum adapted for employer needs, but mostly through intermediaries | Year 6 | Official | Yes |
| *PACE* | | | | |
| Year Up (8 cities) | Strong: Employers directly shape occupational curricula; internship hosts provide feedback | Year 7 | Official | Yes |
| Carreras en Salud (Chicago) | Moderate: Semiannual council discusses "potential curriculum changes," but training is standard nursing program | Year 6 | Official | New |
| Workforce Training Academy (Des Moines) | Moderate: Employer input shaped which certificates to offer, not curriculum content | Year 6 | Official | New |
| VIDA (Rio Grande) | Weak: Employer input selects which college programs to fund, not curriculum | Year 7 | Official | New |

**Table 5. Strength of evidence for employer involvement in curriculum for arguable sector programs**

*Notes: Q = quarter. Aside from CET and its replications, all programs are marked by Peck et al. (2021) as involving employers in curriculum development. Three so marked are dropped for lack of textual support: Green Jobs/AIOIC, PACE/I-BEST, and PACE/PCPP. Sources: Burghardt et al. (1992), Rangarajan, Burghardt, and Gordon (1992), Cave et al. (1993), Miller et al. (2005), Maguire et al. (2010), Martinson et al. (2016), Copson et al. (2016), Rademacher, Bear, and Conway (2001), Roder and Elliott (2021), Betesh et al. (2015), Modicamore et al. (2017), Schaberg and Greenberg (2020), Gardiner and Juras (2019), Copson, Martinson, and Gardiner (2014), Gardiner and Grittner (2022), Rolston, Copson, and Gardiner (2017), Farrell and Martinson (2017).*

This exercise confronts us with a distinction between intention and execution. The evaluators judged that only 6 of 14 CET replication sites achieved high fidelity on “employer involvement in design and training” (Miller et al. 2005, Table 1.1). (Separately, as noted, they judged that only 4 replicated the CET with high overall fidelity.) One can argue that the low-fidelity sites should be deleted from the sectoral strategy evidence record. Loosely speaking, they don’t count because they didn’t do it, or didn’t do it right.[55] This makes sense when the goal is to learn from success. But it is sometimes more meaningful to generalize from what was planned than what was done. For a decision-maker pondering whether to greenlight a project to bring sector-based training to more people, the decision is not exactly whether to do that but whether to back people who *plan* to do it. The proper reference class for predicting impacts is then the set of past projects with similar stated plans.

Since there is value on both sides of this distinction—learning from intended sectoral programs as a group and actual sector programs as a group—Table 5 lists programs fitting the more inclusive meaning (intention) while adducing information as to the more demanding meaning (implementation). In addition, the table supplies facts that situate Katz et al. within the larger body of evidence on sectoral programs. It documents how long subjects were followed, because Katz et al. only includes studies with “medium-term impact estimates (covering 2 years or more after randomization).” And it shows whether outcome data came from self-report or official sources and, if the first, the response rate.

Overall, because we are interested in generalizing from the full evidence base on (would-be) sector programs, there is a good case for adding 10 randomized evaluations of sector programs to the 9 (of 8 programs) in Katz et al. They are the CET and its replications; three of the Green Jobs and Health Care experiments (Martinson et al. 2016); the Accelerated Training for Illinois Manufacturing program (Betesh et al. 2015); the Accelerating Connections to Employment programs in Maryland and Texas (Modicamore et al. 2017); and three PACE programs in addition to Year Up, including VIDA in the Rio Grande region of Texas.

It is of course arguable whether the trait we have singled out as definitional—employer engagement in curricula—truly is. Nevertheless, we believe this exercise expands the reference class of sector programs in a way that is reasonably meaningful and minimally arbitrary.

[55] It is entirely possible that some deviations from the CET model made the replications *more* effective.

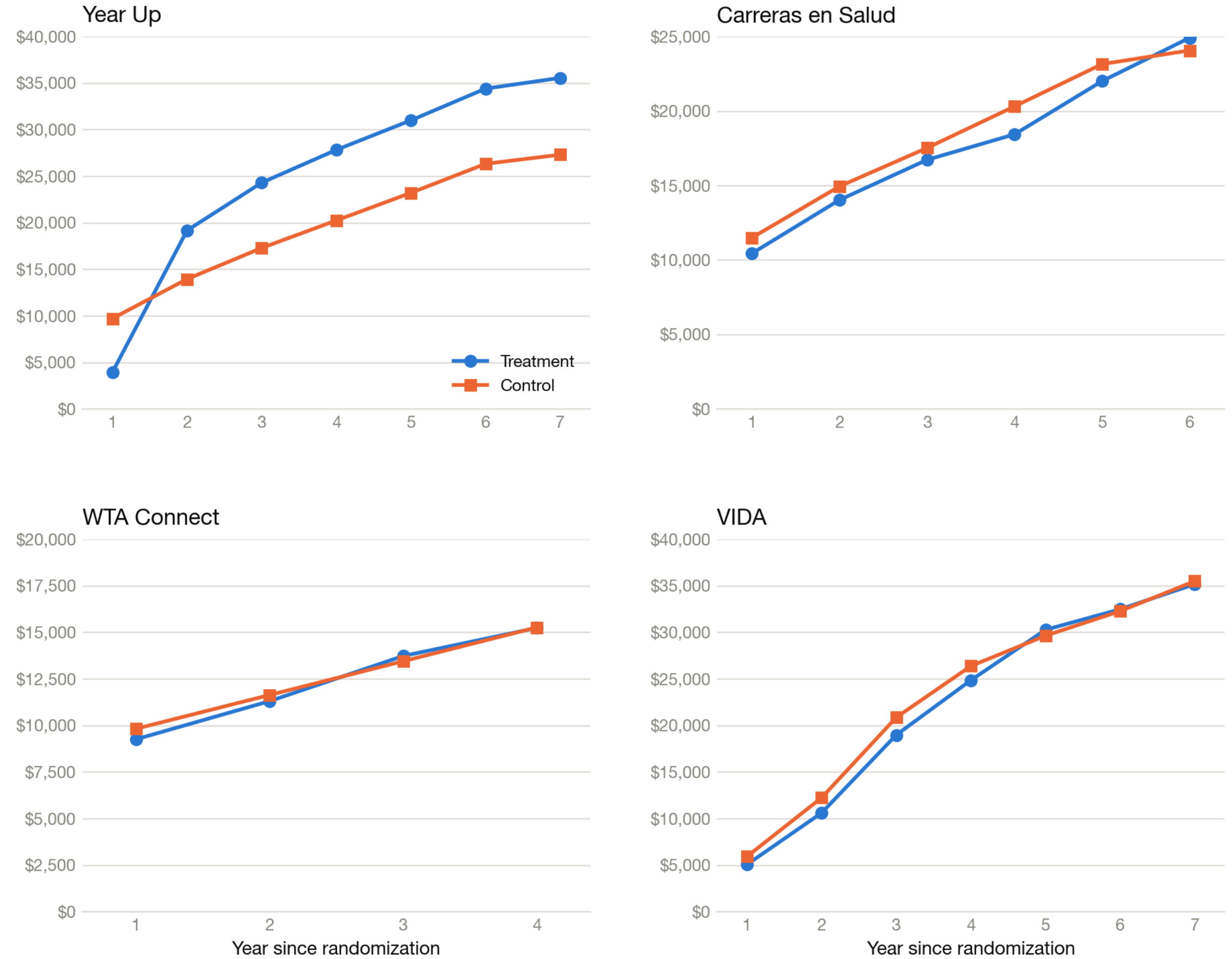

**Figure 12. Annual earnings of subjects in selected PACE-evaluated programs**
*Notes: Adapted from Juras et al. (2022). Shown are results from the four programs in the Pathways for Advancing Careers and Education (PACE) evaluation that best fit the "sectoral strategy" label. Randomization took place in quarter 0.*

To convey the range of performance in this larger family, Figure 12 shows earnings in the treatment and control groups of the four PACE-evaluated ones. One of the plots is not like the others. Students in Year Up earn less during that initial year of classroom work and internship, but then shoot past their control-group peers. Thereafter the gap between the two holds steady at around $2,000 per quarter per treatment group member. The program (on average) lifts students onto a higher-paying career track. The other three PACE programs achieve no such success.

Meta-analytic impact estimates for the larger family of sector programs are reported in Table 6. For context, Figure 13 and Figure 14 collate the ITT estimates from the individual studies. In these forest plots, central estimates are represented with dots and 95% confidence intervals with whiskers; the results are also shown numerically on the right. Impacts on annual earnings are adjusted for inflation to dollars of 2025. A random-effects average appears at the bottom of each plot.

| | Control mean | Impact | N |
|---|---|---|---|
| Program take-up (%) | 13.5<br>(6.5) | 55.5<br>(8.6) | 17 |
| Any training take-up (%) | 53.8<br>(4.6) | 24.5<br>(4.2) | 8 |
| Employment, short-term (%) | 67.5<br>(6.0) | 3.7<br>(1.4) | 12 |
| Employment, medium-term (%) | 73.2<br>(4.8) | 2.9<br>(0.9) | 10 |
| Employment, long-term (%) | 80.3<br>(0.6) | 1.3<br>(0.8) | 6 |
| Earnings, short-term (2025 $) | 16,119<br>(1,048) | 1,169<br>(1,092) | 13 |
| Earnings, medium-term (2025 $) | 19,447<br>(938) | 3,149<br>(1,260) | 12 |
| Earnings, long-term (2025 $) | 25,137<br>(1,664) | 3,703<br>(1,545) | 9 |

**Table 6. Control group means and impact estimates in sector program experiments**
*Notes: Impacts are estimated with restricted maximum likelihood (REML) meta-analysis. Control-group statistics are computed using the corresponding REML weights. Figures in parentheses below each estimate are standard deviations of control-group values and standard errors of impact estimates. Sample sizes vary because of missing data. Short-term outcomes have a ~1-year follow-up, medium-term ~2 years, and long-term ~3–5 years.*

The forest plots deliver bad news and good news, if more good news. In all four plots, point estimates range from negative to positive. Yet the random-effects means are positive, with statistical significance at $p < 0.05$ in all cases except in the fourth (long-term impact on employment). Subject to the caveat that the sample of studies shifts from plot to plot, sector programs on average cause a partially transient employment bump, about 3.0 points, and a more permanent increase in pre-tax pay, $3,000–4,000/year in 2025 dollars.

Factoring in the impacts on take-up leads to LATEs about twice as high. The average sector experiment generated a 56-point differential in participation in the program being evaluated, as shown in Table 6; that implies that each LATE-program is 1.79 times the corresponding ITT. As with training programs more generally, the impact on take-up of any training is about 25%. However, if alternative training programs are substantially less effective than the tested sector programs, then the LATE-training, at 4 times the ITT, should not be interpreted as an estimate of the impact of sector programs.

The control-group means in Table 6 also capture differences in the populations served, reflecting the selectivity of sector programs. For experiments incorporating training, the long-term employment rate among controls is 60% compared to 80% for sector programs. The control groups from sector programs also earn 50–75% more depending on the time horizon. Sector programs draw from populations that are higher paid even without training.

Although sector programs' benefits greatly exceed those of the average training-primary intervention, their costs do not: $11,602 per treatment group member instead of $13,598. As a result, we estimate the benefit-cost ratio of sector programs at 7.18, almost five times the 1.44 for the larger group. (See Table G2 in the appendix.) If the government were to fund a statistically average sector program, it would pay for itself twice over. As a result, the MVPF would be infinite, meaning that positive good would be produced at negative fiscal cost.

In sum, this critically motivated expansion of the collection of studies in Katz et al. fundamentally reinforces the optimism about sector programs. Even when averaging in programs that did not manage to work closely with employers, the earnings impacts are 3–5 times higher than for the average job training program.

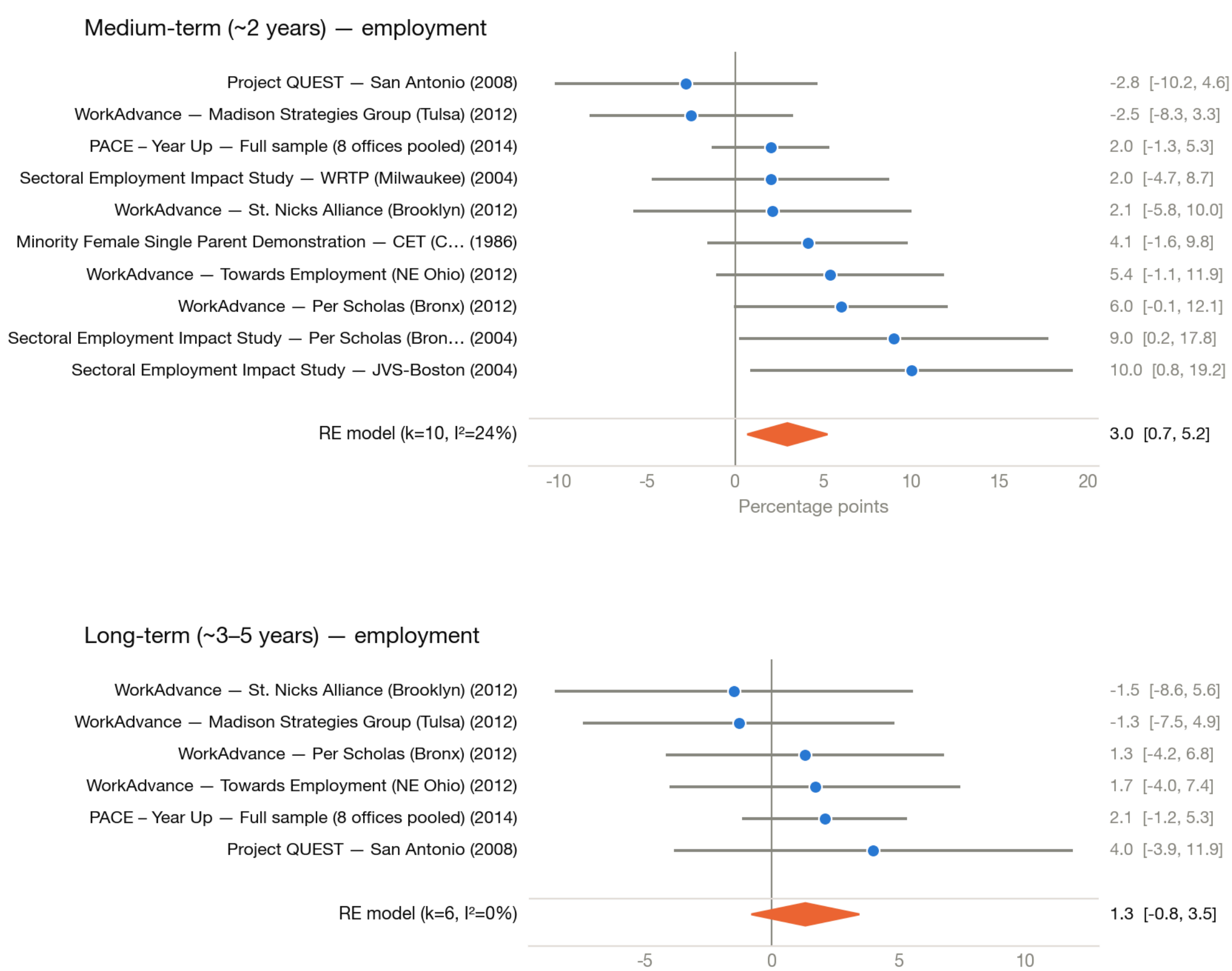

**Figure 13. Employment impact of sector programs**

*Notes: "RE" signifies the mean effect from random-effects meta-analysis. The lengths of whiskers and diamonds depict 95% confidence intervals, which are also reported numerically on the right. All results are extracted from sources cited in Table 5. All are intention-to-treat (ITT) impacts.*

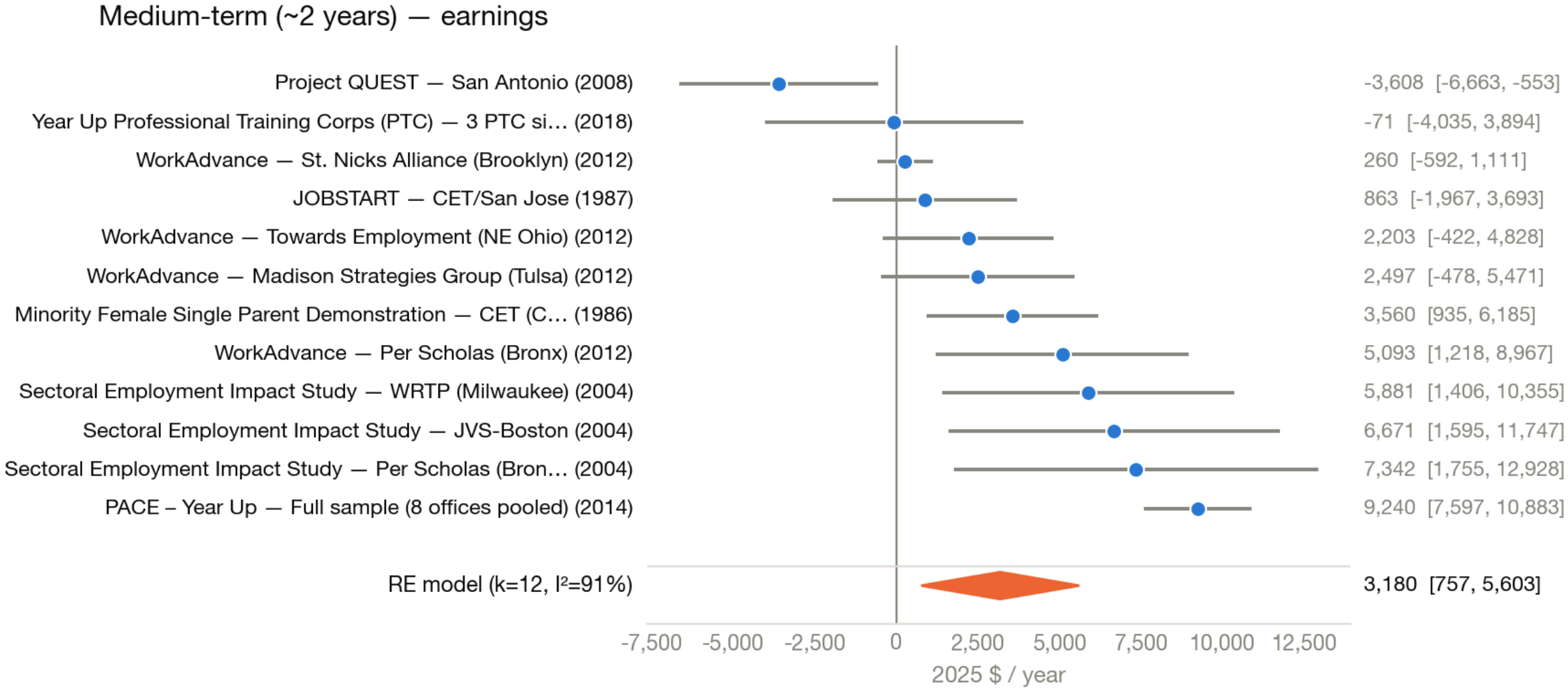

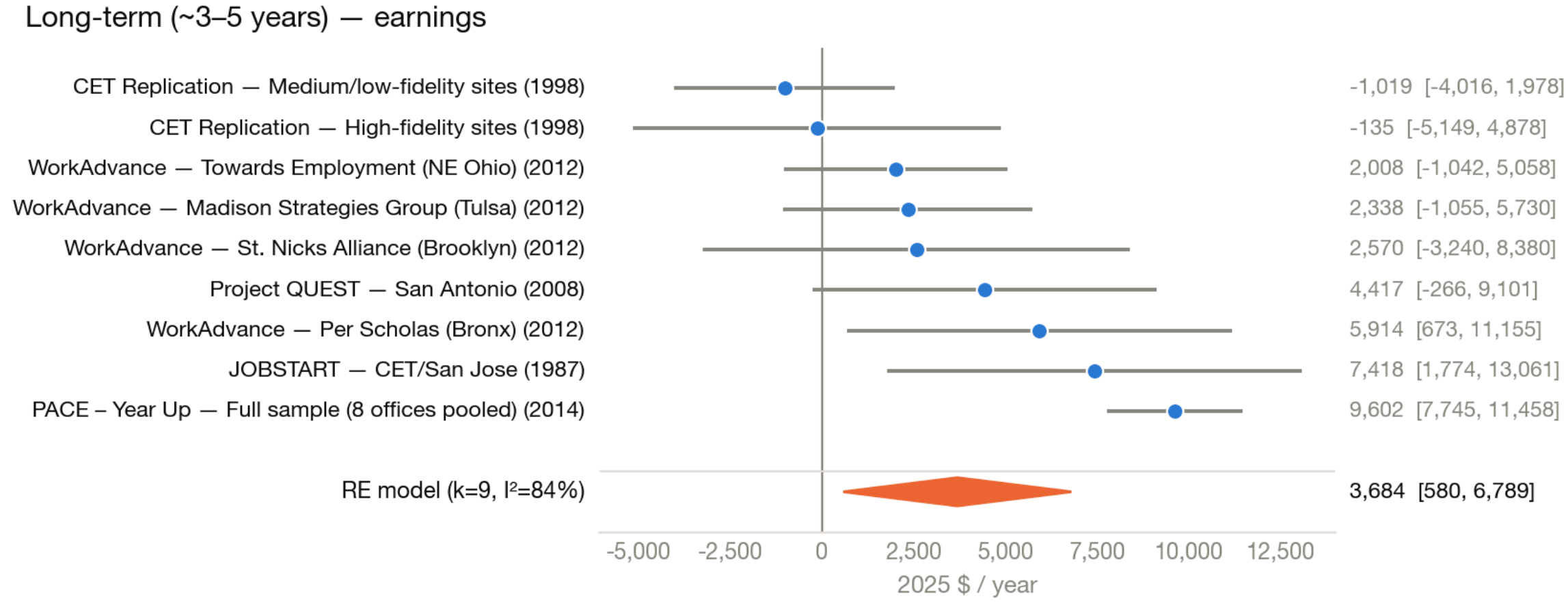

**Figure 14. Earnings impact of sector programs (2025 $/year)**
*Notes: See notes to previous figure.*

## 7.4 Sector programs: Summary

Why do certain sector programs, like Project QUEST, Per Scholas, and Year Up, stand out in the research record? Successful sector programs generally possess these components:

- *Selectivity*. The four WorkAdvance programs performed “intensive screening” of applicants, based on staff judgments about subtle qualities such as commitment and motivation. They accepted about 20% of applicants (Hendra et al. 2016, p. 29). In 2025, 70,000 people applied for 5,000 spots at Per Scholas. The organization used behavioral assessments and other tests, including pre-work, to vet applicants and test their commitment through the burden of the process itself.[56]

- *Training* covering "soft skills" (in behavior during interviews and on the job) and vocational skills, such as in solar panel installation.

- *Individualized coaching* to support students and keep them on track.

- *Graduation into work*. Year Up forges the tightest bond with employers: they cover a substantial fraction of the costs of training and commit to hiring graduates. Per Scholas dedicates staff to maintaining relationships with major employers and tracking regional trends in the demand for skills. Over time, the programs earn the employers' trust in the readiness of their graduates for work.

It is unclear—and a question for further research—exactly how important each of these components is. The second and third are least distinctive. But the intensive engagement with parties on either side of the labor market seems to be of the essence. For the strategy to make a big difference, employers must hire people they would normally have overlooked for certain positions, who in turn may never have seen those jobs as attainable. This bridging launches trainees onto permanently higher professional trajectories.

More broadly, we extract several principles from this history that are not rooted in the quantitative evidence per se, but which inform our extrapolations from it:

- Imbuing humans with skills and equipping them for new work contexts are subtle and complex arts. As administrative tasks, they are far more complicated than disbursing unemployment insurance. Small, nimble, private organizations have achieved the greatest successes. While access to various kinds of training is very much in the public interest, effectively delivering them is harder for public institutions, in part because they are often unable to be selective. In addition, perverse institutional incentives, like the "performance" standards under the JTPA, can unmake the delicate composition that is an effective job training program.

- While sector programs surely have not hit their limits, just as surely those limits exist. One reason Year Up has expanded from Boston to other cities, rather than scaling up where it started, is that its approach depends on deep-pocketed corporate partners, who fund much of the training and commit to hire the graduates. In each city, the supply of such partners is limited.[57] Indeed, Year Up runs for a year—half training, half internship—which is long by the standards of sector programs. So as far as the evidence goes, effective sector programs are confined to skill sets that can be taught in months.

[56] Conversation with Tamara Johnson and Sang Lee, Per Scholas, April 28, 2026.
[57] Conversation with Garret Warfield, former Chief Research Officer, Year Up United, April 24, 2026.

# 8 From past evidence to an AI-disrupted future

We have no research evidence on how well job training helps people adjust when large language models start doing their jobs. In search of partial insight, this report nevertheless reviews the available evidence on job training programs.

Almost all programs subject to research have been subsidized by governments or foundations. Most have targeted low-income people, while a minority have catered to people who have been dislocated by layoffs or who have intellectual disabilities. With the possible exception of training made possible by Trade Adjustment Assistance (TAA) in the US, nearly all training has been short-duration, meaning well under a year.

A key limitation of this report, however, is that it only covers stand-alone job training programs. Those are only one channel through which people gain job skills. Beyond our scope are high school vocational programs, apprenticeships, community colleges, four-year colleges, professional schools, and online education. A vast literature estimates how much other forms of education boost earnings, some using promising research designs like lotteries (e.g., Grosz 2020). In the US, job training programs have been unusually amenable to randomized experiments because they are publicly funded yet not considered a public right. That makes them subject to pressure to prove effectiveness, yet discretionary enough to—for the sake of research—randomly deny or defer access.

The major findings:

- Nearly all the highly credible evidence comes from randomized trials in the US. The evaluations of the multicity National Supported Work Demonstration in the 1970s and the Job Training Partnership Act in the 1980s found benefits for adults, but not youth. The Job Corps, another youth program, also showed no impacts in official data on employment and earnings. The most recent evaluation of the main national program, the WIA Gold Standard study, yielded null results. A major reason may have been lack of statistical power, as many people offered training didn't take it, and many not offered found it elsewhere.

- The handful of randomized and discontinuity-based studies from Europe have not been very informative as to the impacts of training (see Appendices C and D).

- Meta-analyses concur in finding small, average positive effects. There is little consensus on the correlates of impact, such as targeting women.

- In a new meta-analysis of randomized studies in the US, the long-term ITT impacts of job training average 1.7–1.8 points of employment and $700–800/year in income. We estimate that the impacts on pre-tax earnings roughly exceed program costs.

- In the same analysis, the average impact on the rate of participation in any training—including training outside the experiment—is only 28 percentage points. With additional assumptions, this means that the LATE of training, the average impact of training on marginal participants, is nearly four times higher.

- Two sophisticated "judge randomization" studies, in the US and Denmark, find that assistance does help people adapt to job loss, in part by helping them move to new occupations or even geographies. The studies are not as impregnable as randomized trials. But they make serious cases for having achieved causal identification. And they cover programs for which we have little or no randomized evidence.

- Though not mentioned earlier, the meta-analysis includes two studies of pilot programs for dislocated workers: the New Jersey Unemployment Insurance Reemployment Demonstration Project and the Texas Worker Adjustment Demonstration. Within the 12 months of follow-up, job search assistance and training do no better than the former alone (Corson et al. 1989, Table VI.1; Bloom et al. 1990, Table 7.5).

- There is good reason to believe that certain "sector programs" perform an order of magnitude better than the average training-primary program. They appear, however, to have achieved that impact in part by being very selective. There may be other limits to their scope, such as the availability of deep-pocketed employers who can commit to hiring their graduates, and the supply of applicants who are workplace-ready.

What do these findings tell us about how society should help workers displaced by AI?

An article by Manning and Aguirre (2026) frames the challenge. Marshalling seven data sets on American workers and professions, the project estimates 1) the number of people currently working in 356 professions; 2) an AI exposure index of how well AI can do each of these jobs; 3) an "adaptive capacity" index for current jobholders, which factors in liquid wealth, age, jobs/square mile where they work, and the number of jobs in adjacent professions demanding similar skills. Figure 15, copied from the study, conveys the main findings.

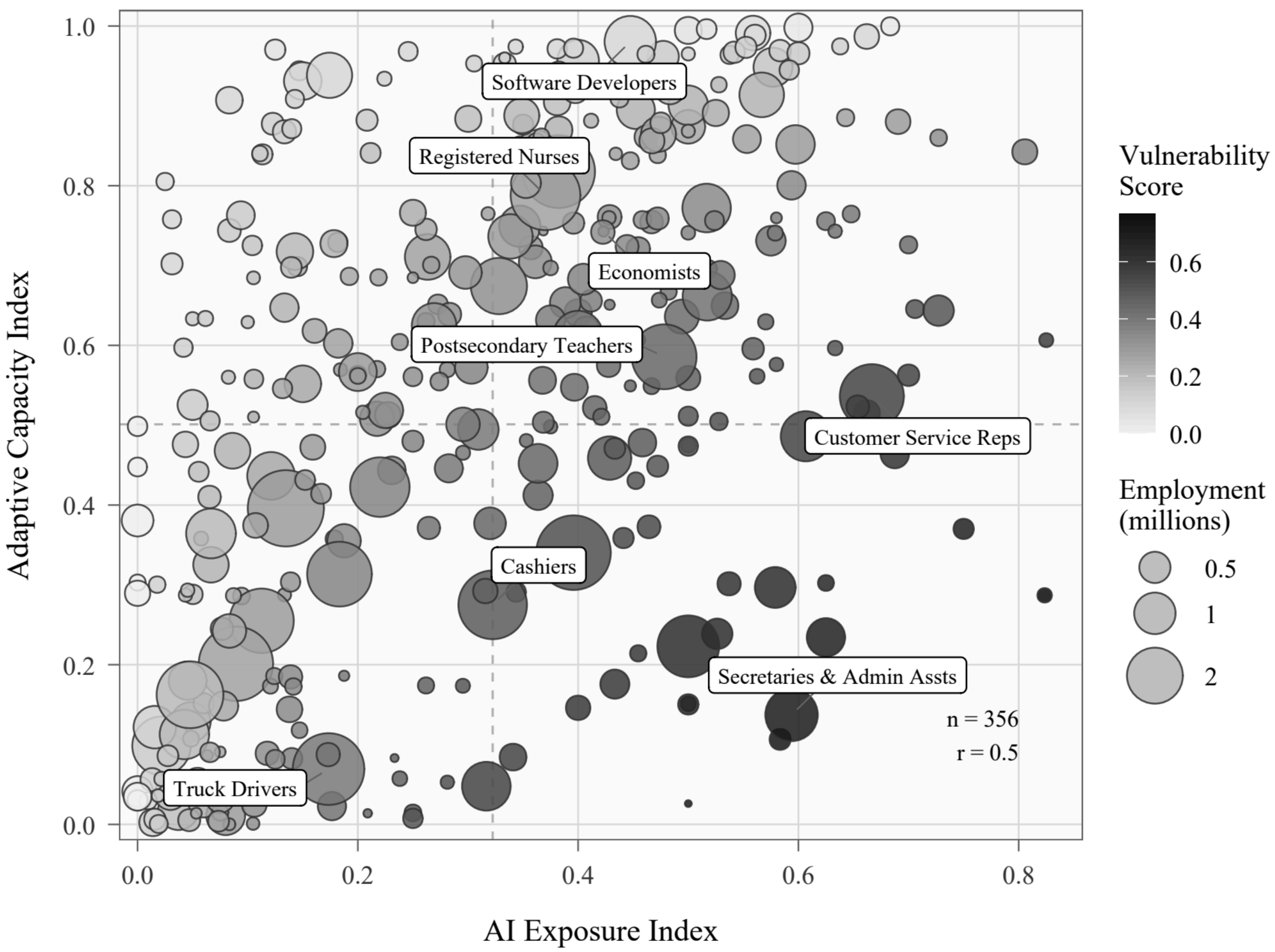

**Figure 15. Adaptive capacity vs. AI exposure for 356 professions in the US, from Manning and Aguirre (2026)**

On average, workers measured as more exposed to AI also have more wherewithal to adjust. That is somewhat reassuring. Still, there are big exceptions. For example, 1.7 million people work as "secretaries and administrative assistants, except legal, medical, and executive"; they score high on AI exposure (59%) and low on adaptive capacity (14%), which puts them in the lower right of the plot. Professions of the same ilk include "payroll and timekeeping clerks" and "office clerks, general"—in general, low-skill, white-collar workers.[58] Manning and Aguirre's methods cast truck drivers as quite unexposed to AI (see lower left of the figure), although broader trends in automation—whether construed as AI or not—could certainly affect those roles. This would make the class of high-exposure, low-adaptability workers even larger.

Meanwhile, even high-skill professionals who have more capacity to adapt may still need to leap wider professional chasms to recover their former income.[59] If a midcareer lawyer with children, a mortgage, and law school loans is displaced out of their profession, they may face a choice between a large and permanent pay cut or a return to school for years of retraining.

[58]On the challenges these workers face, see Molly Kinder, "The Invisible Disruption," published on Substack on June 9, 2026.
[59]Conversation with Molly Kinder, Brookings Institution, May 21, 2026.

With the problem so framed, we offer a few lessons from the evidence we have reviewed.

### 8.1 The sort of short-duration programs that are prominent in this review could help some “low-skill” workers displaced by AI.

One potentially vulnerable professional is the driver—of taxis, or long-haul trucks, or even delivery vehicles. These jobs have mostly not disappeared yet because, from an engineering perspective, they are high-skill. They require prodigious cognition. But from an *economic* perspective, they are low-skilled, for the supply of the ability to drive is high and the price low. And the automation and displacement of low-wage work is not new. For decades, robots and foreign competition have been putting the “rust” in the Rust Belt. Computers have done away with many low-paying jobs too, like messenger and typist. Our estimates of the impacts of sector programs, and training programs generally, come from this historical setting.

In this respect, past may be prologue. Sector programs that impart new skills and certifications in under a year, designed by organizations attuned to local skill demands, could make a difference for many workers.

And there is reason to hope that past performance will *underpredict* future performance. When serving people with good work histories, sector programs might not need to be nearly so selective. And, as explained in §7.3, the average sector program has an impact per complier (the LATE) twice its impact per person offered a spot in an experiment (the ITT). If demand for training surges relative to supply, then more people offered spots may take them, raising the average impact relative to the ITTs achieved in experiments.

### 8.2 Short-duration programs are ill-matched to the needs of more-skilled workers.

Mass displacement of accountants, lawyers, coders, and other white-collar professions would be unprecedented, both in the social classes of people affected, and perhaps in the speed. Short-duration programs would generally not meet their needs, for several reasons:

1. Subsidized training helps people who have enough motivation and agency to grasp the opportunity, but not so much that they don’t need it in the first place. Offering a particular kind of help to someone who doesn’t need it yields zero impact.

2. Deep reskilling of humans takes years. Humlum, Munch, and Plato (2025) tracks Danish workers who suffer a serious injury on the job—so serious that they must leave the skilled trade they once practiced. The study exploits the fact that some vocational certifications in Denmark qualified the holder to continue into higher education while some did not. A subset of injured workers who happened to have this option took it, and ultimately earned 25% more than before. They did so, however, after a typical four years of college.

3. Long training spells not only cost much more; they also break the sector program model. If a sector program detects a rising regional demand for entry-level technicians at data centers, it may be able to produce more graduates qualified for that role in 6–12 months. But if the demand for skills starts shifting rapidly and unpredictably—if it is hard to know what the growth occupations will be in four years—then it will be challenging for any training intermediary to align a yearslong training pipeline to the needs of employers.

### 8.3 The Trade Adjustment Assistance (TAA) approach is worth considering for people who are demonstrably fired because of AI.

The evidence on the effectiveness of TAA is limited. Still, it stands to reason that a crucial resource in adapting to disruption is *time*. Buying time means giving people a safety net, so that the loss of a job, even a profession, does not cascade into a worse catastrophe. In the US, the TAA program extends unemployment insurance from half a year to three years even as it offers funds for reskilling.

Of course, the TAA's trigger-based approach only works for people who can make the case that they lost their jobs to a disruptive force such as AI. When the labor market transforms not through firings, but by firms *not hiring* new people in certain roles when old people leave or when the firms grow, trigger-based aid will not fire.

### 8.4 Public and private funders should aggressively support demonstrations, evaluations, and scaling of sector programs.

The sector program model, and the strong evidence on it, emerged through a slow, ad hoc process. Given the potential urgency of the moment, and the potential for sector programs to help, funders should invest dramatically more in learning which components of the model are essential, and how best to replicate the model in more locations for more kinds of workers. One potential option is a "fire drill" that attempts to rapidly scale one of these programs for a targeted population.

# 9 Discussion

Over the decades, the occupational mix in the US has shifted dramatically (Autor 2015). Many jobs have disappeared. These shifts are not unusual, and they are not always associated with large-scale economic distress. Cavounidis et al. (2023) show, for example, that wages rose for horse-driving teamsters even as employment in the occupation declined. But the rollout of trucks was gradual. A major economic worry with AI is the combination of immense capability and compressed timing.

It's difficult to find evidence of any large impacts of AI on employment yet. But rapid automation would not be outside of the range of possibility. People working as elevator operators, film projectionists, and telephone switchboard operators went through condensed periods of disemployment.

While plenty of observers do not predict increases in unemployment in the near-term, the uncertainty seems great enough to want a firm handle on the policy options. This report studied what we know on one of the most popular ones, worker retraining.

We find that, on average, the programs have clearly positive but small impacts. Our current suite of training programs may be a defensible government expenditure, in that the investments ultimately cover their initial cost over a recipient's working life. But the impacts are small enough that they would not leave a dent in a persistently high unemployment rate. We see promise in the cluster of sector programs, although these have not been tested at scale.

The best response would likely be a combination of partial responses. In that spirit, and in light of the potentially great need, we think it's best for governments and other funders to aggressively explore and develop multiple responses at once. That includes job training, and especially sector programs. Some of these investigations are evergreen: yielding helpful evidence regardless of AI's impact on unemployment. Whether or not AI disrupts the labor market, we will not regret having learned how to best help workers adapt.

# Suggested citation

Roodman, D., & Massenkoff, M. (2026). *An evidence review of worker retraining*. The Anthropic Institute, Working Paper 2026-01.
www.anthropic.com/research/reviewing-the-evidence-on-worker-retraining-programs

## BibTeX

```
@techreport{roodman2026training,
  author      = {Roodman, David and Massenkoff, Maxim},
  title       = {An evidence review of worker retraining},
  institution = {The Anthropic Institute},
  type        = {Working Paper},
  number      = {2026-01},
  year        = {2026},
  url         =
{www.anthropic.com/research/reviewing-the-evidence-on-worker-retraining-programs}
}
```

# Appendix

## Appendix A: Other themes

### A.1 History of US workforce legislation

In the US, the history of federal workforce programs stretches back to the Manpower Development and Training Act (MDTA) of 1962, and from there to the employment programs of Roosevelt's New Deal. The motives for federal involvement in workforce development have always been multiple: enhancing economic competitiveness, ameliorating racial injustice, reducing poverty, helping workers adapt to foreign competition exacerbated by openness to trade. As is typical in US policy, the impulses of fragmentation and unification have competed over the decades, as Congress created numerous programs and occasionally strove to bring many under a single umbrella.

Although the legislative history is an important backdrop for most research reviewed here, the full history is beyond the scope of this report. Table A.1 provides a thumbnail sketch of the major laws.

| Act | Year | What it did |
|---|---|---|
| Manpower Development and Training Act (MDTA) | 1962 | Officially established the federal public employment and training system in the US |
| Comprehensive Employment and Training Act (CETA) | 1973 | Extended the WPA approach by providing work for the long-term unemployed and low-income adults, plus summer jobs for low-income youth, while ceding more administrative control to state governments. |
| Job Training Partnership Act (JTPA) | 1982 | Decentralized administration and expanded the private sector's role in delivering services; eliminated CETA's public service employment and stipends; and introduced the first national outcomes-based performance standards system for federal workforce programs. |
| Workforce Investment Act (WIA) | 1998 | Kept authority largely with states and localities, required State and Local Workforce Investment Boards led by business and other stakeholders, and mandated one-stop career centers in every local area. |
| Workforce Innovation and Opportunity Act (WIOA) | 2014 | Strengthened coordination and co-location across workforce programs, gave states more flexibility to shift funds between Adult and Dislocated Worker programs, required at least 20% of Title I youth funds to go to work-based learning, and established Pay for Performance as an eligible use of formula funds. |

**Table A.1. History of US workforce development laws**
*Notes: Based on Heinrich (2023), pp. 243–44.*

## A.2 Ashenfelter's dip and the credibility revolution

Experience over the last 50 years has led to a consensus that non-randomized study designs are not nearly as reliable as randomized studies. Some of the most important illustrations pertain to job training. Ashenfelter (1978) studies the impacts of the Manpower Development and Training Act, the primary federal training program in the 1960s. Entry into the program was not randomized, so Ashenfelter had little choice but to simply compare people who did get training to a particular control group of people who did not.

The study reveals a peculiar pattern, now called the "Ashenfelter dip." People who entered MDTA training in 1964 had a drop in earnings beforehand, in 1962–63 (Ashenfelter 1978, Table 1). In the same pre-treatment years, the non-participants' earnings rose. Clearly, even before treatment the two groups differed systematically. And if the treatment-control difference in earnings trends beforehand could not be attributed to the training, how could any difference after be confidently ascribed to training? Perhaps any post-treatment rise in the treatment group was just a rebound from hard times, i.e., regression to the mean. This was probably one reason that Orley Ashenfelter and future Nobelist David Card—again studying job training—concluded that "randomized… trials are necessary to reliably determine program effects" (Ashenfelter and Card 1985).

Surely another reason for their valorization of randomization was a generationally significant study that Ashenfelter's student, Robert LaLonde, was just finishing. LaLonde (1986) shows that various reasonable ways of forming a (non-randomized) control group and running the quantitative analysis lead to quite different estimates of the program's impact. It puts a question mark over all evaluations whose control groups are constructed without randomization. Gordon et al. (2023) is a more recent demonstration of the same phenomenon: non-randomized interventions, even with modern machine learning techniques for generating the best possible counterfactual group, give you biased estimates.

In part because of LaLonde (1986), a "credibility revolution" (Angrist and Pischke 2010) has taken over economics. A matter of both culture and technique, it favors research that exploits randomization, as well as certain other techniques that can frame natural experiments that are almost as compelling as actual ones. One example of the latter is the regression discontinuity design (RDD). Often programs impose eligibility cutoffs, such as a maximum income or minimum test score. The people just on either side of a cutoff—whose previous-year earnings are $49,999 or $50,001—probably live statistically similar lives until the day comes when one group can access a program and one cannot.

The credibility revolution strongly influences this report. We devote most of our attention to randomized studies. This decision is made easier by the fact that randomized job training experiments have been carried out in the US for 50 years. We will also cover a pair of RDD evaluations, in Appendix D, as well as two "judge randomization" studies, in §5.

## A.3 Observer and demonstration effects

Experimentation changes that which is being experimented upon.

For example, "Hawthorne effects" can occur. Trainers may train somewhat differently because they know their performance is being measured. Then there are "John Henry" effects: members of the control group may strive harder to compensate for their perceived disadvantage.

Similarly, many evaluations are of "demonstrations," which are stood up expressly to evaluate particular methods of training. Demonstrations generally require special funding from government agencies or foundations. That backing, combined with a desire to do justice to the strategies being tested, can mean that the demonstration is better funded than a going program would be, and more assiduously monitored for fidelity to the model. The funding consideration especially matters for training, because it costs more than helping people write resumes and search for jobs. If a strategy were rolled out nationally on the basis of a successful demonstration, but not given the budget to match, then the training component could be especially squeezed, driving a wedge between theory and practice, original and replication.

## A.4 Externalities and general equilibrium effects

Randomization isolates causal impacts, but there are many effects we can't measure. If someone gets a job thanks to a training program, they are less likely to commit a crime, sparing would-be victims. But the newly hired program participant might have displaced other job applicants, denying them the same benefits. Those effects matter for knowing the net benefit to society even when they are hard to observe and quantify.

One unhappy externality effect is *displacement* or general equilibrium effects (Heckman et al. 1999; Crépon & Van Den Berg 2016). If someone in an experiment's treatment group is trained in welding and gets a great factory job, isn't she just taking that slot from someone else? The someone else would probably not be part of a study, so her fate would be missed when tallying the program's impacts. Worse, if the someone else were in the control group, the benefits of training would be double-counted, as pay rose in the treatment group and fell in the control group.[60]

A key piece of evidence on displacement comes from Crépon et al. (2013), which reports on the results of a national experiment in France. The treatment was a referral to a private placement agency for up to a year of intensive counseling. In each locality, the fraction of people to receive this treatment was first randomly set to 0%, 25%, 50%, 75%, or 100%. Then the requisite fraction was randomly selected and contacted. The study finds that the treatment temporarily increased employment among the treated—the effect showing up about 8 months later but fading by 12—but that even this temporary benefit was offset by lower employment among the untreated, especially in regions where the job market was slack. Gautier et al. (2018) and Cheung et al. (2025) find similar evidence for displacement effects in Denmark and Sweden.

Micro evidence on the program, covering only the *participants*, would have been encouraging. But because of these displacement effects, the authors conclude, "the program seems to have had very little net benefits." This result suggests that studies overestimate the benefit of job search assistance, because they miss the side effects on those outside the study.

Does this extend to retraining? We think displacement effects are probably lower for training programs that increase people's skills, especially in the long term. They expand society's productive capacity, and a dynamic economy tends toward taking advantage of available resources. (For a formal version of this argument, see Albrecht et al. (2009).) And the most effective programs we find in this review target in-demand occupations, i.e., ones where there is apparently less competition for the available jobs.

For perspective, notice that the same issue pertains to immigration (Roodman 2014). Do immigrants take jobs from natives? It makes sense that they might. On the other hand, if the labor market were purely zero-sum, then the 10-fold increase in the American population since 1860 would have led to 90% unemployment. The way out of this paradox is to recognize the role of time. In the very short run, the number of job openings is fixed. In the longer run, which may not be that long, the capital stock—factories and buildings—responds to opportunities created by labor supply expansion and grows too. For example, the decades-long inflow into the United States of immigrants from Latin America—immigrants who are ready to work hard for low wages—probably improved the economics of the restaurant business. In addition to boosting demand for native workers to interact with customers, it helped the economy productively absorb immigrants. How fast the long run arrives—how fast the availability of more workers leads to more jobs—depends on the workers of concern and the economic context.

[60]In the language of Rubin (1980), displacement effects violate the Stable Unit Treatment Value Assumption (SUTVA).

## A.5 The placement of randomized studies is not random

While randomization is a powerful tool for causal inference, even randomized studies need to be read and interpreted with a critical eye. The programs whose randomized evaluations reach our eyes do so only after passing through filters of non-random selection and self-selection (Allcott 2015). As a result, randomized evaluations as a group are not strictly representative of job training programs as a whole. That's not a criticism, just a caution when generalizing.

Many factors determine the distribution of randomized studies. Often they are carried out to test new rather than mature approaches. Social experiments are inconvenient and expensive. Program staff may resist a set-up that forces them to arbitrarily override their professional judgment and withhold the service that gives their work meaning. If program heads who are more confident about their organization's effectiveness are more apt to enter an experiment, and if they are right more often than not, then better programs will get evaluated more. Leaders who are embezzling funds probably won't volunteer. Governments and foundations may be more ready to pay for long-term follow-ups of programs that have performed well in the shorter-term, which would bias the long-term evidence before us.

On the other hand, the training programs that target people who are hardest to help and have the least political power may be subject to more challenges about whether they work at all. This pushes the bias in the other direction: the programs with the toughest assignments would get the most coverage.

## A.6 Signaling vs. human capital

Owing especially to the work of Gary Becker, economists have traditionally viewed investment in education as the cognitive equivalent of buying a tractor, an upfront investment in one's stock of skills that pays for itself by multiplying the value of one's labor. Education and training increase *human* capital.

While recognizing that engineering students really do learn engineering and medical students really do learn medicine, skeptics have argued that little of what passes for education today teaches anything economically useful. Some of what is "taught" is not actually learned. Much of what is learned is soon forgotten. And much of it—French, medieval history, times tables—doesn't make workers more productive.

Why then does getting a high school diploma or even a college degree appear to lift earnings so much? In The Case Against Education, economist Bryan Caplan argues that the main economic function of a degree is to *signal* the bearer's value. Earning a degree requires "intelligence, conscientiousness, and conformity" (Caplan 2018, p. 37). Good students make good employees. One piece of evidence in favor of the signaling theory is the sheepskin effect: people who *almost* reach graduation, but then drop out, earn distinctly less than those who graduate. Yet surely the near-graduates, despite lacking that valuable vellum, are nearly as skilled. They have completed nearly as many courses.

If job training programs also help people mainly by certifying their smarts, grit, and conformity, that sharpens the worry that the gains are zero-sum: certification just gets people to the front of the line in the competition for a fixed supply of jobs. In this view, money spent on training is entirely wasted via knock-on effects.

Yet it is well-recognized that economic agents operate on the basis of information, about the quantity and quality of their inputs and the demand for their outputs, and that this information costs money to obtain. When someone pays for an education, even if it only amounts to a certification of pre-existing traits, that generates information that allows for more productive coupling of the factors of production. With reference to job training for disadvantaged people, completing a curriculum of classroom or on-the-job training demonstrates that one is prepared for the workplace. A training program that injects this information into the labor market effectively expands the supply of labor that is available to employers.

In fact, Caplan, the arch-critic of education, endorses *vocational* training, a category that embraces the programs we study. Caplan (2018, p. 303) guesses that while 80% of the value of a college degree is signaling, 40% of a vocational certificate's value is. The evidence base that Caplan (2018, chapter 8, notes 9–13) assembles in reaching this favorable judgment contains no randomized studies, and arguably only one potentially compelling quasi-experiment.[61] At any rate, it is broadly plausible that vocational training is productive even in a world of educational signaling.

[61] Part of Hanushek et al. (2017) exploits variation from plant closures in Austria.

# Appendix B: Additional figures

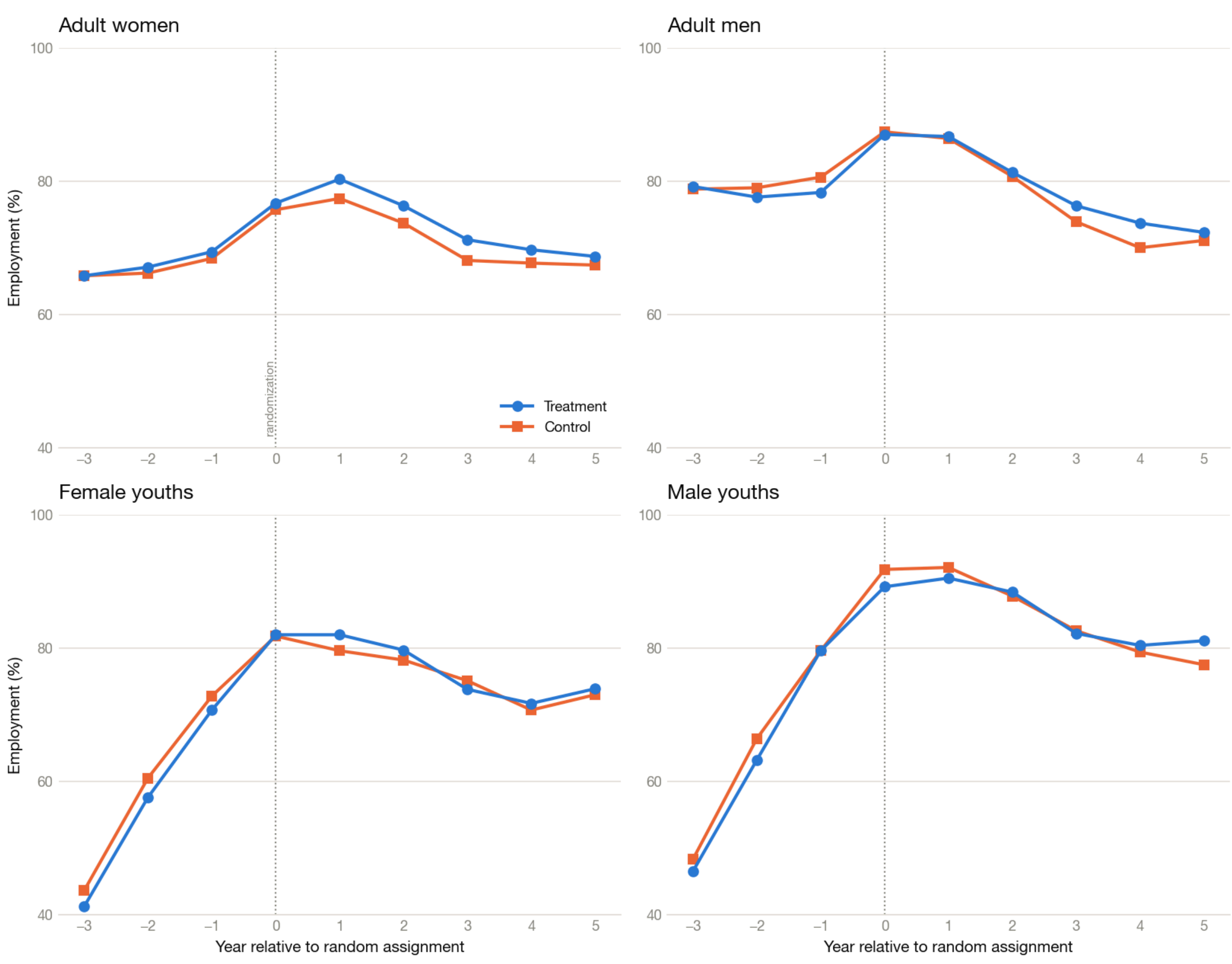

**Figure B1. Wage earnings in National JTPA Study by year before/since randomization**
*Notes: Source is GAO (1996), Tables II.1–II.4.*

# Appendix C: Experiments in Europe

High-quality evidence on training's impacts is much thinner outside the US. In Europe, perhaps one reason social experiments are seldom run is that entitlements such as individualized workforce assistance are viewed more as rights that should not be arbitrarily withheld for research.[62] Here we review all that we found that involve training: three efforts that ultimately do not generate strong and clean experiments, and an American-style evaluation in the UK in which training was a secondary component.

[62] On the case of Denmark, see Humlum, Munch, and Rasmussen (2025), appendix B.

## Raaum and Torp (2002), “Labour market training in Norway—effect on earnings”

Influenced by the 1980s debate in America over the merits of non-randomized evaluations, two Norwegian researchers undertook what looks to be the first randomized job training experiment in Europe. And they too used the randomized results to benchmark the sort of non-randomized methods they would otherwise have applied.

Raaum and Torp (2002) studies the Norwegian government’s labor market training program (LMT). It ran monthslong vocational courses under headings such as “office work,” “transportation,” and “manufacturing.” Demand for the LMT’s services surged as unemployment rose from 3.2% to 5.5% between 1988 and 1994 (Raaum and Torp 2002, p. 210). The government had to ration training spots, which created the opportunity to do so by lottery. Raaum and Torp’s experimental sample consists of 770 people who applied to LMT in August–September 1991 in any of 6 of Norway’s 19 counties. Most of the subjects finished training by early 1992 (Raaum and Torp 2002, p. 221).

The LMT’s rank-and-file staff had reservations about randomizing away their professional judgment (Torp et al. 1993, p. 101). So the researchers agreed to exempt certain courses from the experiment. “These included courses in which co-operating employers insisted on normal assignment procedures, courses aimed at specific target groups, and follow-up courses.” Practice also departed from plan in that after randomization, some staff “inactivated” applications of people they thought were less suited for certain courses. This opened spots to be filled from the (randomized) waiting list (Raaum and Torp 2002).

It is unclear from the text how these deviations from arbitrariness affect the treatment-control comparisons. If the comparisons are between the original, randomization-defined treatment and control groups, then the estimates should still be unbiased for the effects of the experiment as actually run—where “the experiment” includes the staff’s post-randomization filtration.

However, the two groups differ statistically at baseline, which is unusual and concerning. The treatment group is more female than the control (62% vs. 53%), more Norwegian-born (96% vs. 93%), and better educated (38% vs. 30% completed secondary school, 10% vs. 7% had higher education; Raaum and Torp 2002, Table 1). The paper does not report an omnibus test of statistical balance. We find that 13 of the 37 individual tests for equality on individual traits—more than a third—are significant at $p < 0.1$. It is plausible for that many significant differences to occur by chance *if* the various demographic traits are highly intercorrelated, for then the 37 tests are not independent. Absent information on correlations, we default to the view that the randomization somehow went awry.

In the two calendar years after training, 1993 and 1994, the treatment group earned 11,000 and 14,000 kroner more than the control group, or about 17% more (Raaum and Torp 2002, Table 1).[63] A naïve non-randomized assessment, which benchmarks against a sample of people who did *not* apply to the LMT, generates bigger numbers: 20,000 and 22,000 (Raaum and Torp 2002, Table 2, first panel).

Raaum and Torp (2002) provides more evidence that randomized evaluation can produce systematically different results than non-randomized, in this case by cutting the impact estimate by more than 40%. Perhaps if circumstances had allowed for a more perfect experiment, the randomized impact estimates would have been lower still.

### Rosholm and Skipper (2009), "Is labour market training a curse for the unemployed? Evidence from a social experiment"

Another study is set in the mid-1990s in a Nordic nation, Denmark. It too randomizes a going national job training program. And it too runs into trouble effecting the experiment.

As of the mid-1990s, Denmark's Labor Market Training Programmes (AMU) provided short courses to help the unemployed and employed alike transition to new types of work. About 600 courses were offered, with durations ranging from 1 to 7 weeks and averaging 2. Under ordinary, non-experimental conditions, *employed* people got priority admission, which suggests that the courses, in addition to being brief, were not especially designed for people without jobs. Within the experiment reported in Rosholm and Skipper (2009), however, the priority was reversed, so that the results could speak to the programs' ability to return people to work (Rosholm and Skipper 2009, pp. 341–42).

While the randomization-defined treatment and control groups here look statistically balanced,[64] fully 48% of the treatment group skipped the treatment while 22% of the control group somehow got it (Rosholm and Skipper 2009, pp. 338–39).[65] As in the studies of the WIA (§4.4), the 30-point increment in uptake between the groups reduces statistical power.

People entered training in May and June of 1994 and most finished by the end of the year. Rosholm and Skipper (2009, Tables IV–VI, column 1) finds hints of *negative* impacts on employment and earnings in 1995–96. To the extent these effects are real, the authors theorize they reflect a temporary lock-in effect. People *not* busy with courses in the second half of 1994 devoted more time to finding a new job, which on average paid off in 1995–96.

[63] Controlling for all the baseline characteristics, including pretreatment earnings, yields 10,000 and 14,000 (Raaum and Torp 2002, Table 2, first panel).
[64] While this is not immediately clear from the paper (Rosholm and Skipper 2009, Table II), 4 of the 34 cross-group comparisons on pre-determined variables are significant at $p < 0.1$. Two pertain to educational attainment, and so are not independent. This proportion is about what is expected by chance under perfect balance.
[65] "Concerning the high degree of crossing over, no clear explanation is readily available." (Rosholm and Skipper 2009).

At any rate, the authors emphasize, we should expect only small effects on the unemployed from two weeks of training designed for employed people. The moderate sample size (810) and the limited impact on uptake make it impossible to reliably detect small effects.

### Brenninkmeijer and Blonk (2012), “The effectiveness of the JOBS program among the long-term unemployed: A randomized experiment in the Netherlands”

This is a psychology study whose randomized design is marred and sample small.

In the Dutch city of Lelystad, 160 people receiving unemployment benefits were randomly split into three treatment arms: a control group; a group offered vouchers with which to pay for classes, such as in driving; and a group offered the “JOBS” treatment, a set of five half-day classes intended to boost confidence and motivation in job searching. Six months after treatment, 26% of the JOBS subjects were employed, against 9% in the voucher arm and 11% in the control arm (Brenninkmeijer and Blonk 2012, p. 225).

Those results, however, come from a shrunken sample at substantial risk from attrition and selection bias. 35 of the subjects, rather evenly split across the groups, dropped out. Another 7 apparently found a job before treatment, and 21 more who were not assigned to JOBS switched themselves into it. The study discretionarily drops both the latter groups from the analysis. (Brenninkmeijer and Blonk 2012, p. 223.) In the face of such attrition and cross-over, more rigorous studies retain all subjects, and continue categorizing them under their randomized assignment. Only in this way does randomization produce unbiased comparisons.

### Hendra et al. (2011), *Breaking the Low-pay, No-Pay Cycle: Final Evidence from the UK Employment Retention and Advancement (ERA) Demonstration*

This is a rare American-style evaluation of a European program, by which we mean one in which the agency that runs the intervention initiates and fully buys into the randomized experiment, and provides the substantial, sustained funding needed for professional evaluation. Indeed, an American organization led the evaluation of this workforce program: MDRC, which the Ford Foundation created in the 1970s for exactly such work.

The study ultimately provides little information on our interest, job training. As its name implies, the UK Employment Retention and Advancement (ERA) demonstration was designed to go beyond established government programs that, in effect, stop at the employer’s door—programs that help people find jobs and then disengage. The ERA delivered a package of services and incentives to help people stay in their jobs and advance in their careers. Payments for training were one part of the package:

> Once employed, ERA participants could receive at least two years of advice and assistance from an employment adviser to help them continue working and

> advance in work. Those who consistently worked full time could receive substantial cash rewards, called "retention bonuses." Participants could also receive help with tuition costs and cash rewards for completing training courses. (Hendra et al. 2011)

The demonstration targeted three groups of people defined by the existing structure of public benefit programs:

- Unemployed single parents of young children, who had voluntarily enrolled in an established welfare-to-work program called New Deal for Lone Parents (NDLP).

- Single parents of school-age children, who were working part-time and receiving Working Tax Credit (WTC), which was analogous to the Earned Income Tax Credit in the US.

- People 25 and over who had been unemployed for years and were receiving Jobseeker's Allowance, a cash benefit for people actively looking for work. People in this category were already required to participate in a program called New Deal 25 Plus, which gave the group its name: ND25+. (Hendra et al. 2011, p. 1.)

The NDLP and WTC participants were predominantly female while the ND25+ participants were predominantly male. The latter were also significantly worse off on average. 36% had no formal educational qualifications, compared to 23% and 12% of the NDLP and WTC groups. "Health problems, histories of substance abuse, and involvement with the criminal justice system were not uncommon. [The ND25+ group] was widely viewed as difficult to help." (Hendra et al. 2011, p. 8)

Remarkably, ERA had the largest impact on those who had been viewed as hardest to help. The results for the NDLP group (the ones with young children) tentatively point to an initial but transitory boost to earnings, at £308 in the first fiscal year of follow-up. There was essentially no effect on how many months people were employed. (Hendra et al. 2011, Table 4.1.) Results are similarly weak for the WTC group (with older children; Hendra et al. 2011, Table 4.2). In contrast, in the ND25+ group, ERA lifted employment by 0.2 months per year, for an overall employment effect of about 0.2 / 12 = 2 percentage points. And it increased earnings by £296/year over five years, with no sign of diminishment at the end. (Hendra et al. 2011, Table 6.1)

ERA incentivized job training in several ways. Caseworkers encouraged participants to consider it. The program reimbursed up to £1,000 in training costs and followed up with a £1,000 bonus for completing a course, plus £8 per hour of training time. Yet only a minority of participants took advantage of these incentives: about 15% in the NDLP group, 33% in the WTC group, and 10% in the ND25+ group (Hendra et al. 2011, Table 3.5). The low take-up in the ND25+ group was part of a larger pattern. The group's members had for most of their adulthood been detached from societal systems, were often content just to have secure employment, and were pessimistic about the odds of advancement (Hendra et al. 2011, p. 188).

Across the three target groups, training was thus anti-correlated with employment and earnings impacts. There are many ways to explain that pattern, so we do not view it as evidence that training does harm. Still, the evaluation of the ERA provides little evidence that these offers of government training helped. One possible problem is that the caseworkers who advised clients on what to train in knew little about the hiring needs of local employers (Hendra et al. 2011, pp. 71–72; Hendra and Hamilton 2015, p. 421).

## Appendix D: Discontinuity-based studies

Advocates of the credibility revolution in economics are not so parochial that they believe only in randomized trials. A few other analytical strategies (research designs) can frame a natural experiment in such a way that the statistically constructed treatment and control groups are comparable and the allocation of treatment between them effectively arbitrary. This section covers studies that make more or less convincing claims to be studying such "as-if" random variation in treatment.

One potentially compelling strategy is the regression discontinuity design (RDD). We found two RDD-based studies of training, and review both here.

### Battistin and Rettore (2002), "Testing for programme effects in a regression discontinuity design with imperfect compliance"

Between October 1995 and June 1996, the city of Turin ran 600-hour classes in the use of software for accounting, word processing, and other purposes. The classes were intended for the unemployed. Twice as many applied as there were spots, so the administrators rejected applicants who scored below 55 on an "attitudinal test" (Battistin and Rettore 2002).

To follow up on the people on either side of that threshold, Battistin and Rettore sought to interview them by phone 17 months after the courses finished. They achieved a response rate of 89% (Battistin and Rettore 2002, p. 45) and a sample size of 212. Since the methods in the paper put the most weight on the observations closest to the qualifying score and no weight on observations more than 13, 15, or 17 points from it, the effective sample sizes are probably substantially below 212. The paper runs the analyses with all three of those bandwidths in turn, to see how that affects results. The preferred value, 15, is chosen through a "cross-validation" algorithm, which finds the bandwidth that leads to the most accurate predictions of employment outcomes away from the threshold.

The study's preferred impact estimate is 12 points of employment, which comes with a *p* value of 0.3 (Battistin and Rettore 2002, Table 2, row 2). A positive impact is plausible but, perhaps because of the small sample, not clearly present.

### Leuven and Oosterbeek (2004), “Evaluating the effect of tax deductions on training”

While this study takes place in a different context, with different sorts of variables feeding into the analysis, its abstract econometric story resembles the one just above. In 1996, the Dutch government enacted several tax credits to encourage companies to train their employees. One generated a qualification discontinuity: a 40% credit for spending on training of employees over 40 years old. Leuven and Oosterbeek (2004) points out that firms could respond to this incentive in two ways: training over-40s more, or just waiting for under-40s to age and then train them. The latter response reaps the tax credit without increasing total training.

Regardless, if 41-year-olds were trained distinctly more than 39-year-olds, that is a natural experiment. To assess the impacts, these authors too ran a telephone survey, in 1999. They asked people if they had undergone training in the last 12 months, and how much they were now paid (Leuven and Oosterbeek 2004, p. 471). Since the questions looked back 0–12 months, an average 6 months passed between training and follow-up.

Leuven and Oosterbeek (2004, Table 1, column 4) demonstrates that the quasi-experiment took place: 41-year-olds were 17 points more likely to have trained in the last year than 39-year-olds, and with great statistical significance even in a sample of 149 people. (People who were 40 at the time of survey were left out because they would have been on both sides of the threshold within the previous year.) Widening the age ranges on either side to 38–39 and 41–42, or 37–39 and 41–43, etc., smoothly reduces the estimated difference across the 40th birthday. At the same time, the standard errors shrink. This illustrates how widening the bandwidth increases the precision of measurement even as it changes what is measured.

The impact on pay is quite indistinguishable from 0, regardless of which of these bandwidths is used (Leuven and Oosterbeek 2004, Table 4, first panel).

This paper’s research question is, effectively: Did firms who trained people in response to the tax credit pay them more about 6 months later? The answer: not really, as far as we can tell. It is not known whether, in the longer term, the employees received raises or found better-paying jobs elsewhere.

## Appendix E: Conceptual preliminaries for meta-analysis

### Fixed and random effects

When averaging results from multiple studies, one way to proceed is to mathematically merge the individual studies’ samples. This is called the *fixed-* or *common-effect* method. A study of 1,000 people would get 10 times the weight as a study with 100. Common-effect meta-analysis is grounded in the assumption that all the pooled studies estimate the same thing, such as the impact of TAA on under-30 whites who are laid off from auto plants in Ohio in 2002.

It is often more realistic to assume that different studies measure somewhat different things—in our case, the impact of different kinds of training on different kinds of people in different contexts. The *random-effects* approach accepts this complexity but tames it (usually) by assuming that across various studies the average impact is distributed according to a normal or "bell" curve.[66] The center of this curve is what we will call the random-effects average.[67] In practice, random-effects meta-analysis up-weights small studies. The idea is that all studies are informative as to how impact varies across contexts, so that small studies are valuable out of proportion to their size (Borenstein et al. 2010). The study of 1,000 people gets *less* than 10 times the weight as the study with 100—though it still gets more weight.

## Publication bias

Publication bias arises when the studies producing the biggest impact estimates (or in some cases, the smallest) are more likely to be published and thus to enter a meta-analysis.

A standard way to check for publication bias in meta-analysis is to check for a correlation between a study's average impact estimate and the precision thereof, which depends in part on sample size. The reasoning runs as follows. Imagine a large collection of presidential approval polls with samples ranging in size from 20 to 2,000 respondents. This is a common-effect scenario: all the polls ask the same question of the same population at the same time. Suppose the true approval rating is 50%. Without publication bias, we would expect about as many polls of any given size to overestimate as to underestimate. And because of their imprecision, small polls' results will be scattered more widely around the true number. If publication bias *were* present—if, say, poll results below 45% were dropped—this would filter out more small polls. And that would lift the average among the small polls that reached the public's eyes. Across all published polls, the result would become negatively correlated with the precision.

Researchers can check for this correlation after assembling a collection of studies—something one cannot do when examining a single study. They can perform the check graphically, in a "funnel plot," or numerically using various methods.

Publication bias is especially pernicious in random-effects meta-analysis. For the bias surfaces mostly among the small polls, to which random-effects methods give disproportionate weight.

## Cross-group comparisons

In discussing meta-analyses, we will make cross-group comparisons. Do privately run programs work better than public ones? Do they work better in the Midwest? Even when the individual studies are compelling as to causation—thanks, say, to randomization—comparisons across them are not. If by chance effective approaches were evaluated in New York and not in New Jersey, that does not prove that training works better in New York. Worse, the number of studies in most meta-analyses is small from a statistical point of view. A meta-analysis might include three studies from New York and two from New Jersey.

---

[66]Not all random effects methods assume normality.
[67]Names in the literature include "overall effect," "combined effect," "pooled effect," and "average effect."

For these reasons, cross-group comparisons are worth making, and worth making with care.

## Appendix F: Meta-analysis with persistence term

This table shows results from the “survivor” meta-regressions when the previous-period impact estimate is controlled for. Coefficients on the new “persistence” term appear in the top row. The previous horizon for the medium-term meta-regressions (year 2 after randomization) is the short-term (year 1); for long-term (years 3–5), it is medium-term.

| | Employment | | | | Earnings | | | |
|---|---|---|---|---|---|---|---|---|
| | Medium-term | | Long-term | | Medium-term | | Long-term | |
| **Persistence** | | | | | | | | |
| Impact at previous horizon | 0.18* | 0.15** | 0.48*** | 0.32** | 0.45*** | 0.39*** | 0.52*** | 0.25* |
| | (0.10) | (0.08) | (0.10) | (0.15) | (0.10) | (0.09) | (0.12) | (0.14) |
| **Setting (omitted = South, urban)** | | | | | | | | |
| Randomization year | −0.09* | −0.06 | −0.06* | −0.03 | | | | |
| | (0.05) | (0.04) | (0.03) | (0.04) | | | | |
| Northeast | 3.69** | 3.31*** | | | | | | |
| | (1.31) | (1.16) | | | | | | |
| Midwest | 3.73*** | 3.07*** | −0.07 | −0.02 | | | | |
| | (1.13) | (0.99) | (0.74) | (0.69) | | | | |
| West | 1.95* | 2.04* | | | | | | |
| | (0.98) | (1.08) | | | | | | |
| Unemployment over follow-up (%) | 0.80** | 0.85** | | | 180 | 166 | | |
| | (0.38) | (0.36) | | | (156) | (144) | | |
| **Target population (omitted = welfare or low-income adult)** | | | | | | | | |
| Substance abuse | | | 6.77*** | 7.24*** | | | 3315*** | 3931*** |
| | | | (1.05) | (0.77) | | | (271) | (346) |
| Justice-involved | | | | | | | 1584*** | 2162*** |
| | | | | | | | (263) | (356) |
| Dislocated workers | −3.25** | −2.49*** | −1.27* | −1.52* | | | | |
| | (1.21) | (0.93) | (0.71) | (0.80) | | | | |
| Youth | −2.54 | −3.40*** | | | | | | |
| | (1.48) | (1.32) | | | | | | |
| Disability | 5.43** | 5.58* | 3.56*** | 5.20*** | | | | |
| | (2.45) | (2.87) | (1.15) | (1.53) | | | | |
| **Demographics** | | | | | | | | |
| % male | | | | | | | | 1 |
| | | | | | | | | (4) |
| **Treatment (omitted = mandatory)** | | | | | | | | |
| Classroom training | 0.71 | 0.41 | 0.18 | | | | | |
| | (1.02) | (1.03) | (0.82) | | | | | |
| On-the-job training | | | | | | 263 | | |
| | | | | | | (283) | | |
| Voluntary | | −1.01 | | | | | | |
| | | (0.94) | | | | | | |
| Cost per treated ($1,000s) | | 0.00* | | | | | | |
| | | (0.00) | | | | | | |
| **Employer engagement** | | | | | | | | |
| Employer commits to hiring | | | | | 4693*** | 5074*** | 4651*** | 6854*** |
| | | | | | (1349) | (1214) | (1108) | (1204) |
| **Study characteristics (omitted = outcome data source: official records)** | | | | | | | | |
| Outcome data source: survey | | | | | | | 692*** | 26 |
| | | | | | | | (173) | (358) |
| Multiple imputation | | ✓ | | ✓ | | ✓ | | ✓ |
| $R^2$ | 0.54 | 0.67 | 1.00 | 0.56 | 0.87 | 0.85 | 0.98 | 0.91 |
| Number of experiments | 73 | 87 | 51 | 64 | 100 | 121 | 75 | 81 |

Notes: Standard errors in parentheses. * $p < 0.10$, ** $p < 0.05$, *** $p < 0.01$. REML estimates. MI columns (✓) pooled across 10 imputations via Rubin's rules; remaining columns use complete cases.

# Appendix G: Benefit-cost analysis

In §6.5.3 and §7.3, we compare the costs and benefits of subsidized job training in the US, starting from meta-analytical averages for gross cost of training and the net present value (NPV) of the impacts on pre-tax wages—both expressed in 2025 dollars per member of the treatment group. Here we explain the methodology.

As is standard, we distinguish between three perspectives: that of treatment group members; that of governments, who in this exercise are assumed to fund the training; and that of society as a whole. The social perspective is the algebraic sum of the other two. Thus, in the societal perspective, transfers associated with taxation and public welfare programs cancel out. For simplicity, federal, state, and local governments are lumped together.

We perform the analysis for the training-primary programs, and for the set of sector programs identified in Table 5. We closely follow the benefit-cost methodology in the Schaberg and Greenberg (2020) evaluation of WorkAdvance, except in the estimation of lifetime impacts on pre-tax earnings.

A key ingredient for understanding benefits is: how long do the earnings gains last? Only three experiments in our meta-analysis have produced initially positive earnings impacts and then been followed beyond 6 years, which makes the evidence on long-term persistence quite thin.[68] We estimate lifetime impacts by drawing on research on the persistence of income shocks. Karahan and Ozkan (2013) study the persistence of earnings shocks in the Panel Study of Income Dynamics (PSID). Karahan and Ozkan (2013, Table 1, left panel) estimate that after controlling for age, education, and individual fixed effects, the serial correlation in the residual component of log earnings starts at 0.7 at age 24, rises to 0.98 at age 45, and then declines slightly through age 60—all as modeled by a cubic curve.[69] To apply these multipliers, we assume participants start training at age 27, the average in the sample, and work till age 65. The year 3–5 earnings impact, expressed as a fraction of control group earnings, constitutes the shock that propagates forward. We copy Hendren and Sprung-Keyser (2020) and Mountjoy (2026) in modeling the control group's lifecycle earnings profile using a data series on mean earnings by age from the American Community Survey. The means from the 2015 cross-section are adjusted upward for 0.5%/year real wage growth. Then the entire series is rescaled so that the value for age 31—the midpoint of the year 3–5 period relative to the starting age of 27—matches the year 3–5 control-group mean from the meta-analysis.[70]

[68] The female group in the National Supported Work Demonstration (Couch 1992), Project QUEST (Roder and Elliott 2021), and Per Scholas in the WorkAdvance evaluation (Yusim et al. 2025).

[69] Karahan and Ozkan (2013, Table 1, left panel) reports coefficient estimates for a cubic model for the age-specific serial correlation parameter $\rho$. The best fit is $\rho = 0.7003 + 0.2974h - 0.0978h^2 + 0.0095h^3$, in which $h = (\text{age}-24)/10$. The model is fit on data through age 60, but we extend the spline to age 65.

[70] If $Y_{0^*t^*}$ is the control-group mean earnings at age $t$, and $r$ is the proportional impact in years 3–5, the estimated benefit is $Y_{0^*}[(1+r)^{\tilde{\rho}}-1]$ where $\tilde{\rho} = \Pi_{t=32}^{65}\rho_t$ and $\rho_t$ is modeled as described in note 10.

We discount future impacts with the US government's real long-term cost of funds, as proxied by the market yield on a 30-year TIPS bond, currently 2.82%.[71]

Generally following Schaberg and Greenberg (2020), we count these items:

- *Program cost* is the average treatment-control difference in program spending per treatment group member, for studies providing this information. Only spending that is part of the experiment is counted. A few sources provide costs per *participant*; these are converted to cost per treatment group member using reported participation rates.

- *Reduced spending on other training*. Training provided by an experiment will sometimes substitute for training provided by other sources, including public ones, to this extent reducing aggregate expenditure on training. In our data set, for training-primary studies providing both numbers, this broader definition of net cost averages 20% less than the narrower one above. Just for sector programs, the differential is 11%. This line item factors in that adjustment, by applying one of those discounts.

- *Pre-tax earnings* is the NPV of earnings impacts through age 65, as just described.

- *Fringe benefits, including employer-paid payroll taxes.* All of the studies in our meta-analysis estimate impacts on pre-tax wages. This leaves out impacts on fringe benefits, which are part of the payment for workers' product, as well as a source of return flows to the government, in the form of employer-paid taxes. The ratio of pre-tax wages to total compensation varies by year and profession. According to the Employer Costs for Employee Compensation (ECEC) data set of the Bureau of Labor Statistics, pre-tax wages accounted for 72.4% of total compensation in 2025, in the occupational group "office and administrative support occupations."[72] We take this entry-level occupational category as representative of the jobs trainees graduate into. We estimate the impact on the NPV of fringe benefits as $(1 - 0.724)/0.724$ times the NPV of the impact on pre-tax earnings.

- *Payroll taxes (employee- and employer-paid).* Employee-paid taxes for Medicare and Social Security (FICA) are estimated at the standard 7.65% of pre-tax earnings, which assumes that none of the subjects breach the income limit for the Social Security tax, which is currently $184,500/year. Employer-paid taxes, including for unemployment insurance, are estimated at 6.85% of total compensation, again using ECEC data for "office and administrative support occupations" in 2025. Note that this accounting treats payroll taxes as pure transfers to the government. It does not count the insurance and retirement benefits to which they entitle the employee.

- *Income taxes.* Schaberg and Greenberg (2020) estimates the marginal income tax rate in the Per Scholas and St. Nicks programs, both in New York state, at 19.0%. The rates for Towards Employment and Madison Strategies, in Ohio and Oklahoma respectively, are 15% and 17.6%. All figures include state and federal income taxes. We use the average of the four, 17.2%.

[71] fred.stlouisfed.org/series/DFII30, data point for May 18, 2026.
[72] bls.gov/web/ecec/ecec-civilian-dataset.xlsx, accessed May 25, 2026.

- *Sales taxes*. Following Schaberg and Greenberg (2020), we assume that all take-home pay is spent on goods and services subject to sales tax. Weighting by state population, the national average sales tax as of January 1, 2026, was 5.69%.[73]

- *Use of public benefits.* Declines in the reliance on public benefits, such as food stamps and unemployment insurance, are losses for the trainees and gains for the government. The estimation here is rough. In the WorkAdvance project, the year-2 survey asked people which benefits they were receiving. This allowed the researchers to estimate the impact of training on the probability of any use of each public benefit. The researchers then incorporated separate figures on typical per-beneficiary spending to estimate impacts on use of public benefits in dollar terms. Across the four WorkAdvance sites, the average ratio of benefit reduction to earnings increase is 14.8% (Schaberg and Greenberg 2020, Tables 3.1, 3.3, 3.4, 3.5). We use that ratio to estimate the average reduction in our samples of studies.

- *Work-related spending*. Working more sometimes entails spending more on goods and services that are necessary for work, but otherwise of no value, such as uniforms and bus fare. We again copy Schaberg and Greenberg (2020, p. 77), who estimate the ratio of marginal work-related expenses to marginal pre-tax earnings at 8.34%.

- *Unpaid time.* More time in paid work means less unpaid time, which could mean less leisure, or less time for unremunerated work such as childcare. Citing a study finding that certain clerical workers valued their unpaid time at 58% of their actual pay rate, Schaberg and Greenberg (2020, p. 80) estimates the cost of lost unpaid time at fully 50% of the earnings impact of job training. However, this cost depends in principle on the impacts on employment, not earnings. We therefore depart from Schaberg and Greenberg by factoring in the employment effect. We model and discount the employment impacts through age 65 the same way we do the earnings impacts, which means that the ratio of the "NPV" of the impacts to year 3–5 impacts is the same for employment and earnings.

Unlike Schaberg and Greenberg, we do not factor in the deadweight loss from raising government revenue to pay for training. As Hendren and Sprung-Keyser (2020) point out, the use of any particular deadweight loss multiplier implies a theory about how revenue is raised. We sidestep this question, preferring to separate the analysis of job training from the question of how it is paid for.

---

[73]Combined state and local tax rates from Abir Mandal, "State and Local Sales Tax Rates, 2026," Tax Foundation, taxfoundation.org/data/all/state/sales-tax-rates, accessed May 25, 2026. State populations from U.S. Census Bureau, "State Population Totals and Change: 2020-2025," census.gov/data/tables/time-series/demo/popest/2020s-state-total.html, accessed May 25, 2026.

The cost-benefit analysis is reported in Table G1 for all training-primary programs in our sample, and in Table G2 for the sector programs.

We summarize the costs and benefits with five statistics, which appear at the bottoms of the tables:

- We define a narrow benefit-cost ratio as the impact on total compensation—the sum of rows 3 and 4—divided by the net training cost in row 1. It compares the total increase in labor payments (ignoring general equilibrium effects) to the up-front budgetary cost, and ignores the effects on tax take and provision of other public benefits, including other training. This framing may be salient for legislators who are deciding how to allocate this year's budget and who are less concerned with the tax and benefit impacts because they come far in the future or are the province of other committees. As well, the framing is relevant for private funders since they do not collect taxes or distribute food stamps. The narrow B-C ratio is 1.44 for training-primary programs overall. In light of all the uncertainties in benefit-cost analysis, we view this result as a bit better than break-even. The ratio is 7.18 just for sector programs.

- A broader benefit-cost ratio is the same as the narrow one, except that its denominator incorporates the reduced spending on other training. It is 1.80 for training-primary programs and 8.07 for sector programs.

- Another metric, the Marginal Value of Public Funds (MVPF; Hendren and Sprung-Keyser 2020), does plant itself in the government's perspective, and does factor in all fiscal effects. Here, it is computed as the *net* benefit per treatment group member divided by the net cost to government. Both numbers appear at the bottoms of the main tabulations. For training-primary programs, the MVPF is 2.52. For sector programs, the mathematical ratio is *negative* (–1.98) because the programs pay for themselves thrice over through taxes and benefit savings, and thus do positive good at negative cost. Following Hendren and Sprung-Keyser, we label this as infinite in the table.

- We also compute the ratios of net to gross cost for the government. We estimate that the government recoups about three-quarters of its outlay on the average training-primary program—through higher tax receipts, reduced benefit payments, and reduced spending on other training—leaving a net cost of $0.24 per dollar spent. For sector programs, the ratio is –1.88.

- Finally, since the results depend on the debatable choice of discount rate, we compute the internal rate of return (IRR), that is, the break-even discount rate. The IRR is infinite for beneficiaries since they incur no upfront costs. And it is perhaps less relevant for governments than the previous two indicators, so we compute it just from the societal point of view. The societal IRR is 5.97% for training-primary programs and 34.03% for sector programs.

| | Treatment group member | Government | Society |
|---|---|---|---|
| Program cost | | –13,598 | –13,598 |
| Reduced spending on other training | | 2,778 | 2,778 |
| Pre-tax earnings | 14,146 | | 14,146 |
| Fringe benefits, incl. employer taxes | 5,379 | | 5,379 |
| Payroll taxes (employee & employer) | –2,420 | 2,420 | |
| Income taxes | –2,434 | 2,434 | |
| Sales taxes | –572 | 572 | |
| Use of public benefits | –2,096 | 2,096 | |
| Work-related spending | –1,180 | | –1,180 |
| Unpaid time | –2,511 | | –2,511 |
| Total | 8,312 | –3,299 | 5,013 |
| | | | |
| B/C narrow | 1.44 | | |
| B/C broad | 1.80 | | |
| MVPF | 2.52 | | |
| Government net cost / gross cost | 0.24 | | |
| Societal IRR | 5.97% | | |

**Table G1. Cost-benefit analysis for training-primary programs**

*Notes: All dollar figures in 2025 dollars per treatment group member. B/C is the benefit-cost ratio. "Benefits" are gross compensation, which is the sum of pre-tax earnings and fringe benefits, including employer-paid payroll taxes. "Costs" are gross training costs, which are computed as the simple average from studies providing estimates. MVPF is the marginal value of public funds (Hendren and Sprung-Keyser 2020), computed as the net benefit for the participant divided by the net cost for government. Societal IRR is the real internal rate of return from the societal point of view.*

| | Treatment group member | Government | Society |
|---|---|---|---|
| Program cost | | –11,602 | –11,602 |
| Reduced spending on other training | | 1,289 | 1,289 |
| Pre-tax earnings | 60,319 | | 60,319 |
| Fringe benefits, incl. employer taxes | 22,938 | | 22,938 |
| Payroll taxes (employee & employer) | –10,317 | 10,317 | |
| Income taxes | –10,379 | 10,379 | |
| Sales taxes | –2,441 | 2,441 | |
| Use of public benefits | –8,937 | 8,937 | |
| Work-related spending | –5,031 | | –5,031 |
| Unpaid time | –3,099 | | –3,099 |
| Total | 43,052 | 21,762 | 64,814 |
| | | | |
| B/C narrow | 7.18 | | |
| B/C broad | 8.07 | | |
| MVPF | ∞ | | |
| Government net cost / gross cost | –1.88 | | |
| Societal IRR | 34.03% | | |

**Table G2. Cost-benefit analysis for sector programs**
*Notes: See notes to previous table.*

These benefit-cost calculations are subject to several limitations. First, we assume that the income gains from these programs decay slowly, following the literature on income shocks (Karahan and Ozkan 2013). This matters less for sector programs, since they pay for themselves even looking only at the years where subjects are tracked. But for the training-only programs, the bulk of the benefit accrues in the extrapolation beyond the studies' follow-up windows. Without that persistence, they would have a benefit-cost ratio under 1. When available, the evidence does find persistent gains: among the 28 training-primary estimates with a medium-term gain of at least $200 and a longer follow-up (typically years 2 and 4), long-term impacts totaled 110% of medium-term impacts, with a median ratio of 0.90.

Second, jobs have benefits outside of wages. They might decrease crime or improve mental health. Including these would increase the benefits side of the ledger, but we opt not to because the research is so uncertain on this. Finally, we assume away general equilibrium or displacement effects. If trainees displace other workers seeking the same jobs, the benefits we measure are inflated.